\documentclass[12pt,preprint]{aastex}

\usepackage{natbib}
\newcommand{\um}{$\micron$}
\newcommand{\halpha}{H$\alpha$}
\newcommand{\hbeta}{H$\beta$}
\newcommand{\msun}{M$_\odot$}
\newcommand{\sfr}{M$_\odot$\,yr$^{-1}$}

\shorttitle{11 Mpc H$\alpha$ Survey}
\shortauthors{Kennicutt et al.}

\begin{document}


\title{Dust-Corrected Star Formation Rates of Galaxies. I. 
Combinations of H$\alpha$ and Infrared Tracers}


\author{Robert C. Kennicutt, Jr.\altaffilmark{1,2}, 
Cai-Na Hao\altaffilmark{1}, Daniela Calzetti\altaffilmark{3}, 
John Moustakas\altaffilmark{4,5}, Daniel A. Dale\altaffilmark{6},
George Bendo\altaffilmark{7}, Charles W. Engelbracht\altaffilmark{2},
Benjamin D. Johnson\altaffilmark{1}, and Janice C. Lee\altaffilmark{8}
}


\altaffiltext{1}{Institute of Astronomy, University of Cambridge, 
Madingley Road, Cambridge CB3 0HA, UK}
\altaffiltext{2}{Steward Observatory, University of Arizona, Tucson, AZ 85721, USA}
\altaffiltext{3}{Department of Astronomy, University of Massachusetts, Amherst, MA 01003, USA}
\altaffiltext{4}{Center for Astrophysics and Space Sciences, University of California,
San Diego, 9500 Gilman Drive, La Jolla, California, 92093, USA}
\altaffiltext{5}{Center for Cosmology and Particle Physics, New York University, 
4 Washington Place, New York, NY 10003, USA}
\altaffiltext{6}{Department of Physics, University of Wyoming, Laramie, WY 82071, USA}
\altaffiltext{7}{Blackett Laboratory, Imperial College London,  Prince Consort Road
London, England SW7 2AZ, UK}
\altaffiltext{8}{Carnegie Observatories, 813 Santa Barbara Street, Pasadena, CA 91101, USA}


\begin{abstract}

We combine \halpha\ emission-line and infrared continuum measurements of
two samples of nearby galaxies to derive dust attenuation-corrected star formation
rates (SFRs).  We use a simple energy balance based
method that has been applied previously to HII regions in the Spitzer
Infrared Nearby Galaxies Survey (SINGS), and extend the methodology to
integrated measurements of galaxies.  We find that our composite 
\halpha\ $+$ IR based SFRs are in excellent agreement with attenuation-corrected
SFRs derived from integrated spectrophotometry, over the full range of
SFRs (0.01 -- 80 \msun\,yr$^{-1}$) and attenuations (0 -- 2.5 mag) studied.  
We find that the combination of \halpha\
and total infrared luminosities provides the most robust SFR measurements,
but combinations of \halpha\ measurements with monochromatic luminosities
at 24\,\um\ and 8\,\um\ perform nearly as well.  The calibrations differ
significantly from those obtained for HII regions (Calzetti
et al. 2007), with the difference attributable to a more evolved population
of stars heating the dust.  Our results are consistent with a significant
component of diffuse dust (the `IR cirrus' component) that is heated by
a non-star-forming population.  The same methodology can
be applied to [\ion{O}{2}]$\lambda$3727 emission-line measurements, and the radio continuum
fluxes of galaxies can be applied in place of IR fluxes when the latter
are not available.  We assess the precision and systematic reliability
of all of these composite methods.  

\end{abstract}




\section{INTRODUCTION}

Interstellar dust absorbs and reprocesses approximately half of
the starlight in the Universe (Lagache et al. 2005), and this
extinction introduces the largest source of systematic error
into measurements of star formation rates (SFRs) in galaxies
(Kennicutt 1998a, hereafter denoted as K98, and references 
therein).  In present-day 
disk galaxies the typical dust attenuations\footnote{We adopt the convention 
of using the term extinction to describe the combined
effect of scattering and absorption of starlight by intervening dust grains,
and the term attenuation to describe the net reduction of starlight in an
extended source such as a galaxy, in which stars and dust are mixed in a
complex geometry. Attenuation includes the effects of differential embedding of
stars within the dust distribution as well as absorption of starlight by dust,
with part of the scattered radiation being returned to the observer.
} are of order 0--2 mag in H$\alpha$ 
and 0--4 mag in the non-ionizing ultraviolet continuum ($\sim 130 - 250$ nm), the most
commonly used SFR tracers
(e.g., Kennicutt 1983; Brinchmann et al. 2004; Buat et al. 2005).
The resulting systematic
error in the overall SFR scales can be largely removed by
applying a statistical correction for the dust attenuation
(e.g., Kennicutt 1983; Calzetti et al. 1994, 2000).
However the attenuation within and between individual galaxies 
varies from virtually zero to several magnitudes (K98 
and references above), so even if the systematic effect
is removed SFRs for individual galaxies will suffer typical random
errors of a factor of two or more.  Emission-line 
diagnostics such as the Balmer
decrement or the Pa\,$\alpha$/H$\alpha$ ratio correct
the H$\alpha$ SFR measurements for attenuation (e.g., 
Kewley et al. 2002; Brinchmann et al. 2004; Moustakas et al.
2006), but such spatially-resolved spectrophotometry is difficult
to obtain and not widely available.  Likewise at ultraviolet (UV)
wavelengths the slope of the continuum can be
used to estimate an attenuation correction,
but the large scatter in the attenuation vs UV color relation
makes such corrections highly uncertain (e.g., Meurer et al. 1999; Kong et al.
2004; Dale et al. 2007; Johnson et al. 2007a; Cortese et al. 2008).

An alternative approach to estimating the SFR in very dusty 
galaxies is to measure one or more components of the mid-
and far-infrared emission.  In the limit of complete
obscuration the dust will re-emit the bolometric luminosity
of the embedded stars, and if young stars produce most of
the integrated starlight the re-emitted
infrared (IR) luminosity will effectively provide a bolometric
measure of the SFR (K98).  This combination of a high dust
optical depth and a young star dominated radiation field
is often satisfied in the most actively star-forming galaxies
in the Universe, the luminous infrared (LIRG) and 
ultraluminous infrared galaxies (ULIRGs).  Consequently most
of our knowledge of the star-formation properties of 
IR-luminous starburst galaxies comes from measurements
of the IR dust continuum emission.  Although the 
total integrated IR emission of galaxies should
provide the most robust measure of the dust-enshrouded
SFR, calibrations based at specific wavebands such as
the rest 8\,\um\ and 24\,\um\ emission have been derived
by various authors (e.g., Calzetti et al. 2005,2007; Wu et al. 2005; 
Alonso-Herrero et al. 2006; P\'erez-Gonz\'alez et al. 2006; Rela\~no et al. 2007;
Bavouzet et al. 2008; Zhu et al. 2008; Rieke et al. 2009).  

Although the dust emission by itself can provide a reliable
measure of the SFR in the most obscured circumnuclear 
starbursts, its application as a quantitative SFR tracer
in normal galaxies suffers from its own
set of systematic errors.  Since the typical attenuation
of young starlight in normal galaxies is only of order
1 mag, this implies that roughly half of the starlight
is {\it not} processed by dust, and this in turn will tend
to cause the IR emission to systematically under-estimate
the SFR.  The problem is much more
severe in gas- or metal-poor environments such as dwarf galaxies and the outer disks of
galaxies, where the fraction of obscured star formation
can become negligible (e.g., Wang \& Heckman 1996; Bell 2003).  
Another, competing systematic effect is the 
contribution to dust heating by more evolved stars, sometimes
referred to as the ''infrared cirrus" problem (K98
and references therein).  The effects of finite dust opacity
and cirrus contamination tend to be of 
roughly comparable magnitude in typical massive star-forming
spiral galaxies, and thanks to this coincidence the SFRs derived from IR
measures are often surprisingly consistent with those derived
from other methods (e.g., Sauvage \& Thuan 1992; Kewley et al. 2002).
However this coincidence breaks down in other types of
galaxies.

Since the starlight removed at short wavelengths by the
interstellar dust is re-radiated in the IR, it should
be possible to calibrate a much more robust set of 
attenuation-corrected star formation tracers by combining
observations in the UV or visible with those in
the infrared.  In recent years this approach has been 
applied widely using the combination of UV and IR
observations to estimate dust-corrected SFRs 
(e.g., Gordon et al. 2000; Bell 2003;
Hirashita et al. 2003; Iglesias-P\'aramo et al. 2006;
Cortese et al. 2008),
and has been explored for \halpha\ and IR-based
SFR measurements by Inoue (2001), Inoue et al. (2001),
and Hirashita et al. (2001, 2003).
However the absence of an independent set of attenuation-corrected
SFRs has made it difficult to assess the reliability of
these corrected SFRs, and to calibrate the relation between
effective UV attenuation and the ratio of IR to UV luminosities.

The advent of large sets of multi-wavelength observations
of nearby galaxies now provides us with the opportunity
to derive attenuation-corrected H$\alpha$ and UV continuum
luminosities of galaxies by combining these fluxes with
various components of the IR emission.  Moreover the
availability of integrated optical spectra (and in some
cases Pa\,$\alpha$ maps) of the same galaxies allows us
to derive additional attenuation estimates, and
test the precision and systematic reliability of the
respective attenuation-corrected SFR measurements.  

The Spitzer Infrared Nearby Galaxies Survey
(SINGS; Kennicutt et al. 2003) offers an ideal dataset
for testing and calibrating such multi-wavelength SFR
estimators.  The survey includes imaging of a diverse
sample of 75 galaxies within 30 Mpc, with wavelength
coverage extending from the UV to the radio, including
ultraviolet imaging at 150 and 230 nm, H$\alpha$, 
and 7 infrared wavelengths over 3.6 -- 160\,\um.
In addition, drift-scanned spectra over the 
wavelength range 3600 -- 6900\,\AA\ are available, which
complement matching infrared spectra over the range
10 -- 40\,\um.  We first applied these data to 
calibrate and test the combined use of H$\alpha$
and 24\,\um\ infrared fluxes of individual HII regions
to derive attenuation-corrected emission-line fluxes
(Calzetti et al. 2007, hereafter denoted C07; Kennicutt et al. 2007;
Prescott et al. 2007).  These studies revealed that the
Spitzer 24\,\um\ sources were highly correlated in position
and flux with those of optical HII region counterparts.
Kennicutt et al. (2007)
and C07 found that the ratio of 24\,\um\ to \halpha\ fluxes
yielded attenuation-corrected \halpha\ luminosities and attenuation
values that were consistent with those derived from  Pa$\alpha$/H$\alpha$
reddening measurements, extending over ranges of $>$3 mag in A(\halpha)
and more than 4 orders of magnitude in ionizing luminosity and
\halpha\ surface brightness, albeit with a large 
random error (approximately $\pm$0.3 dex) that 
probably arises from variations in stellar population
and dust geometry, as discussed later.

In this paper we now explore whether the same approach
of combining H$\alpha$ and IR observations can be
used to derive reliable attenuation-corrected SFRs of
entire galaxies.  Our analysis combines SINGS observations
with integrated spectra and IR observations of a larger
sample of galaxies from Moustakas \& Kennicutt (2006).
We find that excellent estimates of attenuation-corrected
SFRs can be derived by combining H$\alpha$ emission-line
luminosities with 24\um, total infrared (TIR), and even
8\um\ IR luminosities.  We provide prescriptions for
each of these multi-wavelength SFR tracers, and also 
test other combinations of tracers, including combinations
of [\ion{O}{2}]$\lambda$3727 and IR luminosities, and combinations
of emission-line and radio continuum luminosities.  
A companion paper (Hao et al. 2009, hereafter denoted
Paper II) uses the same galaxies to empirically calibrate
the combination of UV and IR luminosities to measure 
attenuation-corrected SFRs.

The remainder of this paper is organized as follows.
In \S 2 we describe the multi-wavelength dataset that
was compiled for this analysis.  In \S 3 we describe
our method for combining \halpha\ and IR fluxes of galaxies
to estimate the \halpha\ attenuation, and discuss the
assumptions and limitations that underlie the method. 
In \S 4 we use the 
integrated 24\,\um, TIR, and \halpha\ fluxes of the
galaxies in our sample to calibrate the method,
and test for systematic dependencies of
the results on the spectral energy distributions (SEDs),
star formation properties, and physical properties
of the galaxies.  In \S 5 we extend the approach to
other composite SFR indicators, including combinations
of optical emission lines with 8\,\um\ (PAH-dominated)
IR luminosities and radio continuum luminosities.
In \S 6 we use the results to re-examine the systematic
reliability of the methods. We summarize our results in \S 7.

\section{DATA}

\subsection{Galaxy Sample}

Our basic approach is to compare and calibrate
global attenuation estimates for galaxies derived from
the combination of H$\alpha$ and IR luminosities of
galaxies with independently derived attenuation measures 
from integrated optical emission-line spectra.  Our samples
are drawn from two surveys of nearby galaxies,
the SINGS survey (Kennicutt et al. 2003) and a survey
of integrated spectrophotometry of 417 galaxies by  
Moustakas \& Kennicutt (2006; hereafter denoted   
MK06).\footnote{During the early phases of our study our
sample also included the Nearby Field Galaxy Survey (Jansen   
et al. 2000).  However after the sample was vetted
for minimum signal/noise and matching IR data only
a handful of galaxies remained.  As a result we chose to
restrict our analysis to the SINGS and MK06 samples.}
The SINGS sample is comprised of 75 galaxies with distances 
less than 30 Mpc, which were chosen to span wide ranges
in morphological type, luminosity, and dust opacity.
The MK06 survey was designed to include the full range
of optical spectral characteristics found in present-day
galaxies, and includes a subsample of normal galaxies as well as
large subsamples of optically-selected starburst galaxies
and infrared-luminous galaxies.  The combined sample
includes objects ranging from dwarf irregular galaxies
to giant spirals and IR-luminous galaxies 
($-13.4 \ge M_B \ge -22.4$), with SFRs of $\sim$0.001 
-- 100 M$_\odot$\,yr$^{-1}$,
and $\sim$0.01 $< {\rm L}_{IR}/{\rm L}_B < $100.  This diversity
is important for testing the applicability limits for our
methods, and for uncovering any second-order dependences
of our results on properties of the galaxies or their SEDs.
Detailed information about the SINGS and MK06 samples
can be found in the respective survey papers.

Our analysis draws on subsets of these samples which
satisfy a number of further selection criteria.
Galaxies showing no detectable star formation
as measured at H$\alpha$ were excluded; most of those were
early-type E and S0 galaxies.  Likewise we required that
the galaxies had well-measured infrared fluxes or
strong upper limit fluxes from {\it Spitzer} (SINGS sample)
and/or {\it IRAS} (for both samples), as discussed further in
\S 2.4.  We also applied a
signal/noise requirement on the optical spectra, to
ensure an accurate measurement of the Balmer decrement
(H$\alpha$/H$\beta$ ratio).  This was done by a combination
of formal signal/noise estimates and visual inspection
of the continuum-subtracted spectra at H$\beta$.  This
translated to a minimum $S/N \sim 15$ at H$\beta$.
Finally, we separated galaxies with spectra dominated by
star formation from those with bright active galactic
nuclei (AGN-dominated) or composites of star formation and
AGN signatures, so their behavior could be analyzed separately.
This separation was performed using the optical emission-line
spectra, using the criteria of Kewley et al. (2001) and Kauffmann et al. (2003)
criteria as described in Moustakas et al. (2006).  

Application of these minimum signal/noise criteria in
IR photometry, \halpha\ photometry, and spectroscopic
$S/N$ yielded subsamples of 58 SINGS galaxies (including
4 AGN-dominated galaxies, NGC\,3190, NGC\,4550, NGC\,4569, and NGC\,4579),
and 147 galaxies from MK06 (including
3 AGN-dominated galaxies and 31 with composite nuclei).
The galaxy population retains the mixture of the parent
samples, with normal spiral and irregular galaxies and 
especially strong representation of UV, blue, and IR-selected
starburst galaxies.  

This combination of selection criteria (adequate signal/noise in
the infrared continuum and Balmer emission lines, emission-line
spectra dominated by star formation) was applied to provide
an accurate calibration of the attenuation-correction methods that
are developed in this paper.  However it is important to bear in
mind that these criteria tend to favor the selection of galaxies
with significant SFRs per unit mass, typically intermediate to
late-type spiral galaxies and luminous irregular and starburst 
galaxies.  Objects that tend to be excluded include early-type
(E, S0, Sa) galaxies with weak line emission, dwarf irregular
galaxies with little or no detectable dust emission, and the
most highly obscured infrared luminous and ultraluminous galaxies,
which tend to exhibit weak H$\beta$ line emission and (frequently)
AGN signatures in their spectra.  Some of these selection effects
are unimportant, for example the absence of IR-weak galaxies from
the sample is irrelevant because no attenuation correction is 
required in those cases.  However the very early-type galaxies
and extreme starbursts may probe star formation and attenuation 
regimes that fall outside of the valid range of our calibrations.
We discuss the potential effects
of these selection effects further in \S 6.

\subsection{Optical Emission-Line Spectra and Balmer Decrements}

Integrated H$\alpha$ fluxes and H$\alpha$/H$\beta$ flux ratios
for the MK06 sample were taken directly from the 
integrated spectra.  The galaxies were observed with the 
B\&C Spectrograph on the Steward Observatory Bok 2.3\,m 
telescope using a long-slit drift-scanning technique, which
produced an integrated spectrum over a rectangular aperture
that covered most or all of each galaxy.  The resolution
of the spectra ($\sim$8\,\AA) is sufficient to deblend
H$\alpha$ from the neighboring [NII]$\lambda\lambda$6548,6584
lines.  The underlying continuum and absorption-line
spectra were fitted and removed as part of the emission-line
flux measurements, which removes the effects of stellar
Balmer absorption on the H$\alpha$ and H$\beta$ measurements 
(see MK06 for details).

Drift-scanned spectra were also obtained for the SINGS sample,
using the same instrument and setup as MK06 for northern galaxies, 
and the R-C spectrograph on the CTIO 1.5\,m telescope
for a handful of galaxies that could not be reached from Kitt Peak.  
These observations were configured to match the spatial coverage
of the mid-IR spectral maps obtained as part of SINGS.  Mapping
entire galaxies was not practical with the {\it Spitzer} Infrared
Spectrograph (IRS), so we mapped representative sub-regions of
the galaxies instead.  One set of optical observations consisted of
driftscans covering an area of 0\farcm9 $\times$ 3\farcm3
(0\farcm9 $\times$ 7\arcmin\ for CTIO) oriented to 
coincide with the Spitzer IRS low-resolution maps, with multiple 
observations laid end to end to extend from the 
centers of the galaxies to $R \ge 0.5\,R_{25}$
(de Vaucouleurs et al. 1991).  A summed spectrum was
then extracted by integrating over this 0\farcm9-wide
strip.  The areal coverage of these driftscan apertures
ranged from $<$10\%\ of the total projected region within
$R_{25}$ for the largest galaxies in the sample (e.g., M81),
to nearly 100\%\ for the smallest galaxies in the sample.
We also obtained a separate drift-scanned
spectra for the optical centers of each galaxy, using
20\arcsec\ $\times$ 20\arcsec\ apertures, to 
approximately match the coverage of a set of high spectral resolution
{\it Spitzer} IRS observations.  Again the two-dimensional
spectra were collapsed into a single integrated spectrum covering 
the central aperture.  A third set of (pointed) spectra just covering
the galactic nuclei were also obtained.  Those data are not
used in this paper, apart from helping to classify the nuclear
emission types, and checking to ensure that AGN emission does
not dominate the other, larger-aperture spectra.  
The processing and emission-line extractions for these
data followed the procedures described in MK06 and
Moustakas et al. (2009), and the data can be found in 
the latter paper.

For some of the SINGS galaxies the drift-scanned spectra could not
be used to reliably estimate the disk-averaged H$\alpha$/H$\beta$
ratio, either because the H$\beta$ line was too weak or because the 
spatial coverage of the spectra was too limited.  For 23 of
these cases spectra for individual HII regions were available from
the literature, and we used these to estimate the disk-averaged 
H$\alpha$/H$\beta$ ratio and its uncertainty.
We checked the validity of this procedure by comparing the mean
Balmer decrements with those derived from drift-scanned spectra
when both types of data were available, and they yield consistent
results with no systematic bias, though with less precision than
from the integrated spectra.  

Table 1 lists the adopted H$\alpha$/H$\beta$ ratios (and data sources)
for the SINGS galaxies, and Table 2 lists the adopted H$\alpha$
and H$\beta$ fluxes for the MK06 galaxies.\footnote{Full machine-readable
versions of these tables can be found in the electronic version of the ApJ.}  

\subsection{Integrated H$\alpha$ Fluxes}

Integrated H$\alpha$ fluxes for the MK06 sample were taken
directly from the drift-scanned integrated spectra.  
The same applied to the observations of the central 
20\arcsec\ $\times$ 20\arcsec\ regions in the SINGS galaxies.
In many cases integrated H$\alpha$ fluxes for the SINGS
galaxies had to be obtained from other sources, because 
of the incomplete spatial coverage of the drift-scanned spectra.
Fluxes were compiled from the literature or were 
measured from H$\alpha$ narrow band
images obtained as part of the SINGS project.\footnote{SINGS
data products including \halpha\ images can be found at 
http://irsa.ipac.caltech.edu/data/SPITZER/SINGS/.}  Most of
these measurements include emission from the neighboring 
[NII]$\lambda\lambda$6548,6584 lines.  We applied
[NII] corrections on a galaxy by galaxy basis, using
the [NII]/H$\alpha$ ratio measured in our drift-scanned
spectra (26 galaxies), or individual measurements of disk HII regions
in the galaxies from the literature (19 galaxies).  For an
additional 13 galaxies we applied the mean relation between 
average [NII]/H$\alpha$ and M$_B$
in the MK06 sample, as published in Kennicutt et al. (2008).

The accuracy of the individual H$\alpha$ fluxes varies considerably,
depending on the absolute and relative 
strength of the line (relative to continuum) and the source
of the fluxes.  A comparison of MK06 H$\alpha$ fluxes with
high-quality narrow-band imaging measurements by Kennicutt
et al. (2008) shows that the average uncertainty of the
fluxes is approximately $\pm$10\% -- 15\%.  This uncertainty
could increase to as much as $\pm$30\% for galaxies with
weak emission or with a strong underlying continuum.
Table 1 lists the adopted \halpha\ fluxes and  
[NII]/H$\alpha$ ratios for the SINGS sample. 

\subsection{Infrared Fluxes}

For the SINGS sample we have used the integrated {\it Spitzer}
measurements from Dale et al. (2007) at wavelengths of 3.6,
8.0, 24, 70, and 160\,\um.  We also used the 
SINGS images (Version DR5) to measure 8\,\um\ and 24\,\um\ fluxes with
the same 20\arcsec\ $\times$ 20\arcsec\ circumnuclear apertures
that were measured spectroscopically (\S 2.2, 2.3).  We did not
measure 70\,\um\ or 160\,\um\ fluxes for the circumnuclear 
apertures, because the instrumental beam sizes at
those wavelengths were comparable to or larger than the 
apertures.  We refer the reader to Dale et al.
(2007) for a detailed listing and discussion of the
uncertainties in these fluxes.  Aperture corrections
(including extended source corrections for scattering of
diffuse radiation across the IRAC focal plane)
were applied to the circumnuclear measurements, and
amount to approximately 10\%\ at 8\,\um\ and 25\%\ at
24\,\um.

Most of the galaxies in the MK06 sample have not been
observed with {\it Spitzer}, so we compiled integrated
fluxes at 25, 60, 100\,\um\ from the {\it IRAS} survey.
Whenever possible we adopted in order of priority the fluxes measured in
the {\it IRAS} Revised Bright Galaxy Sample (Sanders et al. 
2003), followed by {\it IRAS} Bright Galaxy Sample (Soifer et al. 1989),
then the large optical galaxy catalog (Rice et al. 1988) 
and finally Version 2 of the {\it IRAS} Faint Source Catalog (Moshir et al. 1990).  
For this study we are mainly interested in the 25\,\um--band fluxes and 
the TIR fluxes.  Many galaxies in the sample are 
undetected or have only marginal detections at 25\,\um,
and we restricted our analysis to galaxies with $>$3-$\sigma$
detections at that wavelength.  As a supplement to the
{\it Spitzer} measurements we also compiled the same 
{\it IRAS} data for the SINGS sample.  Of the 58
galaxies in our main {\it Spitzer} sample, 46 were observed
by {\it IRAS} and satisfy our 25\,\um\ signal/noise criterion.

A key SFR index in our analysis is the wavelength-integrated
total infrared (TIR) luminosity.  The {\it IRAS} and 
{\it Spitzer} photometry do not cover enough wavelengths
to uniquely define this flux integral, with the emission longward
of 100\,\um\ and 160\,um, respectively, being especially poorly
constrained.  As a result there are numerous prescriptions in
the literature for computing a TIR flux from the {\it IRAS}
and {\it Spitzer} band fluxes.  For the sake of consistency
we have adopted the definition of TIR flux from Dale \& Helou (2002),
which is the bolometric infrared flux over the wavelength range
3---1100\,\um.  We also adopt the semi-empirical prescriptions 
from that paper for estimating $f(TIR)$ from weighted sums of
MIPS 24, 70, and 160\,\um\ fluxes (equation [4] of their paper),
and from IRAS 25, 60, and 100\,\um\ fluxes (equation [5] of their paper).

It is important to test the consistency of the {\it IRAS} and {\it Spitzer}
flux scales, and since the SINGS sample was measured with both sets
of instruments they provide a direct standard for this comparison.  
We first compared the consistency of the {\it Spitzer} MIPS photometry
at 24\,\um\ with the {\it IRAS} 25\,\um\ fluxes, using the common
definition of monochromatic flux 
f($\lambda$) = $\nu {\rm f}_\nu$.
The left panel of Figure 1 shows the ratio of MIPS/{\it IRAS} fluxes
as a function of 24\,\um\ flux.  Overall the flux scales are in excellent
agreement, with an average ratio $f(MIPS)$/$f(IRAS)$ = 0.98 $\pm$ 0.06
after the 3-$\sigma$ outliers were excluded.  The three outliers are 
faint and/or low surface brightness IR emitters with single published
{\it IRAS} flux measurements.
Apart from these isolated examples we find that the MIPS 24\,\um\ 
and {\it IRAS} 25\,\um\ fluxes can be used interchangeably.
For the sake of simplicity we shall use the terms ``24\,\um\ fluxes"
and ``24\,\um\ luminosities" to refer to measurements made either at 24\,\um\ or 25\,\um.

The right-hand panel of Figure 1 shows the ratio of MIPS and {\it IRAS}
TIR fluxes.  Both sets of fluxes were computed using the
prescriptions of Dale \& Helou (2002).  Here the deviations are larger, with a scatter of 
$\pm$23\%\, and a systematic offset of 24\%\ in flux scales (MIPS larger).
This offset is considerably larger than the uncertainty given by Dale \& Helou (2002).
To better understand this discrepancy we compared our flux ratios 
f(TIR)$_{MIPS}$/f(TIR)$_{IRAS}$ as a function of FIR color  
f$_\nu$(60\,\um)/f$_\nu$(100\,\um), as shown in Figure 2.  
For galaxies with warm IR colors (high f$_\nu$(60\,\um)/f$_\nu$(100\,\um) ratios),
the SED peaks near or shortward of 100\,\um, so the 25, 60, and 100\,\um\
IRAS fluxes provide a relatively reliable estimate of the integrated
TIR flux.  However for galaxies with colder colors [$f_\nu(60\,\um)/f_\nu(100\,\um) < 0.4$],
the peak of the SED is longward of 70\,\um, so there is more
of a danger of systematic uncertainties affecting the TIR flux estimates.
The solid line in Figure 2 shows the approximate magnitude of this
error predicted by the Dale \& Helou (2002) SED models.  The actual
comparison for the SINGS galaxies displays the sense of this predicted
trend, but the systematic differences between MIPS and {\it IRAS}
measurements is more severe.  This comes about because the SINGS
sample included a number of galaxies with colder dust than any of
the galaxies in the template reference sample explored by Dale 
\& Helou (2002).  For that reason we analyze the {\it Spitzer} and 
{\it IRAS} TIR data separately in the analyses that follow. 

\subsection{Radio Continuum Fluxes}

In \S 5.3 of this paper we explore whether the integrated
(mainly non-thermal) radio continuum fluxes of galaxies can
be combined in the same way with optical
emission-line fluxes to provide attenuation-corrected SFRs.
Radio continuum fluxes at 1.4\,GHz for the SINGS sample were taken
from the compilation of Dale et al. (2007).  To obtain a
matching set of radio data for the MK06 sample we cross-correlated
the galaxies with the NRAO VLA Sky Survey (NVSS; Condon et al. 1998).
After experimenting with several choices of matching radii we used
a radius of 20\arcsec, which provided the optimal yield of matched
sources without introducing significant numbers of spurious matches
with background sources.  This produced radio fluxes for 100 of the
113 star-forming galaxies in the MK06 sample.  We also used pointed
observations of 32 galaxies from Condon (1987), and adopted them in
preference to the NVSS when both sets of measurements were available, to 
minimize problems with missing extended flux in the NVSS data
(Yun et al. 2001).  

\section{INTEGRATED H$\alpha$ $+$ IR INDICES}

Our method for using linear combinations of \halpha\ and IR
(specifically, 24\,\um) fluxes to derive attenuation-corrected
\halpha\ luminosities was introduced in previous SINGS papers
by Calzetti et al.\ (2007), Prescott et al.\, (2007), and
Kennicutt et al.\,(2007), and applied to measurements of 
HII regions and HII region complexes.  Here we describe
the physical basis for the method in more depth, and adapt
it for measurements of galaxies, where there are contributions
to dust heating from a much wider range of stellar ages than
in HII regions.

\subsection{Methodology}

Our method uses the IR luminosities
of galaxies to estimate attenuation corrections for their
optical emission-line (or ultraviolet continuum) luminosities.
Its physical basis is a simple energy balance argument
(see Inoue et al. 2001 and Hirashita et al. 2003 for 
similar approaches to this problem).
We assume that on average the attenuated luminosity in 
H$\alpha$ is re-radiated in the infrared, with a scaling
factor that is calibrated empirically, and will differ depending
on the IR emission component used and secondarily on the
nature of the dust-heating stellar population.  Following
Kennicutt et al. (2007), we can construct a linear combination
of the observed H$\alpha$ and IR luminosities that reproduces
the true unattenuated H$\alpha$ luminosity: 

\begin{equation}
L(H\alpha)_{corr} = L(H\alpha)_{obs} + a_{\lambda} L(IR)
\end{equation}

\noindent
where $L(H\alpha)_{corr}$ and $L(H\alpha)_{obs}$ denote the 
attenuation-corrected and observed H$\alpha$ luminosities, 
respectively, $L(IR)$ represents the IR luminosity
over a given wavelength bandpass, and $a_{\lambda}$ is the
appropriate scaling coefficient for that wavelength band
and dust heating population.  The equivalent attenuation
correction for H$\alpha$ is simply:

\begin{equation}
A(H\alpha) = 2.5 \log [{1 + {{a_{\lambda} L(IR)} \over {L(H\alpha)_{obs}}}}].
\end{equation}

Equations (1) and (2) represent simple approximations to a
much more complicated dust radiative transfer process in galaxies
(e.g., Witt \& Gordon 2000, Charlot \& Fall 2000),
but their physical motivation can be readily understood as follows.
We first define a scaling
factor $\eta$ that represents the fraction of the bolometric
luminosity of a stellar population that is reprocessed
as H$\alpha$ emission in a surrounding ionization-bounded HII
region:  

\begin{equation}
L(H\alpha)_{corr} = \eta L_{bol}
\end{equation}

\noindent
where as before $L(H\alpha)_{corr}$ represents the intrinsic
(extinction-free) \halpha\ luminosity and $L_{bol}$ represents
the bolometric stellar luminosity of the same region.
Considering now explicitly the attenuation of the 
emission line, its attenuated luminosity can be expressed as:

\begin{equation}
L(H\alpha)_{obs} = L(H\alpha)_{corr}\, {\rm e}^{-{\tau_{\lambda}}} 
\end{equation}
 
\noindent
The attenuation in magnitudes $A(\lambda) = 1.086\,\tau_{\lambda}$.
The IR luminosity for the region (consider for example the total
wavelength-integrated IR luminosity) likewise can be expressed as:

\begin{equation}
L(TIR) = L_{bol}\, (1 - {\rm e}^{{-\overline{\tau}}})
\end{equation}

\noindent
where the relevant opacity in this case is the effective 
(absorption and luminosity-weighted) opacity of the starlight heating the
dust.  

\begin{equation}
\bar{\tau}= - \ln \frac{\int L_\lambda e^{-\tau_\lambda} d\lambda} 
{\int L_\lambda  d\lambda}
\end{equation}

\noindent
This mean opacity depends on the wavelength dependence of the attenuation,
the spectral energy distributions of the stars that heat the dust, and
the dust covering factors for these stellar populations.  In general
it will {\it not} necessarily be the same as the dust opacity $\tau_\lambda$
for the emission line of interest.  We can substitute eqs.\, (4) 
and (5) into eq.\, (3) to eliminate $L_{bol}$ and express
$L(H\alpha)_{corr}$ in terms of the observable luminosities
$L(H\alpha)_{obs}$ and $L(TIR)$ and the opacities.  It is
convenient to introduce a scaling parameter $\beta$, the ratio 
of the emission-line opacity to effective mean stellar opacity
for the stars heating the dust:

\begin{equation}
\beta \equiv {{\tau_\lambda} \over {\overline{\tau}}}
\end{equation}

\noindent
then eq. (1) takes the modified form:

\begin{equation}
L(H\alpha)_{corr} = L(H\alpha)_{obs} + \eta L(TIR)\, {{(1 - {\rm e}^{-\beta \overline{\tau}}}) \over {(1 - {\rm e}^{-\overline{\tau}}})}
\end{equation}

\noindent
In detail the opacity index $\beta$ depends on several parameters, 
most importantly the wavelength of the optical (or UV) SFR tracer,
but also on the overall stellar population mix and the distribution
of dust opacities for different age populations, all which will
influence $\overline{\tau}$.

These relations take a much simpler form in the special situation where
$\tau_{H\alpha} \simeq \overline{\tau}$ (i.e., $\beta \simeq 1$).  
In that case the opacity
term in eq. [8] drops out, and we are left with the simple result:

\begin{equation}
L(H\alpha)_{corr} = L(H\alpha)_{obs} + \eta L(TIR)
\end{equation}

\noindent
the same as eq. (1) with coefficient $a_\lambda = \eta$, the bolometric
correction term for $L(H\alpha)$.  If this approximation holds then
the attenuated-corrected emission-line luminosity can be derived from
a linear combination of the observed \halpha\ luminosities and
the TIR (or other IR) luminosity.  

Thanks to fortuitous physical circumstances this approximation
is valid for the \halpha\ emission line.  Although the line
is emitted in the red (0.6563\,\um), observations show that the 
dust attenuation of galaxies in the Balmer lines is approximately 2.3 times higher than
the corresponding attenuation in the stellar continuum (see
Calzetti 2001 and references therein).  As a result the attenuation
of the \halpha\ line is comparable to that of the stellar continuum
in the 0.3--0.4\,\um\ range, close to where the peak contribution
to dust heating occurs in most galaxies (Calzetti 2001).  Below in
\S3.3 we use the spectral energy distributions of the galaxies in 
our sample to confirm this quantitatively.

We can also use the results presented above to explore the validity
of applying a linear combination of emission-line and IR fluxes (eq. [1])
in the more general case when the line attenuation is systematically
larger or smaller than the mean opacity to the dust-heating starlight.
This would be the case for emission lines at much shorter or longer
wavelengths than \halpha, or even for \halpha\ itself, if the dust-heating
starlight was dominated by extremely young (i.e., blue) or old (i.e. red)
stellar populations.  In such cases the re-emitted starlight in the
dust continuum will tend to under-compensate or over-compensate for the 
flux attenuated in the emission line.  The magnitude of this effect can
be calculated using eq. (8).  Figure 3
plots the ratio of the attenuation-corrected luminosity estimated
from naive application of eq. (1) to the actual intrinsic
luminosity, for six values of $\beta$ in the range 0.25 -- 2
(the solid line at unity is the case for $\beta = 1$).  The
systematic error reaches its maximum value for optical depths
of 0.7 -- 2 (approximately the same range in magnitudes).  
For most combinations of $\beta$ and opacity the systematic
errors are small, of order 10--20\%, when compared to other systematic
uncertainties in the determination of the SFRs.   
The mismatch in opacities is also unimportant for low 
optical depths, because in those situations the emission 
contribution from the IR is negligible.  
The values of $\beta$ shown in Figure 3 covers a wide range of
potential emission lines from Pa\,$\alpha$ at 1.89\,\um\
($\beta \sim 0.25$) to [\ion{O}{2}]$\lambda$3727 ($\beta \sim 2$).
In this paper we shall apply the linear relations
in eqs. (1) and (2), and use comparisons to independent measurements 
of attenuation-corrected SFRs and attenuations to evaluate the efficacy of the method.

\subsection{Reference SFRs}

Our method in effect uses the ratio of infrared to \halpha\ fluxes
of galaxies to derive attenuated-corrected \halpha\ luminosities
and SFRs.  These need to be calibrated using independent measurements
of the \halpha\ attenuation corrections.  For the latter we have
chosen to use the stellar-absorption corrected \halpha/H$\beta$ 
ratios from our integrated spectra, converted to \halpha\ attenuations
using a Galactic extinction curve.  Specifically, we assume an intrinsic
\halpha/H$\beta$ ratios for Case B recombination 
($I(H\alpha)/I(H\beta)$ = 2.86) at electron
temperature $T_e$ = 10,000\,$K$\ and density
$N_e = 100$ cm$^{-3}$ (Hummer \& Storey 1987).  The dust attenuation
curve of O'Donnell (1994) was used to convert the observed
reddenings to attenuation values at H$\alpha$ ($R_V$ = 3.1).  The 
latter is equivalent to adopting a Galactic dust extinction curve,
and assuming a foreground dust screen approximation for the
attenuation correction in the Balmer lines.  We discuss
the possible limitations of these assumptions below in
\S 3.3 and in \S 6.3.


\subsection{Physical Assumptions and Limitations}

As emphasized earlier our simple energy balance method is an
idealized prescription that approximates a much more complicated
radiative transfer problem in galaxies, and it is important to 
bear in mind and quantify any systematic errors and limitations that
may be introduced by these approximations.  The most important
of these approximations are:  1)  assumption of comparable extinction
in the optical emission-line tracer and in the dust-heating continuum;
2) assumption of isotropic dust geometry;  3)  adoption of an
average dust-heating stellar population mix across the sample;
4) assumption of a reliable set of reference SFRs and attenuation
measurements, which are free of dust-dependent systematic errors.
We defer a full discussion of these assumptions and any associated
uncertainties to \S6, after the main results have been presented.
However we briefly address the points here, to offer reassurance
that the assumptions are reasonable for the current application.

As shown earlier the use of linear combination of \halpha\ and IR SFR tracers
to derive attenuation-corrected SFRs is strictly valid in the case where the
averaged dust attenuations in \halpha\ and the dust-heating stellar continuum
are comparable.  The attenuation laws of Calzetti (2001, and references
therein) show that this is a plausible assumption, but we can test it directly.
We combined measurements of the galaxies in MK06 sample from the Galaxy Evolution
Explorer (GALEX; Martin et al. 2005), the Two Micron All Sky Survey,
and the Sloan Digital Sky Survey (SDSS; Adelman-McCarthy et al. 2006) 
to derive the bolometric
luminosities of the galaxies, exclusive of the re-emitted dust luminosities.
The ratio of these luminosities to the total-infrared luminosities were then
used to derive an estimate of $\overline{\tau}$ for each galaxy, using eq. (5).
These values were then compared in turn to measurements of $\tau_{H\alpha}$
from the Balmer decrements.  Figure 4 shows a histogram of the resulting values
of $\beta = {\tau_{H\alpha}}/{\overline{\tau}}$.  The median value of $\beta =
1.08$, which is consistent with unity within the uncertainties in the optical
depths.  Much of the dispersion in values is physical, and reflects differences
in SED shape within the sample (B.\,Johnson et al.\,2009, in preparation).  The
result justifies the use of the linear approximation, at least for combinations
of \halpha\ and IR fluxes, and of course for combinations of UV continuum and
IR fluxes (Paper II).

Variations in dust geometry will be another source of uncertainty
in the IR-based attenuation corrections.  Our model assumes an
isotropic distribution of dust around the stars, but we expect
few regions to adhere to this idealized geometry.  So long as
there is no systematic bias toward foreground or background
dust in galaxies we do not expect this to introduce significant
systematic errors, but geometry variations may introduce a 
significant random error into individual determinations of 
attenuation.  When C07 applied our method to 220 bright HII regions
in 33 SINGS galaxies, they found excellent consistency between
the derived \halpha\ attenuations and those derived from 
Pa\,$\alpha$/\halpha\ ratios, but with a dispersion of $\pm$0.3 
dex for individual objects.  This factor-of-two dispersion
is consistent with the combined effects of varying dust geometry 
and stellar population in the HII regions.  We might well
expect the dispersions to be lower for galaxies, where we
are integrating over hundreds to thousands of individual
star-forming regions.

As will be discussed further in \S6 variations in dust-heating
stellar populations-- the longstanding ``infrared cirrus" 
problem, is probably the largest source of systematic uncertainty
in this approach.  The dust heating in galaxies arises from a 
much larger range of stellar ages (and dust heating geometries)
than in the HII regions and starbursts studied by C07, so 
the calibration coefficients in eq. (1) derived by C07 are not
expected to apply to the measurements of galaxies as a whole.
Comparing the results for galaxies and HII regions will provide
an indirect measure of the relative contributions of heating 
from very young ($<$10 Myr old) and older stars, and some indication
of the uncertainties in SFRs when different types of galaxies
are compared.

Finally, any systematic error in the reference attenuations 
derived from Balmer decrements will propagate with full weight
into our calibrations.  Our approximation to dust with a
Galactic extinction curve in the foreground screen limit
is the most common convention for
analyses of both HII region spectra and integrated spectra of
galaxies (see Calzetti 2001 and references therein), but
its use raises some legitimate questions.  A large body of 
theoretical studies of dust radiative transfer in galaxies
(e.g., Witt \& Gordon 2000, Charlot \& Fall 2000, Charlot \& 
Longhetti 2001, Tuffs et al.\ 2004, Jonsson et al. 2009) show that the actual
relation between the reddening of an integrated spectrum and
its dust attenuation can vary considerably, depending on the
dust geometry and the relation between extinction and stellar
age.  In galaxies with severe dust extinction, such as in many
ULIRGs, these geometric effects will cause the Balmer decrement
to be biased to the least obscured regions, leading to a systematic
underestimate of the actual attenuation.  The predictions for
more typical disk galaxies are inconclusive, with some studies
suggesting a mild bias in attenuations estimated from the 
foreground dust screen approximation (e.g., Charlot \& Fall 2000),
but with other studies showing little or no bias (e.g., Jonsson
et al. 2009).  However observations at other wavelengths
(Pa\,$\alpha$, thermal radio continuum, and UV$+$IR) allow us to
independently derive \halpha\ attenuation estimates for subsets
of our sample, and test for systematic errors in the zeropoint
of our Balmer decrement based attenuation scale.  As shown
in \S 6.3 these tests confirm the reliability of the 
Balmer attenuation scale at the $\pm$10--20\% level, which
is comparable to other systematic errors in our SFR scales.
On that basis we have chosen to use the Balmer decrements 
with the most commonly applied attenuation correction procedures.
We also publish the observed Balmer line ratios so other workers
may apply alternative attenuation corrections in the future.


\section{APPLICATION TO INTEGRATED MEASUREMENTS OF GALAXIES}

In this section we use the combination of \halpha\ fluxes, 
IR fluxes, and optical spectra for our galaxies to examine
the consistency of SFRs measured from \halpha\ and IR tracers,
and to calibrate composite SFR indices.  Since this work
originally grew out of the discovery of a strong correlation
between \halpha\ and 24\,\um\ emission in HII regions, we 
first extend this calibration to galaxies.  We then calibrate
a similar combination of \halpha\ and TIR luminosities (not
possible for the SINGS HII regions due to the limited spatial
resolution of the far-IR data).  The section concludes with
an analysis of systematic residuals in the SFR calibrations
as functions of various properties of the galaxies and their
SEDs.

\subsection{Combinations of \halpha\ and 24\,\um\ Measurements}

As an introduction it is instructive to compare the consistency
of H$\alpha$ and IR SFR measures before any corrections for
attenuation are applied.  Figure 5 compares the observed H$\alpha$
luminosities of the SINGS and MK06 galaxies with their corresponding
24\,\um\ ($\nu\,f_\nu$) luminosities.  When converting fluxes to luminosities we
have used distances listed for SINGS galaxies in Table 1,
and from MK06 in Table 2.  Both sets of distances assume 
H$_0$ = 70\,km\,s$^{-1}$\,Mpc$^{-1}$ with local flow corrections.
In Figure 5 (only) we do not distinguish SINGS galaxies from MK06 galaxies and we 
code the points by spectral type, with open
round points representing galaxies dominated by star formation,
crosses representing galaxies with strong AGN signatures in their
spectra, and circles with embedded crosses denoting composite AGN
and star formation dominated spectra.  The axis label along the top of
the plot shows the corresponding SFR (uncorrected for attenuation),
using the calibration of K98. Finally, the solid
line shows a linear (unity slope) relation for reference,
with the zero-point set to match the mean relation between L(24\,\um) and 
L(${\rm H}\alpha)_{\rm corr}$ found
by Zhu et al. (2008) for nearby galaxies
in the {\it Spitzer} Wide-Area Infrared Extragalactic (SWIRE) 
survey (Lonsdale et al. 2003).

The H$\alpha$ and 24\,\um\ luminosities show a broad correlation,
as is often the case when absolute luminosities are compared.
Upon closer examination, however, important departures are apparent.
The rms dispersion around the linear relation is $\pm$0.5 dex,
more than a factor of three.  Part of this dispersion is caused
by AGN contributions to the dust emission.  The galaxies with 
significant AGN contributions clearly are displaced in the diagram,
reflecting in part the additional dust heating by the active
nuclei (e.g., Soifer et al. 1987).  Since we are solely
interested in calibrating SFR tracers we shall exclude 
galaxies with strong AGN signatures from most of the subsequent
analysis in this paper, but clearly the issue of AGN contamination
is always an important one when applying any SFR tracer. 
Even among the star-forming galaxies, however, the dispersion
about the linear relation is large, approximately $\pm$0.4 dex.
The dispersion among these galaxies is caused almost entirely
by variations in dust attenuation; the observed \halpha\ flux
underestimates the SFR in dusty galaxies, while the IR emission
underestimates the SFR in less dusty galaxies.   
Moreover, the relation between \halpha\ and IR emission clearly is nonlinear, 
with the mean ratio of IR to H$\alpha$ luminosities 
increasing by a factor of 30 from the faintest to the most
luminous galaxies in the sample.  This nonlinearity is a manifestation
of the well established relation between attenuation and SFR
(e.g., Wang \& Heckman 1996).  Higher SFRs are associated 
with regions of higher gas surface density (Kennicutt 1998b, C07) 
and hence higher dust column densities.  As a result
galaxies with the highest SFRs tend to suffer heavy attenuation,
and the observed \halpha\ flux severely under-represents the
actual SFR, moving points in Figure 5 to the left.  Likewise
many (but not all) of the galaxies with low SFRs tend to
be low-mass galaxies with lower dust contents and column 
densities.  For those objects the attenuation at \halpha\ is
low and the emission line provides an accurate measure of the
SFR, but then the dust emission severely under-represents the
SFR, shifting points down in Figure 5.  

We can remove the part of the nonlinearity in Figure 5 that
is produced by attenuation of \halpha\ by comparing instead
the 24\,\um\ luminosities with the attenuation-corrected
\halpha\ luminosities, as derived from the Balmer decrements
in the integrated spectra (\S 3.2).  This comparison is shown in the
top panel of Figure 6 (now with the AGN-dominated and composite
spectrum galaxies removed).   Here and throughout the paper
integrated measurements of SINGS galaxies are shown as open
circles, and those for the MK06 sample are shown as solid circles. 
Applying the reddening corrections tightens the correlation with 
24\,\um\ emission considerably, but the dispersion about 
the mean relation remains substantial at $\pm$0.3 dex,
and the nonlinearity remains.  Nearly all of the scatter
and nonlinearity in this relation are caused by variations in the fraction
of young starlight that is reprocessed by dust.

Previously several groups have investigated the correlation
between 24\,\um\ IR emission and attenuation-corrected H$\alpha$
and Pa\,$\alpha$ emission, and used them to calibrate the 24\,\um\
emission as a SFR measure (Wu et al. 2005, Alonso-Herrero et al.
2006, Rela\~no et al. 2007, C07).  The relations from Wu et al.
(2005), Rela\~no et al. (2007), and C07 can be directly compared to
our data, as shown 
in the top panel of Figure 6.  Apart from slight deviations at the extremes in
luminosity our data generally follow these relations as well.
Nevertheless it is dangerous to use the IR luminosity by itself
as a quantitative SFR tracer, because in any galaxy other than
an extremely dusty starburst the dust reprocesses only a fraction
of the young starlight, and any linear scaling of a SFR calibration
based on dusty starburst galaxies will tend to systematically
under-estimate the SFR in lower-luminosity normal galaxies.
This is shown by the dotted line in the top panel of Figure 6,
which shows a linear (slope unity) relation fitted to the data.  
The calibrations of Wu et al. (2005) and Rela\~no et al. (2007)
mitigate this effect by fitting a nonlinear relation between
$L(24)$ and $L(H\alpha)$, which effectively builds in a 
decrease in mean attenuation with decreasing SFR.  However 
it represents a crude, unphysical correction at best.  While
most present-day galaxies with low H$\alpha$ luminosities tend
to be dwarf galaxies with low dust contents and optical depths,
this same range of fluxes is occupied by dusty galaxies with
low SFRs (mostly early-type spirals); attenuation does not
correlate monotonically with the SFR itself.  Thus while nonlinear
calibrations of the 24\,\um\ luminosity may provide crude statistical
measures of the SFR, the uncertainties associated with measurements
of individual galaxies remain very high.  One risks even larger,
systematic errors if one were to apply such relations at high
redshift, because there is no reason to expect that the relations
between mean attenuation and SFR observed at $z = 0$ necessarily
apply in the early universe.

However we can mitigate the effects of variable attenuation
between the galaxies by applying eq. (1), using a linear
combination of the observed (uncorrected) H$\alpha$ and 
mid-IR luminosities to estimate the attenuation-corrected
H$\alpha$ luminosity.  The result is shown in the bottom 
panel of Figure 6, which compares the corrected H$\alpha$
luminosities derived from the ratio $L(24)/{L(H\alpha)_{obs}}$ with
those derived from the Balmer decrements.  We fitted for
the value of the scaling factor $a$ in eq. (1) that provides
the best overall agreement with the luminosities corrected
from the Balmer decrements ($a = 0.020 \pm 0.001_{\rm r} \pm 0.005_{\rm s}$),
where the first error term lists the formal (random) fitting
uncertainty, and the second term includes possible systematic
errors in the calibration zeropoint including any uncertainty
in the reference attenuation scale based on Balmer decrement
measurements (see \S 6.3).  We shall adopt this convention of
citing both random and systematic errors throughout the remainder
of this paper.

The consistency between the two independent sets of 
attenuation-corrected H$\alpha$ luminosities is striking,
as is the decrease in the dispersion between the two 
SFR estimates, relative to the comparisons of 24\,\um\
vs H$\alpha$ luminosities alone in the top panel of Figure 6.
This is especially apparent for the galaxies with 
$L(H\alpha) < 10^{42}$ ergs\,s$^{-1}$.  The galaxies
with the lowest luminosities tend to have very little
dust, and hence nearly no reddening and only weak IR emission.
In such galaxies the H$\alpha$ luminosity dominates the
sum in eq. (1).  On the other hand most of the galaxies 
with $L(H\alpha) > 10^{42}$ ergs\,s$^{-1}$ in Figure 6
are LIRGs, with very high attenuations.  In such cases
it is the IR term in eq. (1) that dominates the sum.

Our fitted value of the coefficient $a$ in eq. (1) 
is 35\%\ lower than the value $a = 0.031 \pm 0.006$ derived
by C07 based on measurements of HII regions in SINGS galaxies.
However our value is in excellent agreement with $a = 0.022$
derived by Zhu et al. (2008) for a sample of 
luminous star-forming galaxies observed in common between
the SWIRE survey and the SDSS.  We believe that
the difference in scaling coefficients is consistent with
expectation given the different stellar age distributions in
HII regions and galaxies, as discussed in \S 6.

\subsection{Combinations of \halpha\ and Total Infrared Measurements}

Before examining the implications of Figure 6 in more
depth we can make the same comparison, but combining
instead the H$\alpha$ and TIR luminosities of the galaxies.
The results are shown in Figure 7.  As before the top
panels show the TIR luminosity as a function of $L(H\alpha)_{\rm corr}$, 
while the bottom plots show the linear combination of
TIR and H$\alpha$ luminosities (in this case with the
fitted $a = 0.0024 \pm 0.0001_{\rm r} \pm 0.0006_{\rm s}$).  
Because of the systematic differences between MIPS and {\it IRAS} TIR scales we
separate these data in Figure 7, with the left panels
showing {\it Spitzer} MIPS-based TIR luminosities 
(SINGS sample) and the right panels showing {\it IRAS}-based 
luminosities (SINGS and MK06 samples).  Since MIPS observations
are not available for most of the MK06 sample most of the
TIR data used in the remainder of this paper will be based
on {\it IRAS} luminosities, to maximize the uniformity
of the data.  However the more sensitive MIPS data allow
us to extend the range of luminosities covered by nearly
an order of magnitude.  

The most gratifying result shown in the bottom
panel of Figure 7 is not the linearity of the relation
(which is dictated by the fitting of $a$) but rather the
tightness of the relation.  To illustrate this better
Figure 8 shows the residuals from the fits in Figure 6 
(left panels, 24\,\um\ and \halpha) and Figure 7 (right
panels, {\it IRAS} TIR and \halpha).  The combination of
\halpha\ and IR flux dramatically improves the consistency 
and tightness of both relations.
In both cases these dispersions are less than half of the
rms (logarithmic) deviations when the IR luminosities by themselves are
compared to the Balmer-corrected luminosities.
We suspect that the tighter residuals in the \halpha\,$+$\,TIR
index ($\pm$0.09 dex rms vs $\pm$0.12 dex for \halpha\,$+$\,24\,\um)
reflect the higher quality of the longer wavelength IRAS measurements
as well as lower sensitivity
to local variations in dust heating (see \S 6 for further discussion).
The lack of any systematic residual with total luminosity
(and thus SFR) is also significant, and argues against any
large systematic error in the derived values of the fitting
coefficients $a$.

Figures 5--7 compare absolute luminosities in various wavelengths,
and it is well known that such plots can sometimes mask underlying 
physical trends, because of the large range in galaxy masses that
underlies any comparison of absolute quantities.  These effects can
be removed by analyzing mean surface brightnesses (which scale
with the SFR per unit disk area) instead of luminosities.
For this comparison we defined the disk area as the deprojected area of 
the spectroscopic aperture for the MK06 galaxies, and area
within the $R_{25}$ radius (de Vaucouleurs et al. 1991)
for the SINGS galaxies.  These provide only approximate measures of
the radii of the star-forming disks (e.g., Kennicutt 1989, 1998b),
but they suffice for this statistical comparison.
The top panels of Figure 9 show that the same trend we observed
in disk opacity functions of luminosity are also seen in SFR per
unit area.  This is not surprising, because it is well known
that the SFR per unit area correlates strongly with the average
surface density of gas (e.g., Kennicutt 1998b), and it stands
to reason that it would correlate with the mean column density
of dust as well.  The bottom two panels show that 
our prescriptions for correcting the \halpha\ emission 
for attenuation using IR observations are in excellent agreement
with the Balmer-derived attenuations over the full range 
(more than a factor of 1000) in \halpha\ surface brightness.

Finally, Figure 10 compares 
the \halpha\ attenuation estimates themselves.  The left and right panels
compare the H$\alpha$ attenuations derived using \halpha\ $+$
24\,\um\ and \halpha\ $+$ TIR luminosities, respectively, with
those derived from the spectroscopic H$\alpha$/H$\beta$ attenuation
values.  The solid line in each panel shows the line of equality,
while the dotted lines on either side contain 68\%\ of the points
($\sim$1-$\sigma$; the overall dispersions are identical to those in the previous figure).
Overall there is excellent consistency between the attenuation estimates.
The outliers tend to be SINGS galaxies where spatially undersampled
spectroscopic strips do not provide a representative measure of the
global reddening, along with a few galaxies with deeply embedded central
star-forming regions.  Overall, however, the attenuation
estimates, especially those derived from the combination
of \halpha\ and TIR luminosities, show impressive 
consistency with the spectroscopic reddening-derived
attenuations, given that the methods are entirely 
different.

The consistency between the IR-derived and Balmer-derived
attenuations for the integrated measurements of the
SINGS and MK06 galaxies is also much tighter than for
comparable measurements of individual giant HII regions.
C07 applied the same methodology to Spitzer 24\,\um\ and
\halpha\ measurements of 220 HII regions and star-forming complexes in 33 SINGS galaxies,
compared these to attenuations derived from Pa\,$\alpha$/\halpha\
ratios, and derived a mean dispersion of $\pm$0.3 dex 
(a factor of two) about the mean relation.  A similar comparison 
of a homogeneous set of data for 42 HII regions in M51 by
Kennicutt et al. (2007) yielded a dispersion of $\pm$0.25 dex.
For the integrated measurements in this paper we find
a corresponding dispersion of $\pm$0.12 dex between 
24\,\um\ $+$ \halpha\ and \halpha/H$\beta$ measurements,
and even less for the TIR $+$ \halpha\ and Balmer-derived attenuations.  
Zhu et al. (2008) made a similar comparison of attenuated-corrected
luminosities based on 24\,\um\ $+$ \halpha\ and \halpha/H$\beta$
measurements for their SWIRE$+$SDSS sample.  They do 
not quote a dispersion, but based on their Figure 5
we estimate that it is approximately $\pm$0.14 dex, somewhat
larger than our result, but understandable in view of the
uncertainties in the large aperture corrections that were 
required to match the SDSS fiber fluxes to the Spitzer 24\,\um\
observations.  The larger residuals in the HII
region results may be caused in part by larger measuring
uncertainties in the Pa$\alpha$ and H$\alpha$ photometry
of the HII regions; however we would also expect physical
effects such as variations in attenuation geometry and 
age variations in the HII regions to produce larger residuals.
These latter effects tend to average
out when integrated over an entire galaxy's population
of star-forming regions.

\subsection{Systematic Effects and Second-Order Correlations}

The correlations between the IR-based and optically-based
attenuation measurements are so tight that it is worthwhile
investigating whether the residuals from these relations
vary systematically with the SEDs or physical properties of
the galaxies.  As with any quantitative star formation tracer
it is important to establish the limiting range of SFRs
and galaxy types over which these methods can be applied
reliably, and assess whether they
can be extrapolated to galaxy types outside of our sample,
for example the extremely luminous star-forming
galaxies observed at high redshift.  

We tested for systematic residuals against eight parameters, including
SFR (Figure 8), SFR per unit area (cf.\, Figure 9), and six 
other parameters shown in Figures 11 and 13.
Figure 11 shows the logarithmic residuals of attenuation-corrected
\halpha\ luminosities derived from \halpha\,$+$\,24\,\um\ 
(left) and \halpha\,$+$\,TIR (right), relative to those
derived using the Balmer decrement, plotted as functions of 
FIR 60\,\um\ to 100\,\um\ flux ratio (top panel), Balmer-derived attenuation
(middle panel), and axial ratio or inclination 
(bottom panel).  The only significant trends seen are a systematic
residual in \halpha\ $+$ 24\,\um-based SFRs with $f_\nu(60\,\um)/f_\nu(100\,\um)$
ratio and Balmer attenuation.  This trend with IR color is a
byproduct of the systematic variation in infrared SED shapes
of galaxies; galaxies with higher 60\,\um\ to 100\,\um\ flux ratios
also show higher L(24\,\um)/L(TIR) (or L(25\,\um)/L(TIR)) ratios.
This well-known trend is
illustrated for our sample in Figure 12, with the sequence of
SED models by Dale \& Helou (2002) superimposed.  As a result
we believe that the trend in residuals in the upper left-hand
panel of Figure 11 is an artifact of the variation in SED shapes 
of galaxies, and this contributes considerably to the larger
scatter in this index relative to a linear combination of
\halpha\ and TIR luminosities.  Interestingly the latter
residuals show no systematic trend with galaxy IR SED shape.

The trend in residuals with Balmer attenuation (middle left
panel of Figure 11) is less straightforward to interpret.
Interestingly the sense of the trend is that galaxies with
the highest Balmer attenuations have {\it weaker} 24\,\um\
emission than expected, which is in the opposite sense of
what one might expect if the Balmer decrements systematically
under-estimated the true attenuation in more dusty environments.
The same trend is present in the \halpha\ $+$ TIR residuals
(middle right panel), but is considerably weaker.  This 
suggests that much of the trend in the middle left panel
may be the byproduct of a second-order correlation between
average dust temperature (i.e., mix of warm star-forming
regions and background cirrus emission) and characteristic
opacity of those regions (Dale et al. 2007), but this 
should be regarded as speculation rather than an explanation.

Interestingly neither set of IR-based attenuation and
SFR estimates shows any systematic residual with Balmer
attenuation/SFR as a function of axial ratio.  One would
not expect the \halpha\ $+$ IR (24\,\um\ or TIR) indices to show any systematic
effects with disk inclination, but one might imagine that
the Balmer-based measurements would suffer from such a
systematic effect.  We do know that the \halpha\ surface
brightnesses of disks do decrease systematically with
increasing inclination (Young et al. 1996), and our
results suggest that this increase in attenuation is
largely accounted for in both the IR-based and 
spectroscopic attenuation corrections.  
This offers some support for the robustness of the Balmer-based
attenuation measurements.

Figure 13 shows the same residuals as Figure 11, but plotted
this time as functions of three measures of the stellar
populations in the galaxies, the integrated \halpha\ 
equivalent width (EW) of the galaxies, the 4000\,\AA\ spectral
break strength (Bruzual 1983, as applied by Balogh et al. 1999),
and the gas-phase oxygen abundance ($12 + \log(O/H)$) estimated
from the nebular lines (Moustakas et al. 2006, 2009).
The \halpha\ EWs (top panels) roughly scale with the SFR
per unit stellar mass (specific SFR).  The residuals show
no systematic change with EW over a very large range
(5\,\AA\ $-$ 230\,\AA), which includes
galaxies with Hubble types Sb--Irr (K98).  The
bottom panels of Figure 13 show that there
is no significant trend in attenuation residuals with
metallicity.

The $D_n(4000)$ break provides a reddening-insensitive
measure of the relative age of the stellar population in
the galaxies (e.g., MacArthur 2005; Johnson et al. 2007b), 
and we might expect it to correlate with
the attenuation residuals, if variations in heating of
dust from evolved stars is important.  A trend is seen
in the \halpha\ $+$ 24\,\um\ residuals (left middle panel
of Figure 13), but we believe that this is mainly a second-order
consequence of the IR SED dependence discussed earlier.
There is no discernible trend in \halpha\ $+$ TIR
attenuation residuals with $D_n(4000)$.  One might have
expected some deviation for the reddest galaxies with larger
values of $D_n(4000)$, as the result of dust being
heated primarily by evolved stars.  Evidence for such
deviations has been seen in comparisons 
analyses of UV and IR-based SFR tracers (e.g., 
Johnson et al. 2009; Cortese et al. 2008).  We believe that
two effects contribute to this negative result.  One
is the limited range in stellar population probed by this
sample ($D_n(4000) \le 1.4$).  The other reason is a
compensation between two physical effects; in galaxies
with redder stellar populations there will be a stronger
radiation component from evolved stars, but the radiation
will also be shifted to longer wavelengths where the dust
opacity is lower.  In terms of eq. (8) in \S3.1, there
are changes in the coefficients $\eta$ and $\beta$ which
tend to compensate for each other.

An alternate way to examine the residuals is to 
calculate for each individual galaxy a value of the IR/\halpha\
scaling factor in eq. (1) that forces the derived reddening
to match that derived from the Balmer decrement:

\begin{equation}
a_{\lambda} = {{L(H\alpha) \over {L_{\lambda}(IR)}}} \ {[{10^{0.4 A_{Balmer}(H\alpha)} - 1}]}
\end{equation}

\noindent
We expect these individual values of $a$ to show a 
considerable dispersion, especially in galaxies with
relatively low attenuation, where the uncertainties
in single attenuation estimates may be comparable to
the estimates themselves.  Figure 14 shows histograms
of $a_{24}$ and $a_{TIR}$ for the integrated measurements
of SINGS and MK06 samples, with the adopted average
fits for the combined sample shown by vertical lines.   
There are a few prominent outliers in both cases,
and these provide valuable examples of when our 
attenuation measures break down.  For the outliers
with anomalously high values ($a_{24} > 0.05$ and 
$a_{TIR} > 0.004$) the IR emission is considerably
weaker than expected from the Balmer decrement.  These
tend to be either galaxies with low attenuation and IR
emission, where the predicted IR emission is sensitive
to small uncertainties in the Balmer decrement, or 
cases where the limited spatial sampling of the SINGS
spectroscopy does not provide an accurate measure of 
the global reddening.  The points with anomalously low
values of $a$ include metal-poor galaxies with low 
attenuations, objects with deeply
embedded star-forming regions, or other cases where
the SINGS spectra probably under-estimate the global
reddening of the galaxy, because the optical emission
does not penetrate into the main IR-emitting region.  


\section{Composite SFRs with Other Emission Lines and IR Bands}

We have shown that \halpha\ fluxes of galaxies can be combined
with either 24\,\um\ or TIR fluxes to provide robust 
attenuation-corrected \halpha\ luminosities and SFRs.  Can
other combinations of optical emission lines and infrared or radio
continuum bands be used in the same way?  In this section we extend
our approach to the use of the 8\,\um\ (PAH-dominated) band to derive
attenuation-corrected \halpha\ luminosities, the use of 
[\ion{O}{2}]$\lambda$3727 $+$ IR indices in place of \halpha\ composite measures,
and the combination of optical emission-line and radio
continuum measurements.  The combination of ultraviolet
and IR luminosities is a mature subject in its 
own right and is analyzed separately in Paper II.

\subsection{\halpha\ and 8\,\um\ Indices}

With the advent of large-scale galaxy surveys with 
{\it Spitzer} there is growing interest
in calibrating the rest 8\,\um\ mid-IR luminosities of galaxies
as quantitative SFR tracers.  Most galaxies exhibit strong
emission in the rest 8\,\um\ band, and this wavelength region
redshifts into the
Spitzer 24\,\um\ band at $z \sim 2$.  In most galaxies
the primary emission mechanism at 8\,\um\ is molecular band
emission from aromatic ``PAH" grain species, as distinct from
the thermal grain emission that dominates in the FIR,
so its reliability as a quantitative SFR tracer must be
tested empirically.  In massive galaxies with high SFRs the 8\,\um\ 
emission correlates reasonably well with the TIR emission
sufficiently so that it (or neighboring regions observed
with the Infrared Space Observatory) has been calibrated as an IR
SFR tracer of its own (e.g., Roussel et al. 2001, Boselli et al.
2004, F\"orster Schreiber et al. 2004, Wu al. 2005, Farrah et al. 
2007, Zhu et al. 2008).  However the dispersion in 8\,\um\
luminosities at a fixed SFR tends to be higher than for
longer wavelength tracers such as the 24\,\um\ emission, with
systematic dependences of the emissivity on metal abundance, 
and local radiation field strength and hardness
(e.g., Madden 2000, Peeters et al. 2004,
Engelbracht et al. 2005, 2008; Dale et al. 2005,
Wu et al. 2006, Smith et al. 2007, C07).  The large dispersion
in PAH strength (more than an order of magnitude across all
galaxy types and luminosities) means that the rest 8\,\um\
luminosity of a galaxy provides a crude measure at best of
its SFR.  Can these effects be mitigated by constructing a
composite H$\alpha$ $+$ 8\,\um\ SFR index?

To carry out this test we compiled 8\,\um\ fluxes of the 
SINGS galaxies, both integrated fluxes and those with 20\arcsec\
central apertures as described in \S 2.  The emission in
this bandpass is a composite of a (normally) dominant PAH
band emission as well as an underlying continuum from dust and
evolved stars.  The pure dust emission at 8\,\um\ was obtained
by subtracting a scaled 3.6\,\um\ emission from the
measured 8\,\um\ flux. A scale factor of 0.255 was used here,
following C07.  To provide a meaningful
standard of comparison we also compiled MIPS 24\,\um\ fluxes
for the same objects.  Stellar continuum in the 24\,um\ band
(apart from foreground stars, which were removed in the
Dale et al. 2007 data) is negligible,
so no attempt was made to remove it.

The top panels in Figure 15 show the correlation between 8\,\um\
(left) and 24\,\um\ luminosities and Balmer-corrected \halpha\
luminosities for the sample of star-forming SINGS galaxies.
Open circles show integrated measurements of the galaxies, while
open squares show measurements of the central 20\arcsec\ $\times$
20\arcsec\ star-forming regions.  We have also overplotted in small
dots the HII regions from the SINGS sample measured by
C07. 

The correlation between 24\,\um\ and corrected \halpha\ luminosities
in the top right panel of Figure 15 is the same as seen earlier 
(Figure 6), but with a clear offset between the relations for entire galaxies
and the HII regions, which we will return to later.  The scatter
in the relation between 8\,\um\ emission and \halpha\ emission is
much higher, confirming trends seen in the HII regions by C07.
Most of this difference reflects the presence of  
galaxies (and HII regions) with 8\,\um\ luminosities up to 30
times weaker than the main relation; most of these are metal-poor
dwarf galaxies, which are already known to have strongly suppressed
PAH emission.  The top left panel of Figure 15 aptly illustrates
the perils of applying the 8\,\um\ emission of galaxies indiscriminantly
as a quantitative SFR tracer (cf. Smith et al. 2007, 
Engelbracht et al. 2008, and references therein).

The two bottom panels of Figure 15 show the results of constructing
composite H$\alpha$ $+$ 8\,\um\ and H$\alpha$ $+$ 24\,\um\ estimates
of the attenuation-corrected \halpha\ luminosity, compared as before
with the spectroscopically-corrected \halpha\ luminosities.  
For the 24\,\um\ index we used the same scaling factor $a$ as
for the integrated measurements earlier, to compare how well
the central 20\arcsec\ measurements and HII region measurements
are fitted by the integrated relation.  For the 8\,\um\ $+$ \halpha\
composite we derived the value of $a = 0.011 \pm 0.001_{\rm r} \pm 0.003_{\rm s}$ (integrated 
and central 20\arcsec\ data combined).  For 24\,\um\ the calibration from
the SINGS and MK06 integrated measurements provides an excellent
fit to the SINGS central aperture measurements as well, which
is not surprising.  What may be more surprising is the excellent
consistency of the attenuation-corrected \halpha\ luminosities
derived from H$\alpha$ $+$ 8\,\um\ with those derived from the
Balmer decrement attenuation corrections.  The metal-poor outliers
in the upper left panel of Figure 15 join the main relation in
the weighted sum of H$\alpha$ $+$ 8\,\um\ (lower left panel),
because these galaxies have very low reddening and very low
8\,\um\ emission, so both measured quantities in the plot
are essentially (and identically) the observed \halpha\ luminosity.  

Interestingly the residuals
using the 8\,\um\ fluxes are lower than those derived using 
24\,\um\ fluxes, $\pm$0.11 dex vs $\pm$0.14 dex, respectively.
This can be understood if the 8\,\um\ luminosity is more tightly
coupled to the TIR dust luminosities of galaxies than the 
24\,\um\ emission, as shown for example by Mattila et al. (1999),
Haas et al. (2002), Boselli et al. (2004), and Bendo et al. (2008).  Upcoming 
high angular resolution observations of the FIR continuum
of the SINGS sample, currently planned with the {\it Herschel Space Observatory},
will allow us to test the coupling of the various IR emission
components with higher spatial resolution and sensitivity.

\subsection{Composite SFR Indices Using [\ion{O}{2}] Emission Lines}

In principle our approach can be applied to estimate 
dust attenuation corrections for any optical emission
line that is used as a SFR diagnostic.  The most commonly
applied visible-wavelength line, especially for observations
of galaxies at intermediate redshift, is the [\ion{O}{2}]$\lambda$3727 
forbidden line doublet.  The chief advantage of this feature is that
it is accessible to ground-based telescopes and CCD spectrometers
out to redshifts $z \sim 1.7$, whereas \halpha\ redshifts beyond
the optical window above $z \sim 0.5$.  As a result [\ion{O}{2}]-based
SFR estimates are available for tens of thousands of 
galaxies at $z \sim$ 0.1--1.5 (e.g., Franzetti et al. 2007,
Ly et al. 2007, Cooper et al. 2008).

The calibration and reliability of the [\ion{O}{2}] feature as a 
quantitative SFR tracer has been discussed by several authors
(e.g., Hopkins et al. 2003, Kewley et al. 2004, 
Moustakas et al. 2006).  Unlike the Balmer lines,
the luminosity of the collisionally-excited [\ion{O}{2}] doublet 
is not fundamentally coupled to the ionizing flux, so its
accuracy is limited by excitation variations, which in
turn are systematically correlated with the metal abundance
and ionization of the gas.  However the typical variations
in intrinsic [\ion{O}{2}]/\halpha\ are of order a factor of two
or less over a wide range of abundances and galaxy environments,
so the index can be useful, especially in applications to
large samples.  Dust attenuation however is a much more
severe problem, with typical attenuations in normal galaxies
of nearly an order of magnitude, and large variations
between objects.  Kewley et al. (2004) and Moustakas et al.
(2006) provide empirical schemes for correcting for 
this attenuation as functions of [\ion{O}{2}] luminosity and 
$B$-band luminosity, respectively, but these are crude
approximations at best.  Here we investigate whether 
the combination of [\ion{O}{2}] and IR luminosities can provide
more robust attenuation-corrected SFRs.

Our results are summarized in Figure 16.  The upper left
panel compares the observed [\ion{O}{2}] luminosities and 
reddening-corrected \halpha\ luminosities for the MK06
sample (solid circles) and the inner 20\arcsec\ $\times$ 20\arcsec\
regions of the  SINGS galaxies (open squares).
The integrated measurements of the SINGS sample are not
plotted because we do not have full-galaxy [\ion{O}{2}] luminosities
for that sample.  This shows the worst case of 
applying the [\ion{O}{2}] luminosity with no attenuation correction 
at all.  By coincidence the mean attenuation-corrected
luminosity of [\ion{O}{2}] in the MK06 sample is nearly
identical to that of \halpha\ ($<$[\ion{O}{2}]/\halpha$>$ = 0.98),
so any difference in luminosities translates identically to
a deficit in the estimated SFR.  On average the [\ion{O}{2}]
luminosities are suppressed by about 0.6 dex, with a
range (excluding the two outliers) of 0.0--1.7 dex 
(a factor 50 at worst).  

In the other three panels of
Figure 16 we apply a weighted sum of [\ion{O}{2}] luminosity
and TIR, 24\,\um, and 8\,\um\ luminosities.  As before
we derived the coefficients $a_{\lambda}$ from eq. (1)
which best fit the mean relations in Figure 16.
Combining observed [\ion{O}{2}] luminosities
of galaxies with any of the three IR luminosities can
provide a credible attenuation correction to the [\ion{O}{2}]
luminosities, with dispersions that are comparable to their \halpha\
$+$ IR counterparts, when differences in sample are taken
into account.

In \S3.1 we pointed out that our linear combination method
will be subject to modest systematic errors if the effective
attenuation in the emission line of interest deviates 
significantly from the mean dust opacity of the stellar
continuum radiation that heats the dust.  In most normal
galaxies the mean emission-line attenuation at [\ion{O}{2}]
is significantly higher than in the mean dust-heating 
stellar continuum (Calzetti 2001), so we might expect
our residuals to show a mild dependence on attenuation,
as illustrated theoretically in Figure 3.  We examined 
the residuals for such an effect, and observe a qualitative
trend in the expected sense, but we cannot reliably 
separate this from trends introduced by systematic
variations in the intrinsic strength of [\ion{O}{2}]
relative to the Balmer lines.  However 
there is other evidence for this effect from the
fact that the best fitting values of $a$ in Figure 16
are about 40\%--50\% higher than the corresponding values
for \halpha, even though the intrinsic luminosities of 
[\ion{O}{2}] and \halpha\ are nearly identical.
This difference in fitting coefficients (see Table 4)
is a direct 
result of the higher  dust attenuation at [\ion{O}{2}].

\subsection{Composite SFRs Using Radio Continuum Emission}

All of the composite SFR methods described until now 
require flux measurements in at least one IR band.
It is well known that star-forming galaxies show a
tight, linear correlation between their FIR and radio continuum
luminosities (e.g., Condon et al. 1991, Yun et al. 2001),
so it should be possible to use the radio continuum
emission in combination with optical and UV star formation
tracers to derive attenuation-corrected SFRs.  In this
section we extend our tests to 1.4 GHz radio continuum
fluxes, to evaluate the reliability of such hybrid SFR measures.

The results of our comparison are shown in Figure 17.  
The top panel shows the relationship between 
observed 1.4 GHz (21\,cm) radio continuum luminosity and 
Balmer-corrected \halpha\ luminosity, while the bottom panel
shows the correlation between a weighted sum of uncorrected 
\halpha\ and radio luminosity with the same Balmer-corrected
\halpha\ luminosities luminosities.  These can be compared
directly to Figures 6 and 7, which shows similar comparisons 
but with 24\,\um\ and TIR luminosities, respectively.  
Qualitatively the radio continuum fluxes and composite SFR
indices show that same behavior as we found for their IR counterparts.
The 1.4\,GHz radio luminosities by themselves show a
nonlinear dependence on the attenuation-corrected \halpha\
luminosities (Bell 2003), with a power-law slope of 1.28, even steeper
than that seen in TIR luminosity (1.10) and 24\,\um\
luminosity (1.19).  However combining 
the \halpha\ and radio luminosities removes most of the
nonlinearity and much of the dispersion, confirming that the 
radio luminosities can be used to correct the optical lines for attenuation
as well.  The mean dispersions around the fits ($\pm$0.10 dex
for \halpha\ $+$ 1.4\,GHz and $\pm$0.12 dex for [\ion{O}{2}] $+$ 1.4\,GHz)
are 11\%--33\%\ larger than the corresponding \halpha\ $+$ TIR and
[\ion{O}{2}] $+$ TIR indices, but again these dispersions are small
when compared to the random and systematic errors in 
SFRs derived from any of the individual SFR tracers by themselves.

It is interesting that the sense of the nonlinearity in the
radio--FIR relation is in the opposite sense that one might
naively expect.  At 1.4\,GHz the radio continuum is dominated
by non-thermal (synchrotron) emission, which presumably originates
ultimately from supernova events; this interpretation forms the
physical basis for using the non-thermal radio emission as a SFR tracer
(e.g., Condon 1992 and references therein).  If this scaling of radio continuum
luminosity with the SFR strictly held over all types and
luminosities of star-forming galaxies, then one would expect the
correlation in the upper panel of Figure 17 to be strictly linear,
while the slope of the radio vs TIR correlation shown in Figure 18
would be considerably shallower than a linear relation, because
we already have seen that the dust emission systematically 
underestimates the SFR in low-luminosity, low-opacity galaxies.
Instead Figure 17 shows a nonlinear dependence of radio emission
on attenuation-corrected SFR (also see Bell 2003), and the slope 
of the radio vs TIR correlation in Figure 18 is {\it steeper}
than a linear relation.  Apparently the radio
emissivity of galaxies declines even more steeply at low luminosity
than even the dust emission, producing the residual nonlinearities
in Figure 17 and 18.  A discussion of the physical explanation
for this result is beyond the scope of this paper.  One possibility
is a systematic change in the cosmic ray lifetimes and/or
the magnetic field strengths of galaxies with changing mass
and SFR.  Another possibility is the increasing role of radio
emission from a nuclear accretion disk in more massive galaxies.
Whatever the explanation, our results demonstrate that the
radio continuum luminosities of galaxies can be combined with
emission-line luminosities to provide reasonable measurements
of the attenuation-corrected SFRs.

\section{Discussion:  Recommendations and Limitations}

The results of \S4 and \S5 are very encouraging in demonstrating that
linear combinations of optical emission-line luminosities with
a number of IR and radio continuum luminosities can be used to
produce attenuation-corrected luminosities and SFRs.  In Table 4
we summarize the best fitting coefficients $a$ (equations [1] and [2])
for each combination of optical and IR or radio continuum indices,
along with its formal uncertainty, and the mean dispersion of
individual galaxy luminosities around the best fitting relation.
For completeness we also list the corresponding values for the 
SINGS and MK06 samples measured separately.

 
Among the many alternatives, the combination of H$\alpha$ and
TIR luminosities offers the most robust attenuation corrections,
with a precision that rivals or exceeds that obtainable from 
de-reddening high quality integrated optical spectra, and 
the weakest systematic dependences on the SEDs and 
star formation properties of the parent galaxies.  However
for many applications reliable wavelength-integrated TIR luminosities
are not available, and in such situations single-band measurements
of the rest-frame 24\,\um\ or 8\,\um\ infrared fluxes, or alternatively
the 1.4\,GHz radio continuum fluxes appear to perform nearly
as well as combinations using the TIR fluxes.

\subsection{Composite Multi-wavelength SFR Calibrations and Extinction Estimates}

Throughout this analysis we have compared the different SFR
tracers referenced to the total attenuation-corrected \halpha\
luminosity, rather than the SFR itself.  We chose this convention
to anchor our results firmly in terms of observable quantities,
and to circumvent the additional systematic effects that enter into
the conversion of \halpha\ luminosities into SFRs.
However one can readily use the coefficients listed in Table 4 to 
construct SFR calibrations using these composite indicators.
Apart from their dependence on the coefficients in Table 4,
these absolute SFR calibrations also scale with the zero-point
of the reference 
SFR vs L(\halpha) calibration.   The latter is dependent on
the assumed slope and mass limits of the IMF, and on the
stellar synthesis models used in deriving this calibration.
For convenience we provide two such calibrations, one 
on the zero-point of the widely-applied calibration of K98,
and the other using a more realistic ``Kroupa" IMF (Kroupa \& Weidner 2003), as used in the
current version of the Starburst99 synthesis models (Leitherer et al. 1999).  

The calibration of K98 assumed for simplicity a single 
Salpeter (1955) slope power-law, with $\xi(m) \propto m^{-\alpha}$ 
with $\alpha=2.35$ between 0.1--100\,\msun, where $\xi(m) \equiv dN/dm$
is the number of stars with masses between $m$ and $m+dm$.  Ionizing 
luminosities for that calibration were taken from Kennicutt et al. (1994).
For this zero-point the composite SFR calibrations
take the form:

\begin{equation}
{\rm SFR [K98]} (M_\odot \,yr^{-1}) = 7.9 \times 10^{-42} [L(H\alpha)_{obs} + a_{\lambda}\,L_{\lambda}] \ ({\rm ergs\,s^{-1})},
\end{equation}

\noindent
where L(H$\alpha$)$_{obs}$ is the observed \halpha\ luminosity
without correction for internal dust attenuation, 
$a_{\lambda}$ is taken from Table 4 for the IR or radio
luminosity of interest (8\,\um, 24\,um, TIR, or 1.4 GHz), and
$L_{\lambda}$ is the luminosity in the respective wavelength band.
As an example, the calibration for the combination of \halpha\
and 24\,\um\ luminosities is:

\begin{equation}
{\rm SFR [K98]} (M_\odot \,yr^{-1}) = 7.9 \times 10^{-42} [L(H\alpha)_{obs} + 0.020 L(24)] \
({\rm ergs\,s^{-1})},
\end{equation}

\noindent
where $L(24) \equiv \lambda$\,L$_{\lambda}$ at 24\,um\ (or 25\,\um).
For this example the dust attenuation at \halpha\ would be
given by:

\begin{equation}
{\rm A(H\alpha) (mag)} = 2.5 \log {[1 + {{0.020\,L(24)} \over {L(H\alpha)_{obs}}}]}
\end{equation}

\noindent
The analogous calibration for indices using measurements of the
[\ion{O}{2}]$\lambda$3727 doublet is:

\begin{equation}
{\rm SFR [K98]} (M_\odot \,yr^{-1}) = 8.1 \times 10^{-42} [L([{\rm OII}])_{obs} + a{^{\prime}_{\lambda}}\,L_{\lambda}] \
({\rm ergs\,s^{-1})}.
\end{equation}

\noindent
Here we have designated the scaling coefficients $a{^{\prime}_{\lambda}}$
with a prime symbol to emphasize that these coefficients are different
from those derived for \halpha.  Again, as an example, the corresponding
calibration for combining [\ion{O}{2}] and 24\,\um\ luminosities is:

\begin{equation}
{\rm SFR [K98]} (M_\odot \,yr^{-1}) = 8.1 \times 10^{-42} [L([{\rm OII}])_{obs} + 0.029 L(24)] \
({\rm ergs\,s^{-1})},
\end{equation}

The second IMF, which we refer to for convenience as
a ``Kroupa" IMF, has the slope $\alpha=2.3$ for 
stellar masses 0.5--100\,\msun, and a shallower slope  
$\alpha=1.3$ for the mass range 0.1--0.5\,\msun.
The Kroupa IMF is more consistent with recent observations
of the Galactic field IMF (e.g., Chabrier 2003, 
Kroupa \& Weidner 2003).  We also have recalibrated the
ionization rate from this stellar population using 
Version 5.1 of 
Starburst99\footnote{URL: http://www.stsci.edu/science/starburst99/}.
For these models the zero-point of the SFR is lower than
for the K98 calibration by a factor of 1.44.  As a result
the calibrations in eqs. (11) and (14) become:

\begin{equation}
{\rm SFR [Kroupa]} (M_\odot \,yr^{-1}) = 5.5 \times 10^{-42} [L(H\alpha)_{obs} + 
a_{\lambda}\,L_{\lambda}] \ ({\rm ergs\,s^{-1})},
\end{equation}

\begin{equation}
{\rm SFR [Kroupa]} (M_\odot \,yr^{-1}) = 5.6 \times 10^{-42} [L([{\rm OII}])_{obs} + 
a{^{\prime}_{\lambda}}\,L_{\lambda}] \ ({\rm ergs\,s^{-1})},
\end{equation}

\subsection{Applicability Limits:  Where Do These Methods Break Down?}

As with any empirical ``toolbox" of SFR calibrations, these 
composite indicators have been calibrated over limited ranges
of galaxy properties, SFRs, and physical environments, and it
is important to understand both the range of observations beyond
which the methods are untested, and any systematic errors that
may have been built into the zero-points of the methods.
We refer readers to the review in K98 for a discussion
of the assumptions and systematic uncertainties underlying the
individual SFR tracers that have been considered in this paper.
However there are a few additional cautions and caveats that
apply to this new set of composite indicators.

The parent SINGS and MK06 galaxy samples cover essentially
the full range of galaxy types found in the
local universe, ranging from virtually dust-free dwarf galaxies to
ULIRGs and luminous AGN hosts.  However we needed to restrict
our analysis to star-forming galaxies with high signal/noise 
multi-wavelength data, and this vetting process narrowed the 
coverage of galaxy properties covered by these calibrations.
In particular our galaxies
cover restricted ranges in \halpha\ attenuation (0 -- 2.5 mag) and
corresponding observed $L(TIR)/L(H\alpha)$ ratio (45 -- 3150), as well as 
in underlying stellar population and age ($D_n(4000) \le 1.4$),
corresponding roughly to the transition between lenticular (S0)
and spiral galaxies.  The highest IR luminosities in our final
sample were $\log L(TIR)/L_\odot \sim 11.9$, just below the threshold
for a ULIRG, and the highest attenuation-corrected SFRs are
$\sim$100\,\sfr.  This range readily encompasses all normal
galaxies in the present-day universe and 
most star-forming galaxies out to redshifts $z \sim 1$.
However it does not encompass the most luminous star-forming 
galaxies found at high-redshift, or the most dust-obscured 
LIRGs or ULIRGs found in the present-day universe.  At the
other extreme, our methods can break down for early-type red
galaxies with very low SFRs, where the dominant IR dust emission
can arise from evolved stars (e.g., Cortese et al. 2008,
Johnson et al. 2009).  Both red dusty galaxies and very
dusty starburst galaxies will manifest themselves by very
red colors in the visible and/or by anomalously large $L(IR)/L(H\alpha)$
ratios, and application of the limits above can flag such
potentially problematic cases.  We still
believe that these composite methods provide more robust 
SFRs than any of the single-wavelength methods, but one should
attach larger systematic errors (up to a factor of two) for
systems that lie outside of the bounds of our calibrations.

As an additional cautionary note, we emphasize that the relations
in Table 4 were derived from integrated measurements of entire
galaxies, or of regions covering several square kiloparsecs
in area in the centers of galaxies, and as such the SFR calibrations
given here only apply in regions that encompass a representative
sampling of the integrated light of a galaxy.
In particular, the coefficients $a$ differ significantly
from those of individual HII regions and HII region complexes.
When we compare the relation between \halpha\ and 24\,\um\ luminosities
of SINGS HII regions (C07) with that for the integrated measurements
of galaxies, we observe a significant offset (see Figure 15),
with a best fitting value $a_{24} = 0.031 \pm 0.006$ for the
HII regions, compared to $a_{24} = 0.020 \pm 0.001_{\rm r} \pm 0.005_{\rm s}$ for the
galaxies.  This difference is almost certainly due to the
different stellar age distributions in the HII regions and
the galaxies as a whole.  The ionizing lifetimes of O-stars
are nearly all less than 5 Myr, so in HII regions the massive early-type
stars dominate both the gas ionization and the heating of
the dust that produces the 24\,\um\ dust emission.  
On the other hand when one measures entire galaxies that
ionized gas emission is still dominated by massive O-stars,
but the dust heating includes an additional component from
stars older than 5 Myr.

We can estimate the importance of this change in stellar
populations quantitatively, by comparing the evolution in
ionizing UV, far-UV, and bolometric luminosities of 
star-forming populations with age.  We used Version 5.1
of the Starburst99 package of Leitherer et al. (1999) 
to trace the evolution of \halpha\ luminosity, 1500\,\AA\
UV continuum luminosity, and bolometric luminosity for
continuously star-forming populations with ages of
5 Myr (appropriate to an HII region), and 0.1--1 Gyr
(appropriate for galaxies, the ratios do not change
significantly at larger ages).  For either a Salpeter
or Kroupa IMF as defined in \S 6.1  
the ratio of bolometric luminosity to  
\halpha\ luminosity increases by a factor of $\sim$2.0
between ages of 5 Myr and 100 Myr (and $\sim$2.5 when compared
to age 1 Gyr).  We expect the ratio of
dust luminosity to ionizing luminosity to be less sensitive
to these age differences, however, because the attenuation
of a given population decreases with increasing stellar
age (e.g., Zaritsky et al. 2002).  
If we apply the prescriptions of Charlot \& Fall (2000)
or Calzetti et al. (2000), we expect the HII regions to 
experience $\sim$2--3 times the dust attenuation as
the average starlight in a starburst population.  Thus
the additional component of UV/bolometric luminosity
in an older population should increase the dust emission 
(for a given fixed ionizing luminosity) by a factor 
1.4--1.6, similar to the difference in 24\,\um/\halpha\
ratios we observe between the HII regions and the galaxies.

This ``older" dust emission component can be directly
observed in our {\it Spitzer} 24\,\um\ images of the 
SINGS galaxies.  Figure 19 compares 24\,\um\
and \halpha\ images of M81, a nearby, well-resolved
spiral galaxy that clearly illustrates the different
IR emission components.  Virtually all of the 24\,\um\
point sources are associated with bright optical HII
regions (Prescott et al.\,2007), which validates our
association of the 24\,\um\ emission with the dust
attenuation in the HII regions.  However upon closer
examination a diffuse IR emission component can also
be observed, which extends between the HII regions 
and exhibits a more filamentary morphology, roughly
tracing that of the HI gas and the diffuse 8\,\um\ PAH
emission (e.g., Gordon et al. 2004; P\'erez-Gonz\'alez et al. 2006).
This diffuse component is the ``infrared cirrus" 
component of the dust emission and has been identified
in many other galaxies observed with {\it ISO} and
{\it Spitzer} (e.g., Hippelein et al. 2003; 
Hinz et al. 2004; Popescu et al. 2005).  According
to the measurements of Dale et al. (2007) this diffuse
component contributes $\sim$50\%\ of the total 
24\,\um\ emission in M81, similar to the average
fraction in the SINGS sample as a whole.  These
observations tend to support the suggestion by 
Zhu et al. (2008) that the difference in the 24\,\um\
vs \halpha\ correlations between galaxies and HII 
regions are due to the presence of diffuse IR emission.

These results should serve as a stern warning that 
the composite SFR indicators calibrated in this paper,
C07, and Zhu et al. (2008) {\it cannot be applied to
map the spatially-resolved SFR in galaxies} without
the risk of introducing significant and possibly large
systematic errors in the resulting SFR maps.
Our results show that these methods can be applied
reliably to individual HII regions and to galaxies
as a whole (albeit with different scaling factors
between optical and IR emission).  However when observed
at high spatial resolution galaxies show diffuse components
of both \halpha\ and IR emission which may be completely
unassociated with any star formation at the same position.
The diffuse IR component is especially problematic, because
this dust may be only partially heated by young stars,
or possibly not by young stars at all.  As a result
blind application of the relations in Table 4 to 
multi-wavelength images of galaxies will tend to 
produce large regions with spuriously high ``SFRs"
where little or no star formation actually is taking
place.  Methods incorporating other information 
will be needed to extend this type of analysis to
make reliable spatially-resolved maps of star formation
in galaxies.

\subsection{Systematic Uncertainties}

We conclude by re-examining the possibility of systematic
errors in our attenuation scales (\S3.3), now informed by 
the results presented in the previous sections.  The most
important assumptions are the adoption of reference attenuation
corrections based on Balmer decrements with a Galactic extinction
curve and a foreground screen dust geometry, and the assumption
of a constant intrinsic ratio of ionizing luminosity to bolometric
luminosity.  

One of the surprises in the results (to us), has been the 
relatively tightness of the relations between the 
attenuation-corrected \halpha\ luminosities,
surface brightnesses, and attenuations themselves derived
from linear combinations of \halpha\ and IR (or radio) luminosities,
when compared to the same corrected quantities derived from
Balmer decrement measurements.  The agreement in the mean
attenuations is not significant at all, of course, because
we calibrated the scales to match, but the relatively low
dispersions in the comparisons shown in Figures 6--13, 
along with the absence of significant nonlinearities suggests
that any systematic effects from variations in dust geometry
and stellar population variations are likely to be of
secondary importance.  

Although the consistency of our results is encouraging, it
does not rule out the possibility of a systematic error that affects the
entire attenuation and SFR scales, or a systematic error that
scales smoothly with the SFR itself.  Perhaps the most vulnerable
assumption is the adoption of the stellar-absorption corrected Balmer decrement in
the integrated spectrum (using a foreground dust screen
approximation) as the reference for calibrating our composite
SFR and attenuation scales.  Galaxies clearly display point-to-point
variations in line-of-sight attenuation, with the integrated spectrum
representing a flux-weighted average, including effects of scattered
light.  Although recent models suggest that the foreground dust
screen approximation does not introduce significant systematic
errors when applied to normal star-forming galaxies (Jonsson et al. 2009),
it is important to place limits on the magnitude of any systematic
errors in our calibrations and attenuation scales resulting from
these simplifying assumptions.  

Ideally one would use as references measurements of ionizing fluxes
of galaxies that are less susceptible to dust attenuation, such 
as the Paschen or Brackett emission lines, or the free-free thermal
radio continuum, which scales directly with the ionizing flux.
Unfortunately such measurements are not available for large samples
of galaxies, or contain such large uncertainties that they cannot
be applied individually to galaxies.   That is why we have chosen
to base our analysis on Balmer decrement measurements.  However
some of these measurements are available for subsets of our sample,
and we can use these to estimate \halpha\ attenuations and compare
them to those we derived from the Balmer decrements.

Our first empirical test comes from Pa\,$\alpha$ and \halpha\ measurements for 29 of
the SINGS galaxies in our sample (C07).  The Pa\,$\alpha$ observations
were made with the Near-Infrared Camera and Multi-Object Spectrometer
(NICMOS) on the Hubble Space Telescope, and are restricted to the
central 50\arcsec\ $\times$ 50\arcsec\ regions of the galaxies
(and a 144\arcsec\ field for M51).  We used the Pa\,$\alpha$
fluxes to derive attenuation-corrected \halpha\ luminosities,
using the theoretical recombination ratio as described in C07,
and using the observed ratio of Pa\,$\alpha$/\halpha\ fluxes to
correct for the (weak) attenuation in Pa\,$\alpha$ ($\lambda$ = 1.89\,\um).
We do not have optical spectra with matching coverage to these
regions, but we can measure \halpha\ and 24\,\um\ fluxes over
the same regions, and derive attenuation-corrected \halpha\ luminosities,
using eq. (1) with $a = 0.020$, as calibrated from our Balmer decrement
measurements.  Figure 20 compares the resulting \halpha\ surface brightnesses
(cf. Figure 9) using the two methods.  There is a considerable
scatter in the comparison, which probably reflects the measuring
uncertainties in the Pa$\alpha$ photometry.  The
two scales are in good agreement, with an average difference of 0.024$\pm$0.036 dex
(attenuation from  Pa\,$\alpha$/\halpha\ higher
by 0.06$\pm$0.09 mag) and no evidence for nonlinearity, over a
full range of 3 mag in A(\halpha).  

Estimated thermal fractions of radio continuum fluxes are available for
27 galaxies in our sample from the study of Niklas et al. (1997).
The thermal radio component arises from free-free emission of
thermal electrons in the ionized gas, and its brightness scales
directly with the unattenuated \halpha\ emission, with a weak
dependence on electron temperature (Niklas et al. 1997):

\begin{equation}
S_{th} (mJy) = { ({{{2.24 \times 10^9}\,S_{H\alpha}} \over {ergs\,s^{-1}\,cm^{-2}}}}) {({T_e \over K})
^{0.42}} {({\ln {({0.05 \over {\nu/GHz}})} + 1.5\,\ln{(T_e/K)}})} .
\end{equation}

Since the dust extinction in the radio is negligible the ratio of
thermal radio to \halpha\ flux provides a measure of the \halpha\
attenuation that is unaffected by any geometric effects in the Balmer extinction.
For most galaxies the thermal component of the radio emission represents
only a small fraction of the total emission at centimeter wavelengths,
with the dominant contribution arising from non-thermal synchrotron emission.
Niklas et al. (1997) obtained integrated radio fluxes at several (typically
6--7) wavelengths, in order to decompose the thermal and non-thermal components.
The resulting thermal radio fractions and fluxes for individual galaxies
have an average uncertainties of $\pm$0.33 dex (0.83 mag), but this
is sufficient to test for a systematic difference
in \halpha\ flux scales.  Figure 21 shows a histogram of the differences
between the \halpha\ attenuations estimated from the radio/\halpha\ ratios using eq. (18)
under the assumption of $T_e=10^4 K$,
and those derived from the \halpha/H$\beta$ ratios.  The median 
radio continuum attenuation (more reliable than the mean given the
large scatter in measurements) is 0.02 $\pm$ 0.13 mag lower than that
estimated from the Balmer decrements.  In view of the large uncertainty
in the radio measurements we cannot rule out a small systematic bias
in the Balmer decrement measurements, but this test, as with the Pa\,$\alpha$
measurements, appears to rule out a significant systematic error.

For our final empirical test we compared the attenuations estimated
from the Balmer decrements with those which can be derived by 
comparing the resulting \halpha-based SFRs
with attenuation-corrected SFRs from UV and IR measurements.
This test is less direct than those based on radio continuum and
Pa\,$\alpha$ measurements described above, because the intrinsic
ratio of \halpha\ and UV continuum luminosities of a galaxy are
dependent on other factors including the star formation history
and the slope and upper mass limit of the IMF.  We address this
in detail in Paper II, but apply a simple test here to place
limits on any systematic errors in the Balmer decrement based 
attenuation measurements.

Several authors have published prescriptions for using the
combination of UV and IR fluxes of galaxies to derive
attenuation-corrected UV luminosities and SFRs, based on
energy-balance arguments that are similar to those presented
in this paper (e.g., Gordon et al. 2000; Bell 2003; Hirashita
et al. 2003; Kong et al. 2004; Buat et al. 2005;
Iglesias-P\'aramo et al. 2006; Cortese et al. 2008).
We have applied the prescriptions of Kong et al. (2004) and
Buat et al. (2005), which lie in the mid-range of published
calibrations, to FUV and TIR luminosities of the SINGS and MK06
galaxies (Paper II), to derive attenuated-corrected FUV luminosities,
and converted these to SFRs using the most recent Starburst99
models and the Kroupa IMF described in \S6.1.  These yield:

\begin{equation}
SFR  (M_\odot\,yr^{-1}) = 8.8 \times 10^{-29} L_\nu  (ergs\,s^{-1}\,Hz^{-1})
\end{equation}

\noindent
We then compared these SFRs to those derived from the observed (uncorrected)
\halpha\ luminosities, for the same synthesis models and IMF:

\begin{equation}
SFR  (M_\odot\,yr^{-1}) = 5.5 \times 10^{-42} L(H\alpha)  (ergs\,s^{-1})
\end{equation}

\noindent
The ratio of the attenuated-corrected FUV-based SFR to the uncorrected
\halpha-based SFR provides an indirect estimate of the attenuation in
\halpha.  We find that these attenuation estimates are slightly
higher than those derived from the Balmer decrements in the foreground
screen approximation, by 0.08 mag in the median for the Kong et al. (2004)
calibration and by 0.18 mag for the Buat et al. (2005) calibration.
Since the average \halpha\ extinction for the sample is 0.8 mag, the
UV-based attenuation estimates are higher by 10\% and 23\% respectively.
Our own calibration in Paper II (which do not depend explicitly on
the Balmer attenuation scale) yields attenuations that are close
to those derived by Kong et al. (2004).  As a check, one can
also use the UV and IR fluxes to derive the FUV attenuation directly
using the published calibrations by those authors, and apply the
Calzetti (2001) attenuation law to estimate A$_{\rm H\alpha}$ 
from A$_{\rm FUV}$.  This yields \halpha\ attenuations which 
are 0.10, 0.15, and 0.07 mag higher than the Balmer-derived attenuations,
for the calibrations of  Kong et al. (2004), Buat et al. (2005), and
Paper II, respectively.  These are very close to the values derived
from a comparison of FUV and \halpha-based SFRs above.  Based on
the dependence of these results on the calibration used, as well
as the systematic dependences on stellar age mix and IMF we
conservatively estimate in the uncertainties in the comparison
of UV$+$IR and Balmer-line attenuation estimates to be $\pm$20\%.  
Therefore if we average
the results based on Kong et al. (2004) and Buat et al. (2005)
we find a difference in attenuations of $0.13 \pm 0.16$ mag.

Taken together, the Pa\,$\alpha$, thermal radio continuum, and
UV+IR based estimates of \halpha\ attenuations give average
values that are 0.06 $\pm$ 0.09 mag higher, 0.02 $\pm$ 0.13
mag lower, and 0.13 $\pm$0.16 mag higher than those derived
from the Balmer decrements.  This may hint at a slight tendency
for the Balmer decrements to under-estimate the actual \halpha\
attenuations, but within the uncertainties the scales are 
consistent.  It is also notable that other widely adopted 
schemes such as the modified Charlot-Fall attenuation law
(Wild et al. 2007; da Cunha et al. 2008) produce attenuations 
that are within 0.1 mag on average of those derived
here.  These results rule out large systematic errors
in our attenuation scale, with a very conservative upper
limit of 20\%.

The other important assumption underlying our analysis is
the implicit adoption of a constant dust-heating stellar
population for all of the galaxies in our sample.  This is the same
``IR cirrus" problem that has bedeviled efforts to calibrate
the IR emission of galaxies as a quantitative SFR tracer
(e.g., Lonsdale Persson \& Helou 1987, Sauvage \& Thuan 1992).
We already have seen evidence for the effects of varying 
stellar population in the comparison of calibrations of
eq. (1) between disk HII regions and integrated measurements
of galaxies (\S6.2).  The results suggest that although
the coefficients $a$ derived in this paper appear to apply
over a wide range of normal galaxy populations, they may
systematically under-estimate the SFR by as much as 60\%\
in systems that are dominated by young stars, such as 
emission-line starburst galaxies and possibly luminous
and ultraluminous starburst galaxies with young dust-heating
stellar populations. 

This difference in calibrations helps to account for a curious
inconsistency between our results and previous calibrations of
SFRs for starbursts by K98 and other authors.
The relationship from that paper:

\begin{equation}
SFR  (M_\odot\,yr^{-1}) = 4.5 \times 10^{-44} L_{bol}  (ergs\,s^{-1})
\end{equation}
yields SFRs that can be up to 2.2 times higher than those
given by eq. (11) and the coefficients in Table 4, when
applied to very dusty galaxies, where the TIR term in eq. (11)
dominates.  Most of this difference can be readily accounted
for by the different assumptions underlying the two calibrations.
The K98 relation is entirely theoretical, and
it assumes dust heating by stars with ages less than 30 Myr,
whereas the relations in this paper were calibrated empirically,
and automatically incorporate any contributions to dust
heating from stars older than 30 Myr.  If one adopts instead
a bolometric based SFR calibration using a 
dust-heating continuum that is appropriate for normal
spiral galaxies (continuous star formation over 1--10 Gyr)
the inconsistencies with the empirical calibrations in this
paper largely disappear.  Taken at face value
this would suggest that approximately half of the dust heating
in the galaxies in our sample arises from stars older than
30 Myr (Cortese et al. 2008; Johnson et al. 2009).

To investigate this quantitatively we have applied the 
simple dust radiative transfer model of C07
to calculate approximately the expected TIR
emission from a disk galaxy with a constant star formation
history over the past 10 Gyr.  The C07 models were
parameterized in terms of the SFR per unit area, and
we converted these approximately to luminosities by
using the median area of the \halpha\ disks of the 
SINGS galaxies (200 kpc$^2$).  
The model relation is plotted with the observations in
the right panel of Figure 22, for the fitted value
$a = 0.0024$, while the left panel shows for comparison
the expected relation if only stars younger than 30 Myr
heated the dust.  The consistency of the model with the
fitted attenuation corrections suggests that any
bias in the SFR scales from systematic errors in the
Balmer-based attenuation corrections is small.


Our focus on possible systematic effects should
not obscure the immense improvements in the reliability of SFRs
that are afforded by these composite tracers.  Many current 
measurements of SFRs are based on observations at a single
wavelength region, with perhaps a crude statistical correction
for dust attenuation.  As shown in the upper panels of Figures 6--8,
the random uncertainties in SFRs produced by these single-tracer
methods is typically of order a factor of two, and systematic
errors can easily reach a factor of ten or higher.  The methods
developed here reduce the random errors in the attenuation 
corrections from factors of a few
to of order $\pm$15--30\%, and systematic errors to of order
$\pm$10\% in most galaxies.  As discussed above systematic
errors of up to a factor of two may be present in the worst
cases, but those errors are due to uncertainties in stellar
populations in the galaxies, {\it not} in the dust attenuation
scales.  Other factors such as IMF variations may introduce
larger errors in the SFR scales than discussed here, but at least
one should be able to remove dust
attenuation as the dominant source of uncertainty in the extragalactic
SFR scale.

\section{Summary}

We have used two samples of galaxies with measurements of 
the optical emission lines and infrared and radio continuum to
explore the combination of optical and IR (or radio) fluxes
to derive attenuation-corrected SFRs.  Our main conclusions can
be summarized as follows.

1.  Linear combinations of \halpha\ and total infrared (TIR) luminosities
provide estimates of \halpha\ attenuation (and thus attenuation-corrected
\halpha\ luminosities) that are in good agreement with those derived
from measurements of the stellar absorption-corrected \halpha/\hbeta\ 
ratios (Balmer decrements).  The corrected luminosities agree to
within $\pm$23\%\ rms, with essentially no systematic dependence on
luminosity, surface brightness, or spectral properties of the stellar
population, over the range of galaxy types, and SFRs tested.

2.  Similar linear combinations of \halpha\ luminosities with 
single-band IR luminosities (e.g., 24\,\um, 8\,\um) provide 
attenuation corrections that are nearby as consistent with Balmer
decrement measurements as those derived from combinations of 
\halpha\ and TIR measurements.  Combinations of \halpha\ and 
24\,\um\ measurements show small systematic variations which
correlate with IR SED shape, and can be understood as correlating
with the fraction of dust luminosity that is radiated in the
24\,\um\ band.

3.  The same methodology can be applied to other optical emission-line
tracers (e.g., [\ion{O}{2}]$\lambda$3727) or using the 1.4 GHz radio continuum
emission in lieu of infrared continuum measurements.

4.  For a given attenuation in the \halpha\ line the ratio of IR/\halpha\ 
luminosities is systematically higher for integrated measurements of
galaxies than it is for individual HII regions or star-forming complexes.
We attribute most or all of this effect to differences in the mean
ages of the stars heating the dust in galaxies and HII regions.  
This excess dust heating from a non-star-forming population is in
good agreement with expectations from evolutionary synthesis models.

5.  A corollary conclusion of the previous result is that a considerable
fraction of the total infrared radiation in normal star-forming galaxies
is heated by stars older than 100 Myr.  Comparisons with evolutionary
synthesis models suggests that up to 50\%\ of the TIR emission could be
from dust heated from this evolved stellar population.  Part of this excess
emission is probably associated with the diffuse ``infrared
cirrus" dust emission in galaxies, but not necessarily uniquely associated,
because it is possible that young stars may also partially heat the
diffuse dust.

6.  We present prescriptions for measuring SFRs of galaxies from linear
combinations of \halpha\ or [\ion{O}{2}]$\lambda$3727 emission lines with 
TIR, 24\,\um, or 8\,\um\ infrared measurements, or 1.4 GHz radio continuum
measurements.  These appear to be reliable for normal star-forming galaxies
with SFRs in the range 0--80 \msun\,yr$^{-1}$ and A(\halpha) = 0 -- 2.5 mag.
The calibrations may be less reliable for early-type galaxies with
UV--optical radiation fields dominated by evolved stars, or for 
very highly obscured starburst galaxies, where very young stars 
dominate dust heating.  

7.  We carefully examine the systematic reliability of our SFR scales.
The use of Balmer line ratios from optical spectra to anchor the
attenuation scale may introduce a small systematic bias in the SFR
scales, but comparisons to independent tracers show that these errors
are probably less than 15\% on average.  Variations in dust heating
stellar populations in galaxies are probably a larger source of
systematic error.  In particular our prescriptions yield
SFRs for dusty IR-luminous starburst galaxies that are
approximately a factor of two lower than
those given by the widely used calibration in Kennicutt (1998a),
possibly due to an inapplicability of our algorithms in this extreme
star formation regime.  Overall we expect that these algorithms should
provide attenuation-corrected SFRs accurate to $\pm$0.3 dex or better 
for nearly all star-forming galaxies, excluding possible effects of variable IMFs.

\acknowledgements

This work is based in part on observations made with the Spitzer Space
Telescope, which is operated by the Jet Propulsion Laboratory, California
Institute of Technology, under a contract with NASA.  Support for this work was
provided by NASA through an award issued by JPL/Caltech.  C. Hao gratefully
acknowledges the support of a Royal Society UK--China Fellowship.  C. Hao also
acknowledges the support from NSFC key project 10833006. We also gratefully
acknowledge a critical reading of the paper by an anonymous referee, who led us
to undertake a much more rigorous analysis of the systematic uncertainties in
our methods.

\pagebreak

\begin{figure}
\plotone{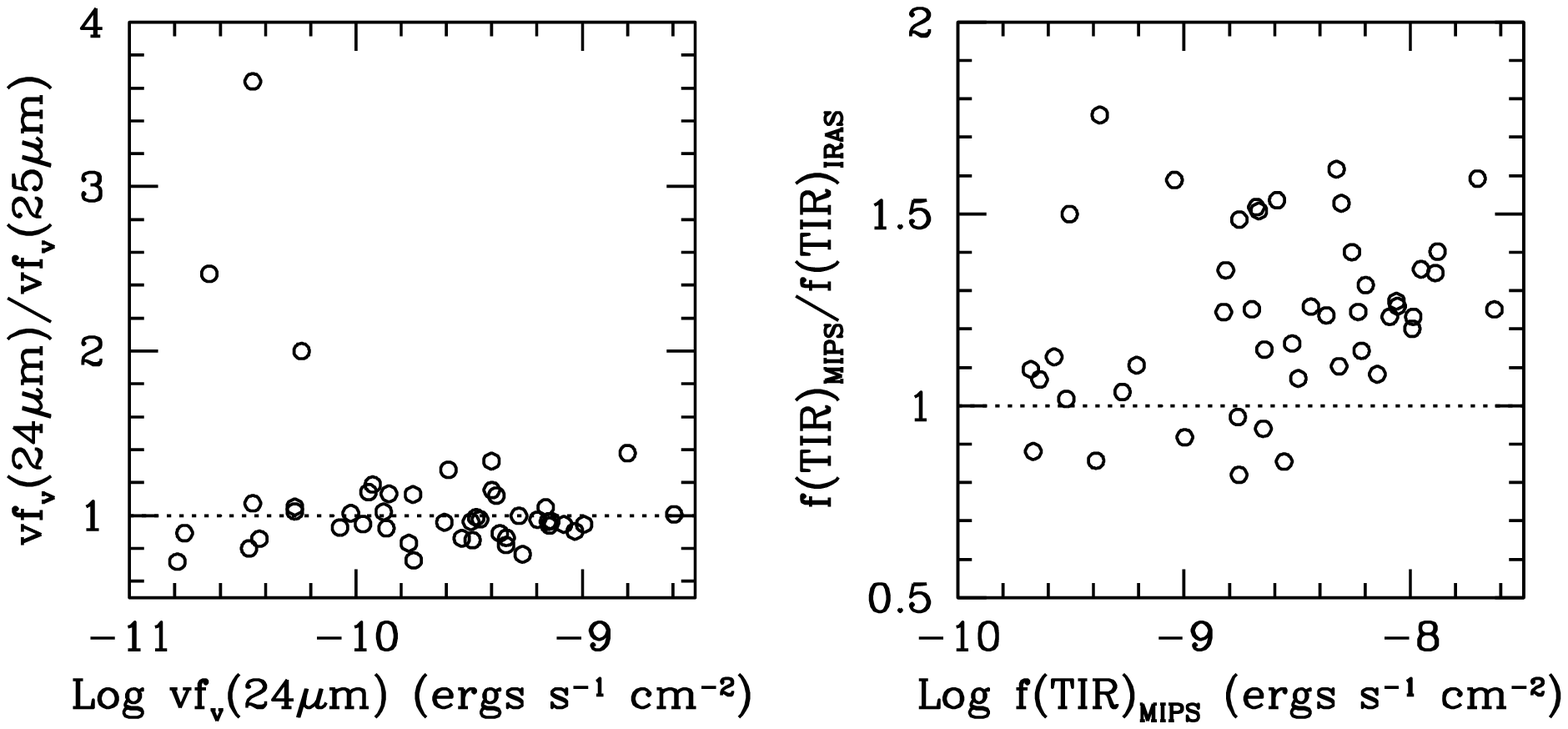}
\caption{{\it Left:}\ Ratio of MIPS 24\,\um\ flux to {\it IRAS} 25\,\um\
flux for SINGS galaxies, plotted as a function of MIPS (apparent) flux.
{\it Right:}\
Ratio of MIPS total infrared (TIR) flux to {\it IRAS} TIR flux for
galaxies in the same sample.}
\end{figure}

\begin{figure}
\plotone{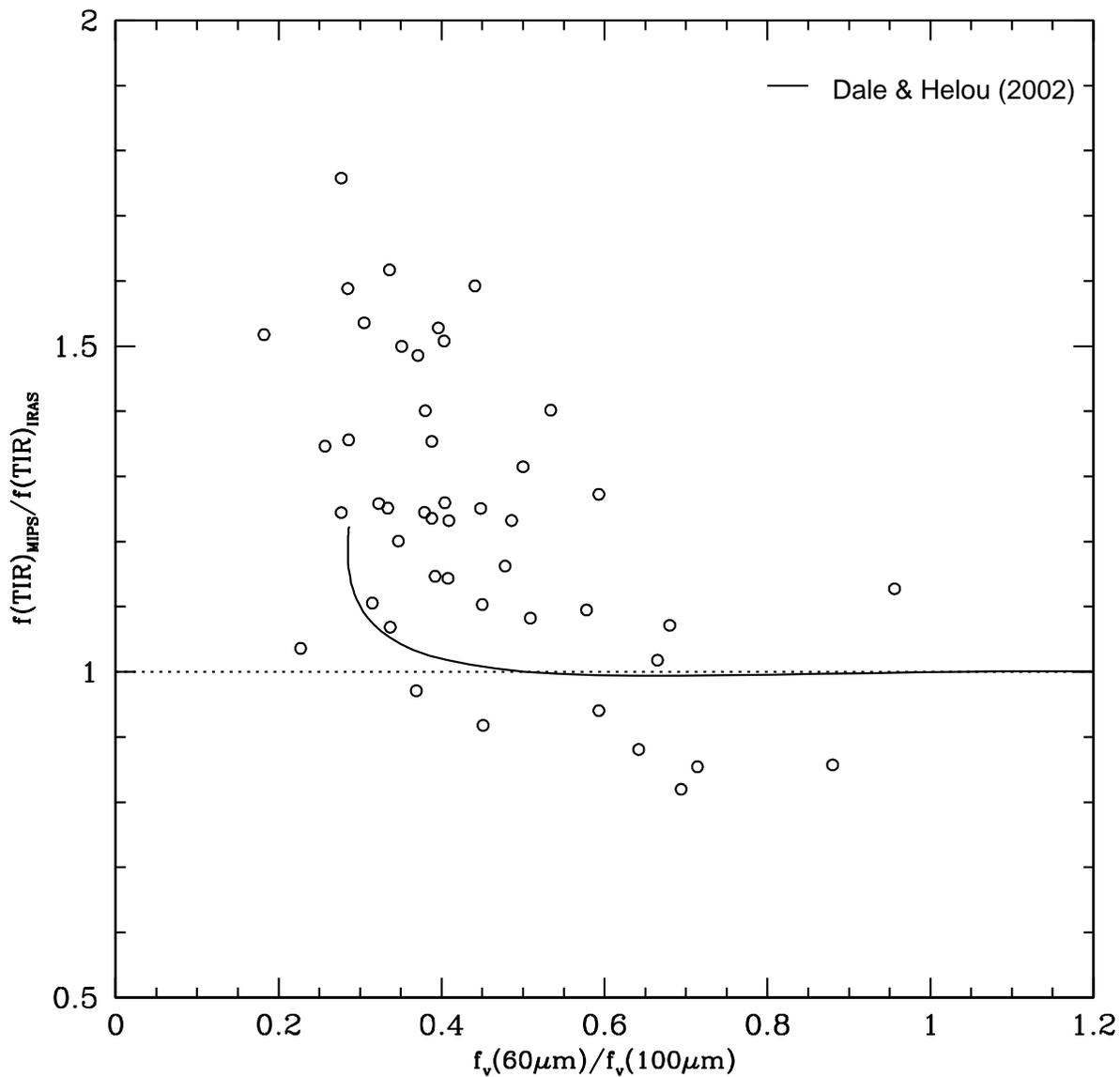}
\caption{Ratio of MIPS total infrared (TIR) flux to {\it IRAS} TIR flux for
SINGS galaxies, calculated using the prescriptions of Dale \& Helou (2002),
plotted as a function of {\it IRAS} 60\,\um\ to 100\,\um\ flux ratio.
Note that the galaxies with the largest discrepancies possess relatively cold
dust temperatures (low $f_\nu(60\,\mu {\rm m})/f_\nu(100\,\mu {\rm m})$ ratios).  The solid line shows the expected
relation based on SED models in Dale \& Helou.  Many of the SINGS galaxies with
the coldest $f_\nu(60\,\mu {\rm m})/f_\nu(100\,\mu {\rm m})$ colors appear to show even larger discrepancies.}
\end{figure}

\begin{figure}
\epsscale{0.9}
\plotone{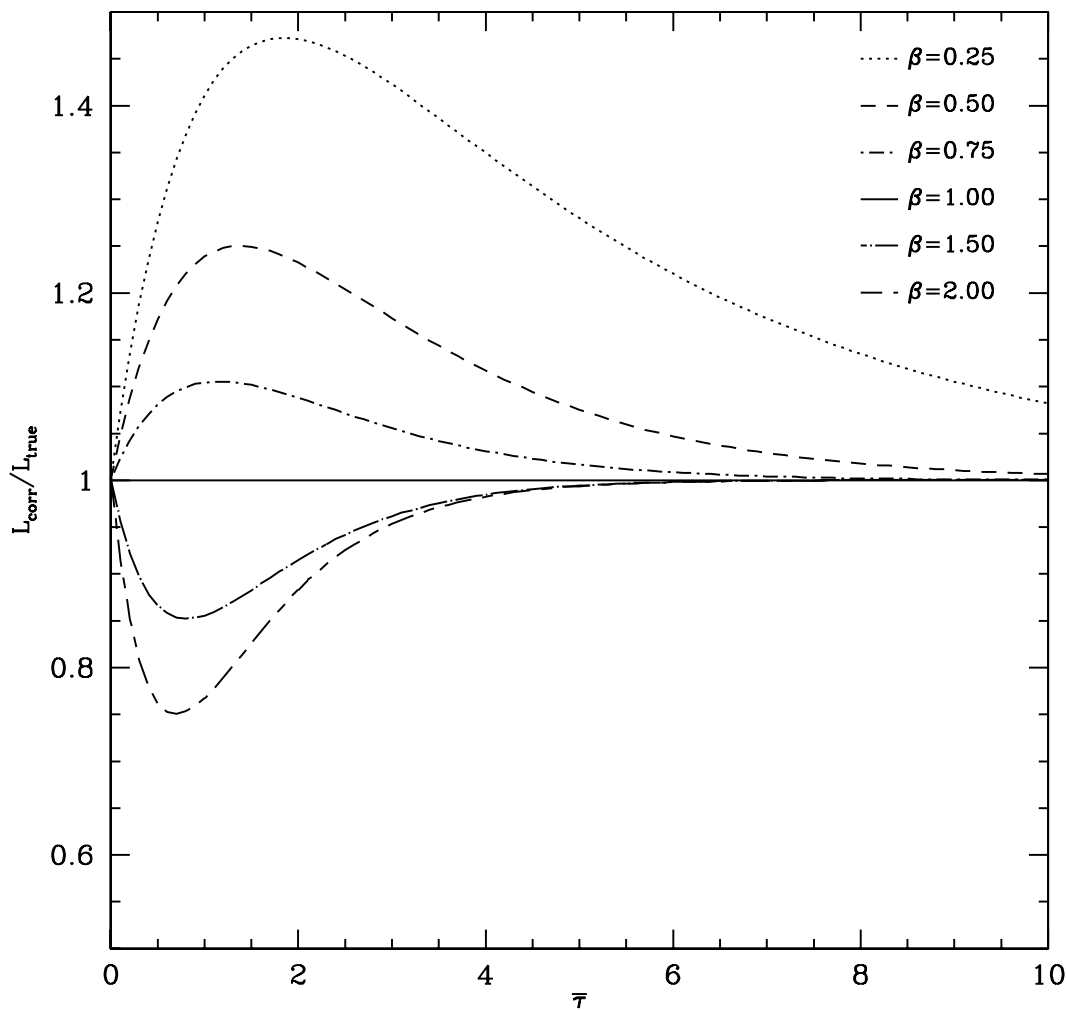}
\caption{These curves show the approximate magnitude of the systematic error
in attenuation-corrected SFRs introduced by applying a linear
combination of optical emission-line and IR continuum fluxes of galaxies to
estimate the mean attenuation.  The curves are parameterized by the ratio of
the attenuation in the emission-line to that of the average dust heating
continuum ($\beta$).  The linear combination is an exact relation when the
two opacities match ($\beta = 1$). } 
\end{figure}

\begin{figure}
\epsscale{0.9}
\plotone{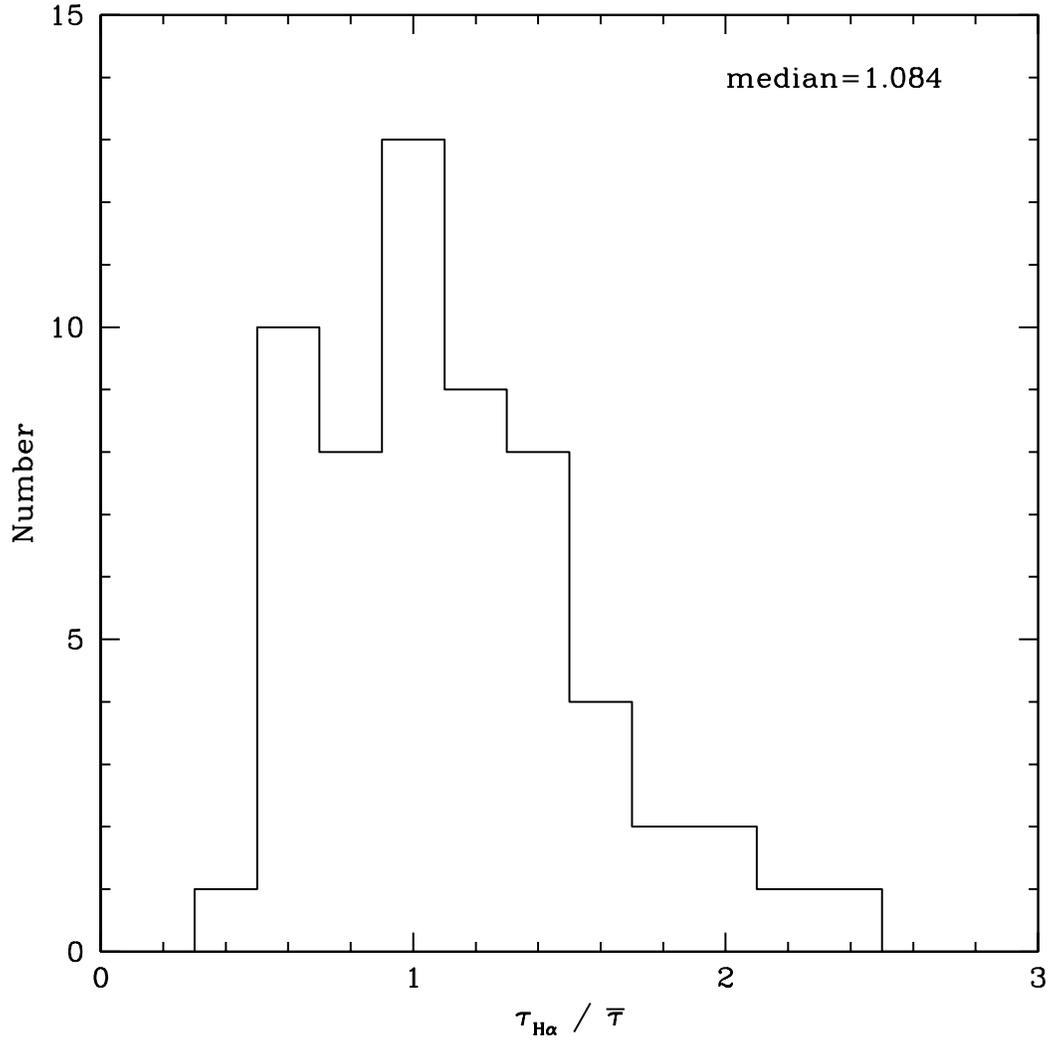}
\caption{Histogram of ratios of dust attenuation in the \halpha\
emission line, as estimated from the Balmer decrement as described
in the text, to the mean continuum dust opacity, estimated from the
bolometric fraction of IR emission in the SED of each MK06 galaxy.
The median value shows that the galaxies lie close to the $\beta = 1$ case.}
\end{figure}

\begin{figure}
\epsscale{0.9}
\plotone{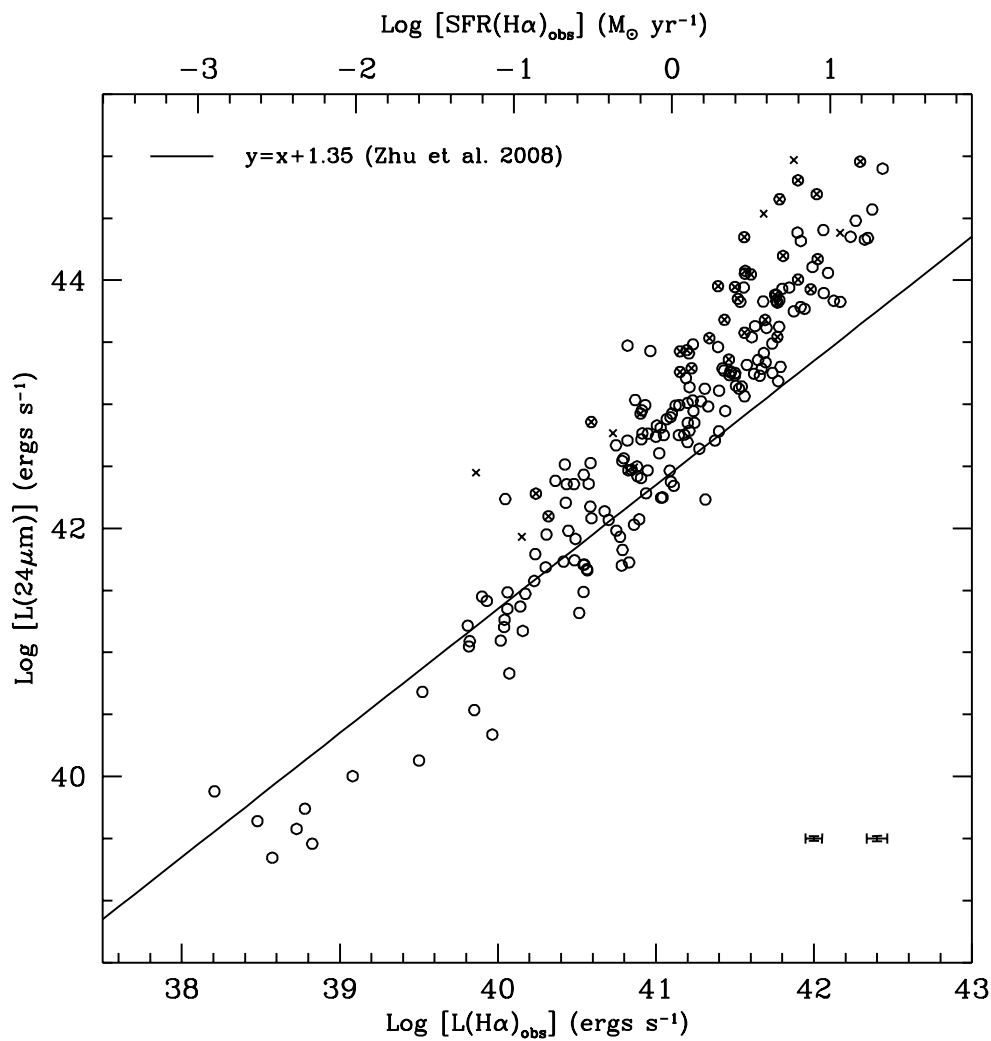}
\caption{Relation between observed 24\,\um\ infrared luminosity and observed 
\halpha\ luminosity (uncorrected for attenuation) for galaxies in the SINGS
 and MK06 samples.  Symbol shapes are coded by
dominant emission-line spectral type: circles for HII region like spectra,
crosses for spectra with strong AGN signatures, and crossed circles for 
galaxies with composite spectra.  The solid line shows a relationship
with linear slope, to illustrate the strong nonlinearity in the observed relation.
The axis label at the top of the diagram shows the approximate range of
SFRs (absent a correction for dust attenuation at \halpha) for reference. }
\end{figure}

\begin{figure}
\plotone{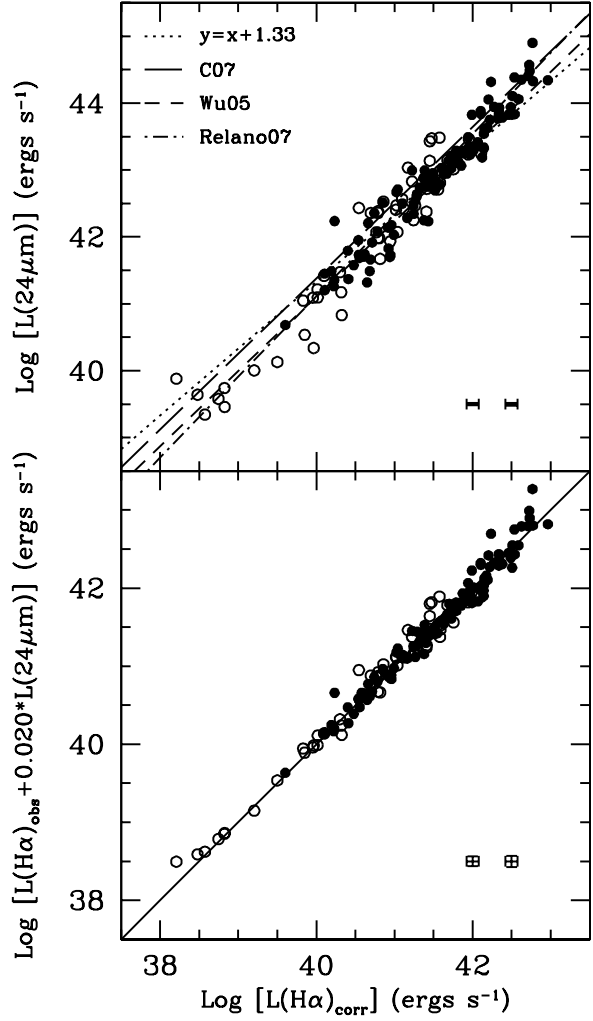}
\caption{{\it Top:}\ Relation between observed 24\,\um\ IR luminosity and 
attenuation-corrected \halpha\ luminosity for star formation dominated
SINGS galaxies (open circles) and MK06 galaxies (solid circles).  The 
attenuation corrections were derived from the absorption-corrected 
\halpha/\hbeta\ ratios in the optical spectra.  The dotted line shows
a linear relation for comparison, while the solid line shows the best
fitting nonlinear relation for HII regions in C07.  The other two lines show published fits to other
galaxy samples.  The error bars in the lower
right show typical 1-$\sigma$ uncertainties for individual measurements of SINGS sample (left) and MK06 sample (right).
{\it Bottom:}\ Linear combination of (uncorrected) \halpha\ and 24\,\um\ luminosities
compared to the same Balmer-attenuation-corrected \halpha\ luminosities,
with the scaling coefficient $a$ (eq. [1]) derived from a median fit
to the linear relation.  Note the tightness of the relation over nearly
the entire luminosity range.}
\end{figure}

\begin{figure}
\plotone{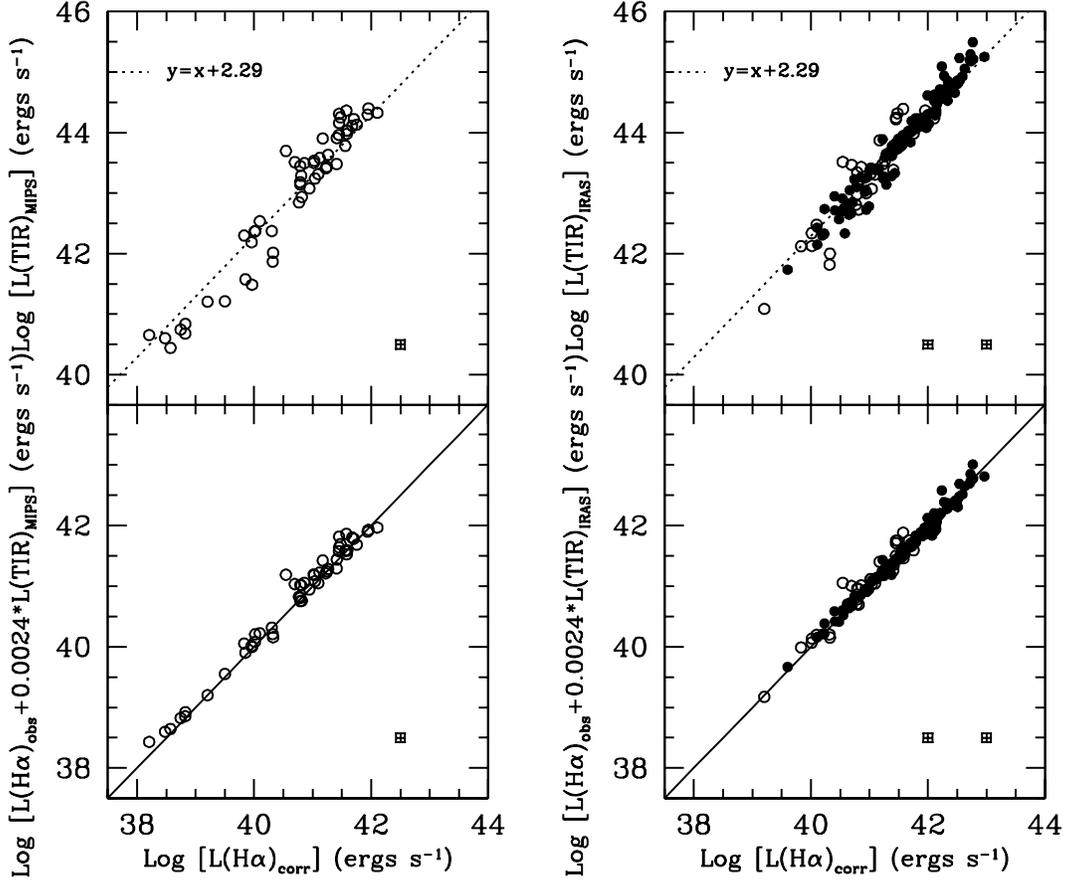}
\caption{The same comparison as in Figure 6, but comparing total infrared
(TIR) and \halpha\ luminosities, in place of 24\,\um\ vs \halpha\ luminosities.
The left-hand pair of plots are for {\it Spitzer} MIPS TIR measurements of 
the SINGS sample, while the right-hand pair of plots show corresponding
{\it IRAS} TIR measurements of the SINGS and MK06 samples.  The SINGS
samples are slightly different in the two cases because not all SINGS
galaxies were detected by IRAS.  As in Figure 6 the top panels compare
the observed TIR and Balmer-corrected \halpha\ luminosities, while the
bottom panels compare linear combinations of (uncorrected) \halpha\ and
IR luminosities with the Balmer-corrected \halpha\ luminosities.  Open
circles denote SINGS galaxies and solid circles MK06 galaxies.}
\end{figure}

\begin{figure}
\plotone{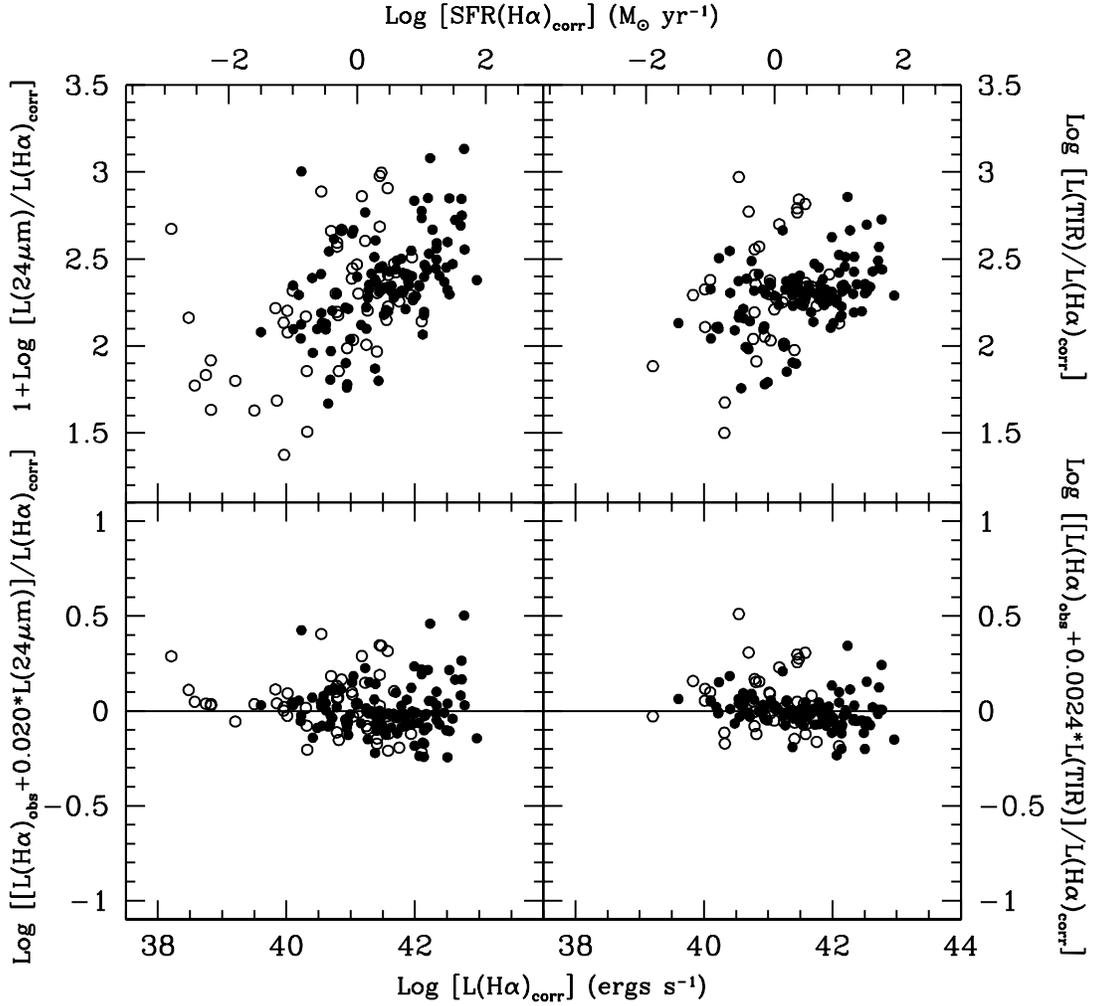}
\caption{Residuals from the comparisons shown in Figure 6 (left panels,
24\,\um\ and \halpha) and Figure 7 (right panels, {\it IRAS} TIR and
\halpha).  Open circles denote SINGS galaxies and solid circles denote
MK06 galaxies.}
\end{figure}

\begin{figure}
\epsscale{0.9}
\plotone{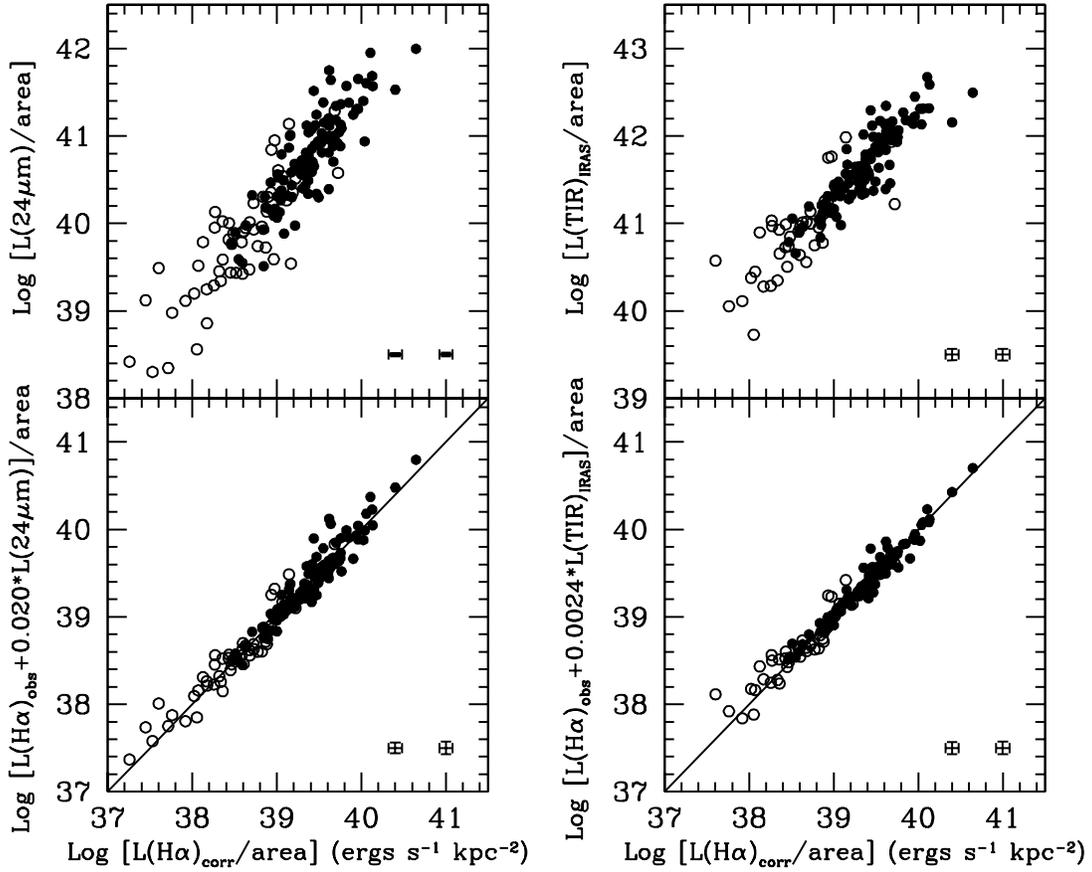}
\caption{Similar to Figures 6 and 7, but comparing galaxy surface brightnesses
rather than luminosities.  The top panels show 24\,\um\ (left) and TIR (right)
surface brightnesses as a function of Balmer-corrected \halpha\ surface 
brightness, for SINGS galaxies (open circles) and MK06 galaxies (solid circles).
The bottom panels compare attenuation-corrected \halpha\ surface brightnesses
as derived from linear combinations of \halpha\ $+$ 24\,\um\ fluxes (left)
and \halpha\ $+$ TIR fluxes (right) with those derived from the absorption-
corrected Balmer decrements.}
\end{figure}

\clearpage

\begin{figure}
\plotone{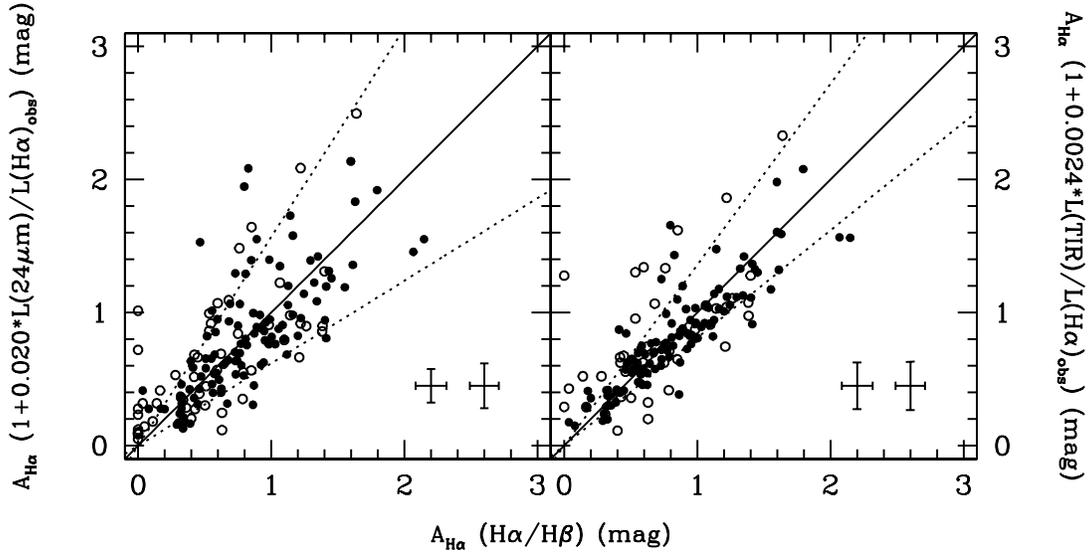}
\caption{Comparison of \halpha\ dust attenuations estimated from 
eq. (2) using 24\,\um\ and uncorrected \halpha\ fluxes (left panel)
and {\it IRAS} TIR and uncorrected \halpha\ fluxes (right panel),
in each case compared to attenuations estimated from the 
absorption-corrected \halpha/\hbeta\ ratio.  Open circles denote 
SINGS galaxies and solid circles denote MK06 galaxies.  The dotted
lines contain 68\%\ of the galaxies, corresponding approximately to
1-$\sigma$ dispersions unit-slope relation.}
\end{figure}

\begin{figure}
\plotone{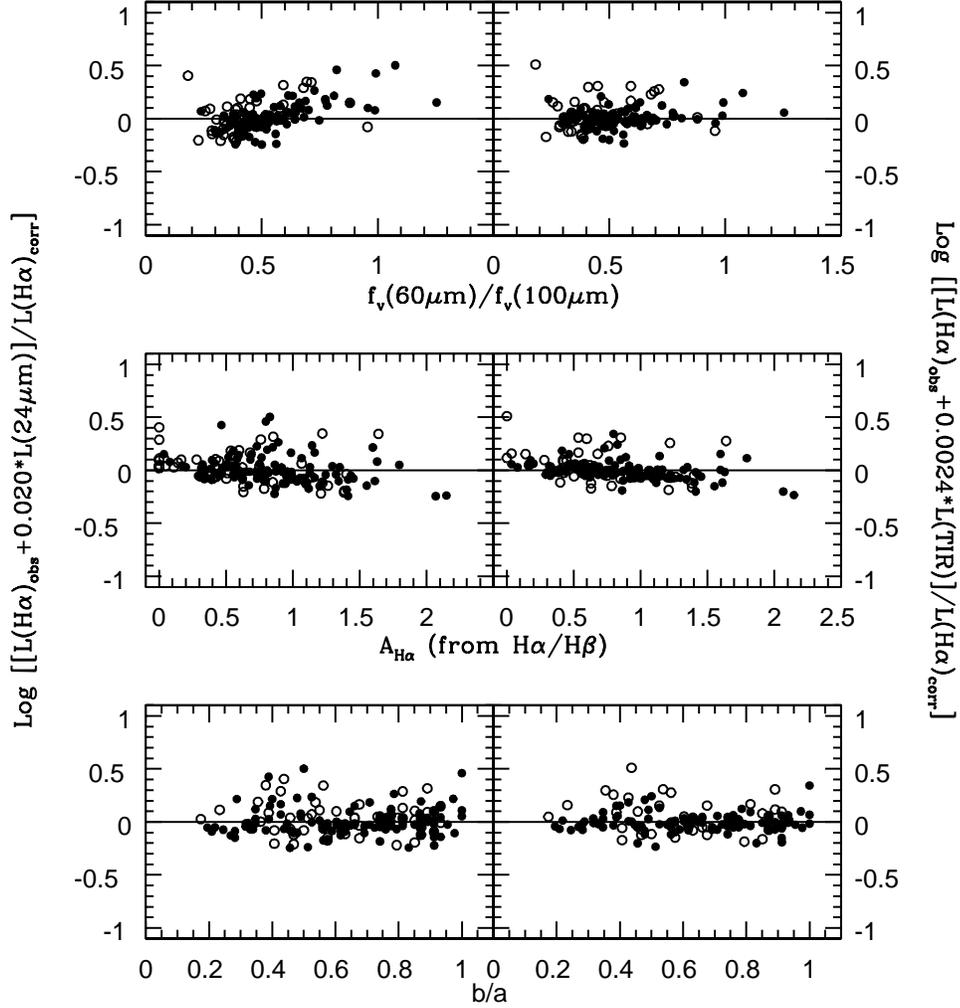}
\caption{Ratios of attenuation-corrected \halpha\ luminosities derived from 
the combination of observed \halpha\ and IR fluxes, to those derived
from the Balmer decrements, plotted as functions of infrared color (top
panels), Balmer attenuation (middle panels), and galaxy axial ratio (bottom panels).
The left panels show residuals using 24\,\um\ $+$ \halpha, while the right
panels show the corresponding residuals using TIR $+$ \halpha\ combination.
Open circles denote SINGS galaxies and solid circles denote
MK06 galaxies.}
\end{figure}

\begin{figure}
\epsscale{0.9}
\plotone{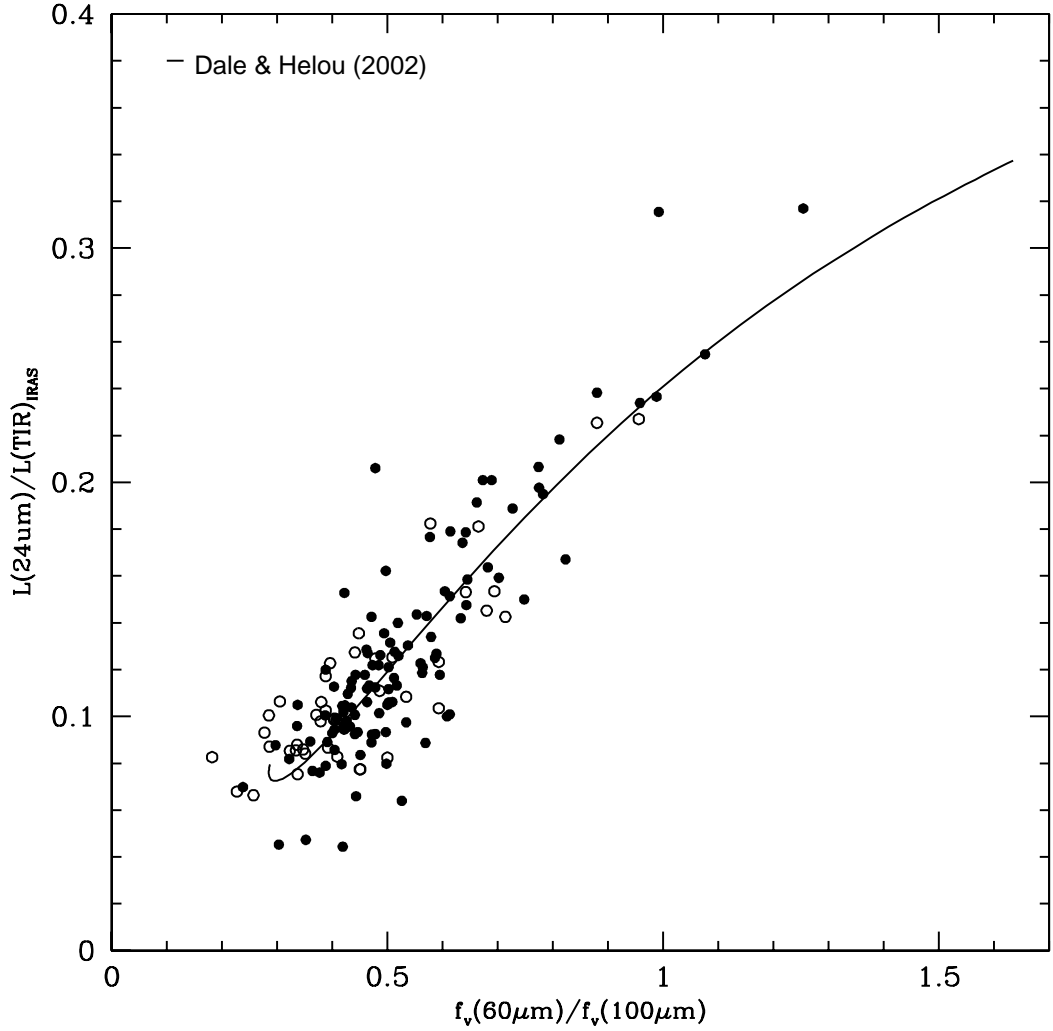}
\caption{Ratios of 24\,\um\ dust luminosity ($\nu {\rm L}_\nu$) to total infrared
(TIR) luminosities of galaxies in our sample, plotted as a function of
60\,\um/100\,\um\ color.  Open circles denote SINGS galaxies and solid circles 
denote MK06 galaxies.  The solid line shows the SED template sequence from Dale 
\& Helou (2002).}
\end{figure}

\begin{figure}
\plotone{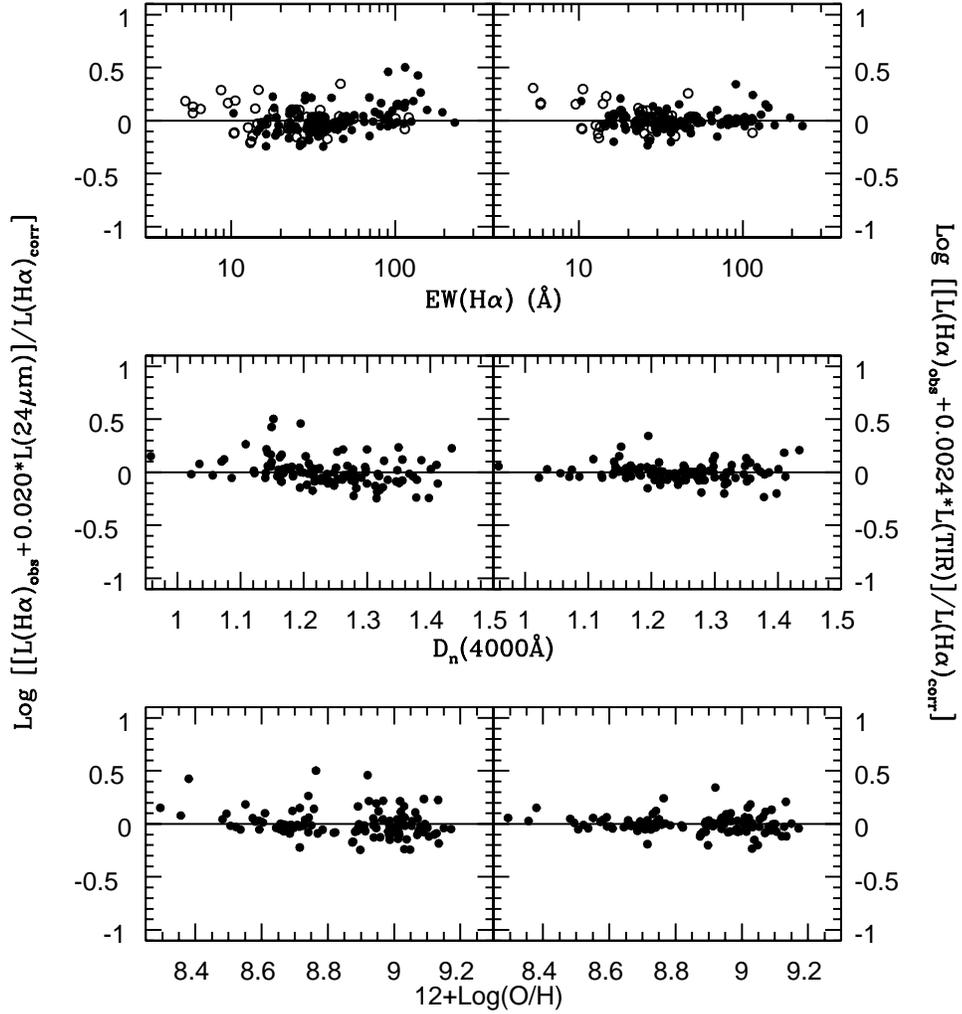}
\caption{Similar to Figure 11, but with residuals plotted as functions
of integrated \halpha\ emission-line equivalent width (top panels),
4000\,\AA\ discontinuity (middle panels), and average gas-phase
oxygen abundance (bottom panels).  See Figure 11 for explanation of
methods and symbols.  }
\end{figure}

\begin{figure}
\epsscale{0.9}
\plotone{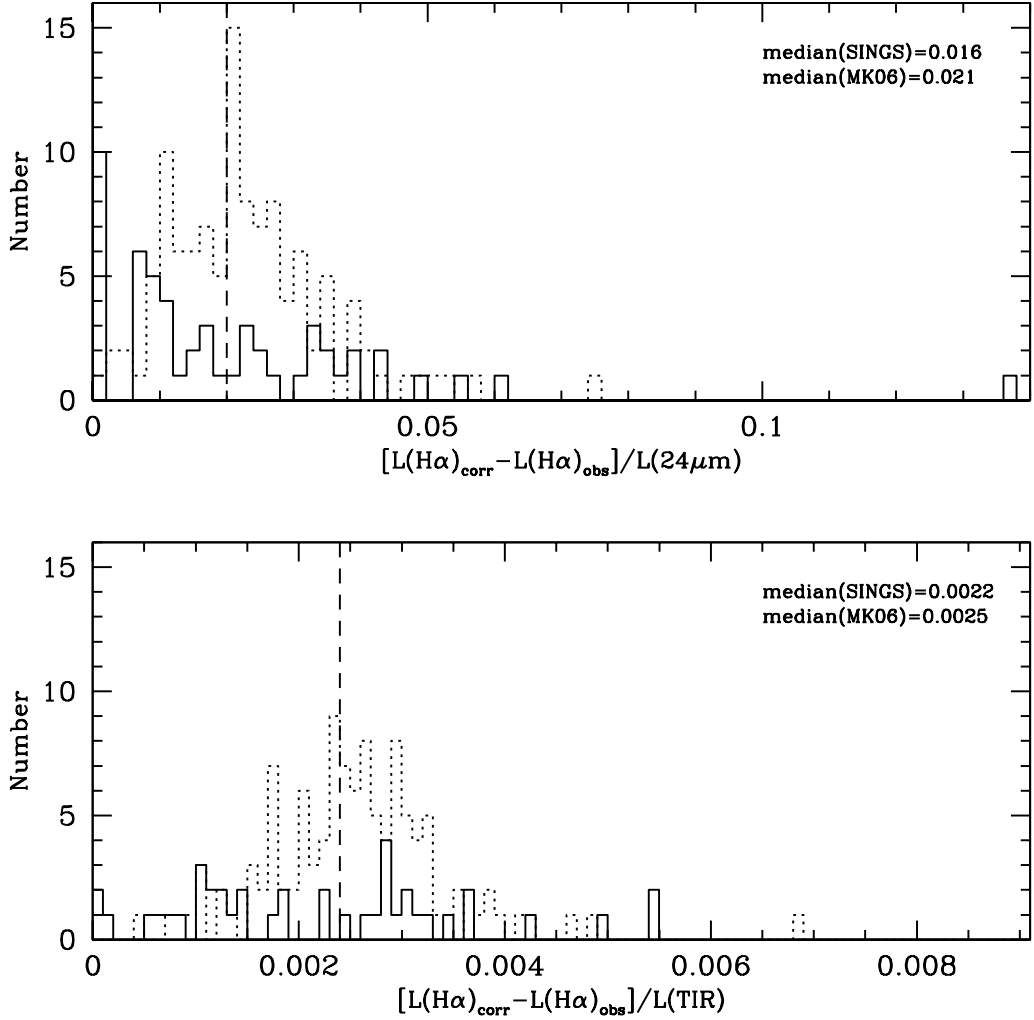}
\caption{Histograms of the scaling constant $a$ in eqs. (1--2), which forces the
attenuation derived from the ratio of 24\,um/\halpha\ fluxes (top panel) and ratio 
of TIR/\halpha\ fluxes (bottom panel) to agree with that derived from the 
\halpha/\hbeta\ ratio in the integrated spectrum.  Solid lines show the
distribution of $a$ values for the SINGS sample, while dotted lines show
the corresponding distributions for the MK06 galaxies. The adopted values of $a$ for
the combined sample (see Table 4) are shown by vertical dashed lines.}
\end{figure}

\begin{figure}
\epsscale{0.9}
\plotone{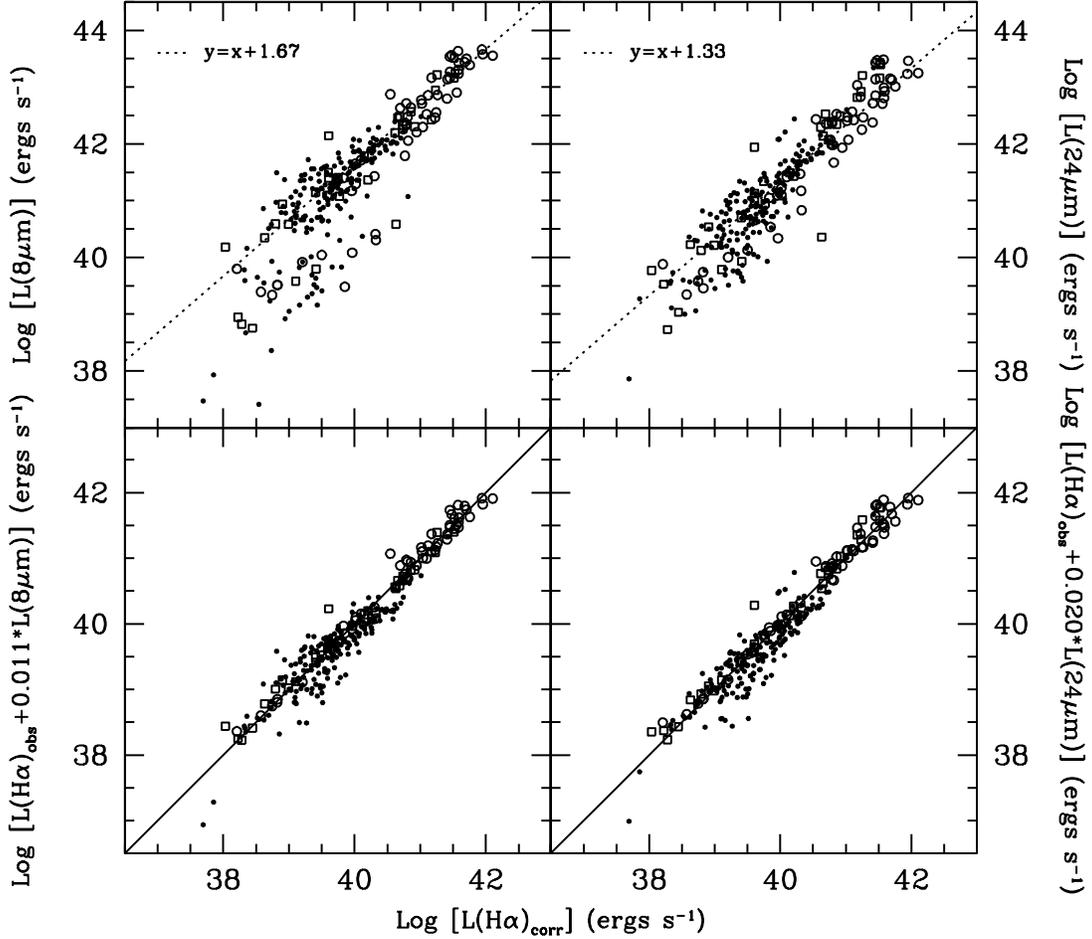}
\caption{{\it Top Panels:}\  Observed 8\,\um\ PAH luminosities 
{\it (left)} and 24\,\um\ luminosities {\it (right)}  of SINGS 
galaxies and subregions,
plotted as a function of Balmer-corrected \halpha\ luminosities,
as in previous figures.
Open circles denote integrated measurements of SINGS galaxies, open
squares denote measurements of the central 20\arcsec\ $\times$ 20\arcsec\
regions, while small solid points denote individual HII regions from C07.
{\it Bottom Panels:}\  Best fitting linear combinations of uncorrected
\halpha\ and 8\,\um\ {\it (left)} or 24\,\um\ {\it (right)} luminosities,
as a function of Balmer-corrected \halpha\ luminosities.}
\end{figure}

\begin{figure}
\epsscale{0.9}
\plotone{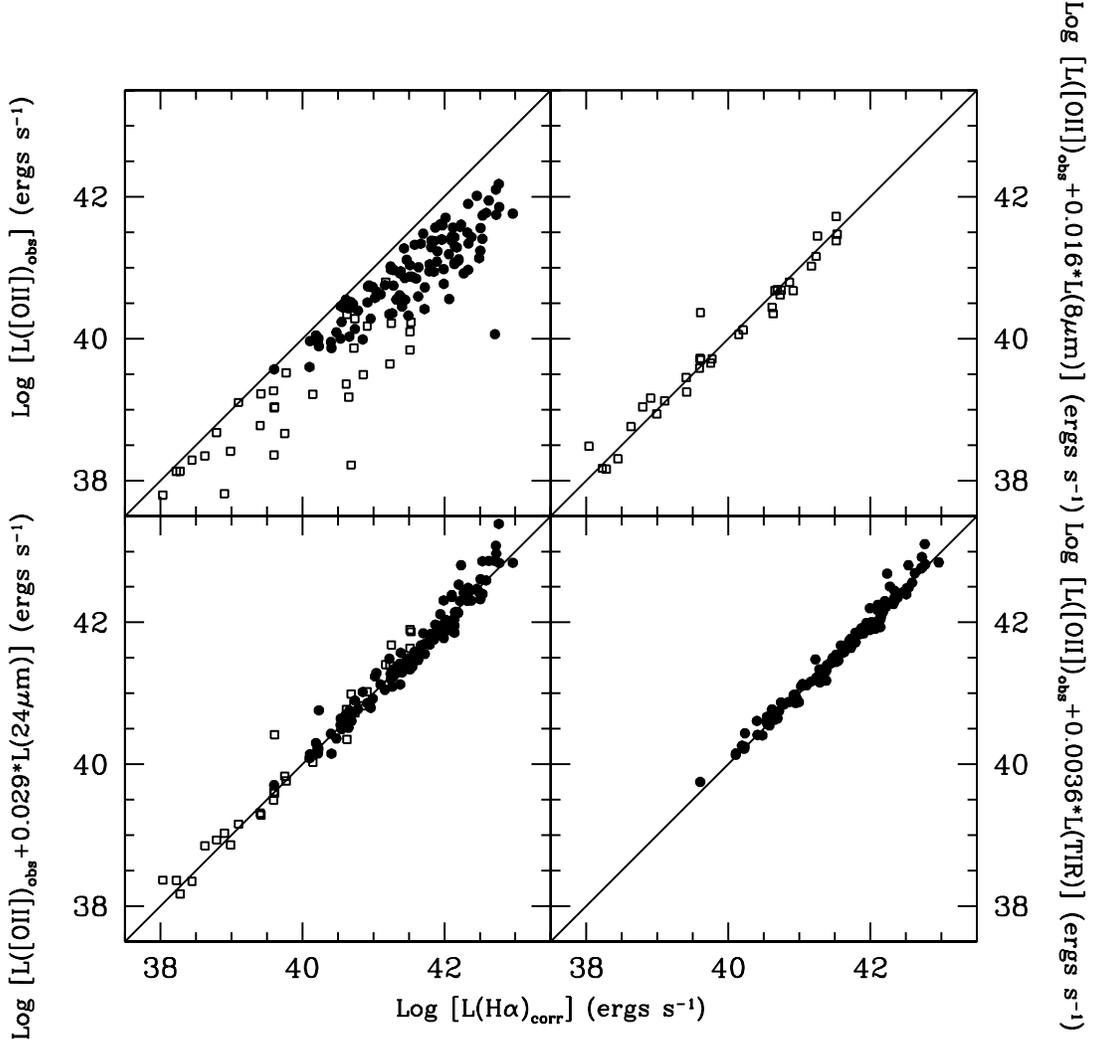}
\caption{{\it Top Left:}\  Observed [\ion{O}{2}]$\lambda$3727 emission-line
luminosities of SINGS central 20\arcsec\ $\times$ 20\arcsec\ regions 
(open squares) and MK06 (solid circles) galaxies,
plotted as a function of Balmer-corrected \halpha\ luminosities,
as in previous figures.  {\it Other Panels:}\ Best fitting
linear combinations of observed [\ion{O}{2}] luminosities and IR
band luminosities, as functions of Balmer-corrected \halpha\ luminosities.}
\end{figure}

\begin{figure}
\epsscale{0.9}
\plotone{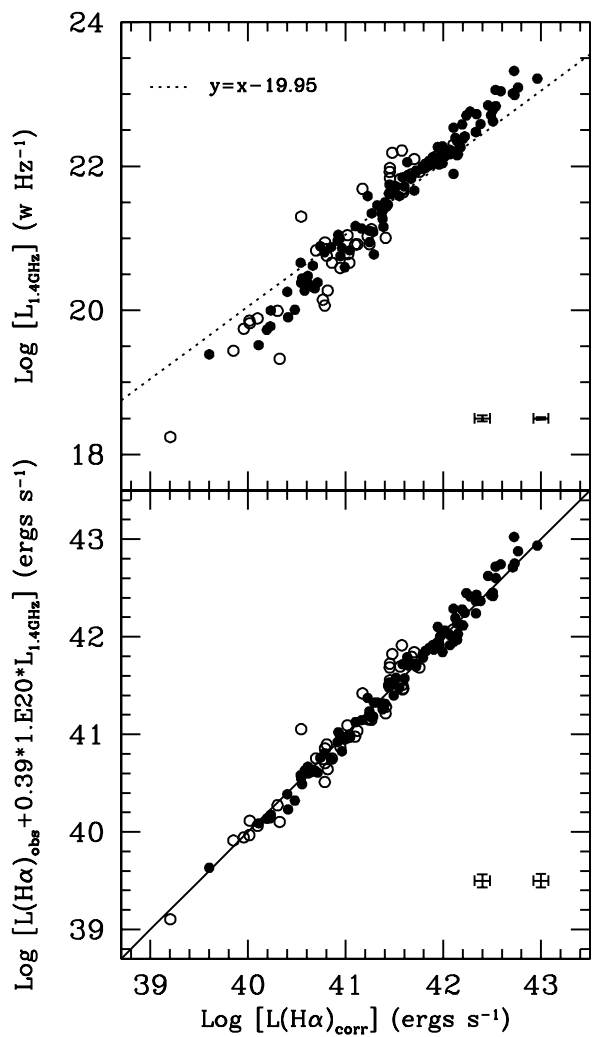}
\caption{{\it Top:}\  Integrated 1.4\,GHz radio continuum luminosities 
of SINGS galaxies (open circles) and MK06 galaxies (solid circles) 
plotted as a function of Balmer-corrected \halpha\ luminosities.
The dotted line shows a linear correlation for reference.
{\it Bottom:}\ Best fitting linear combinations of uncorrected
\halpha\ and radio continuum luminosities, as a function of 
Balmer-corrected \halpha\ luminosities.}
\end{figure}

\begin{figure}
\epsscale{0.9}
\plotone{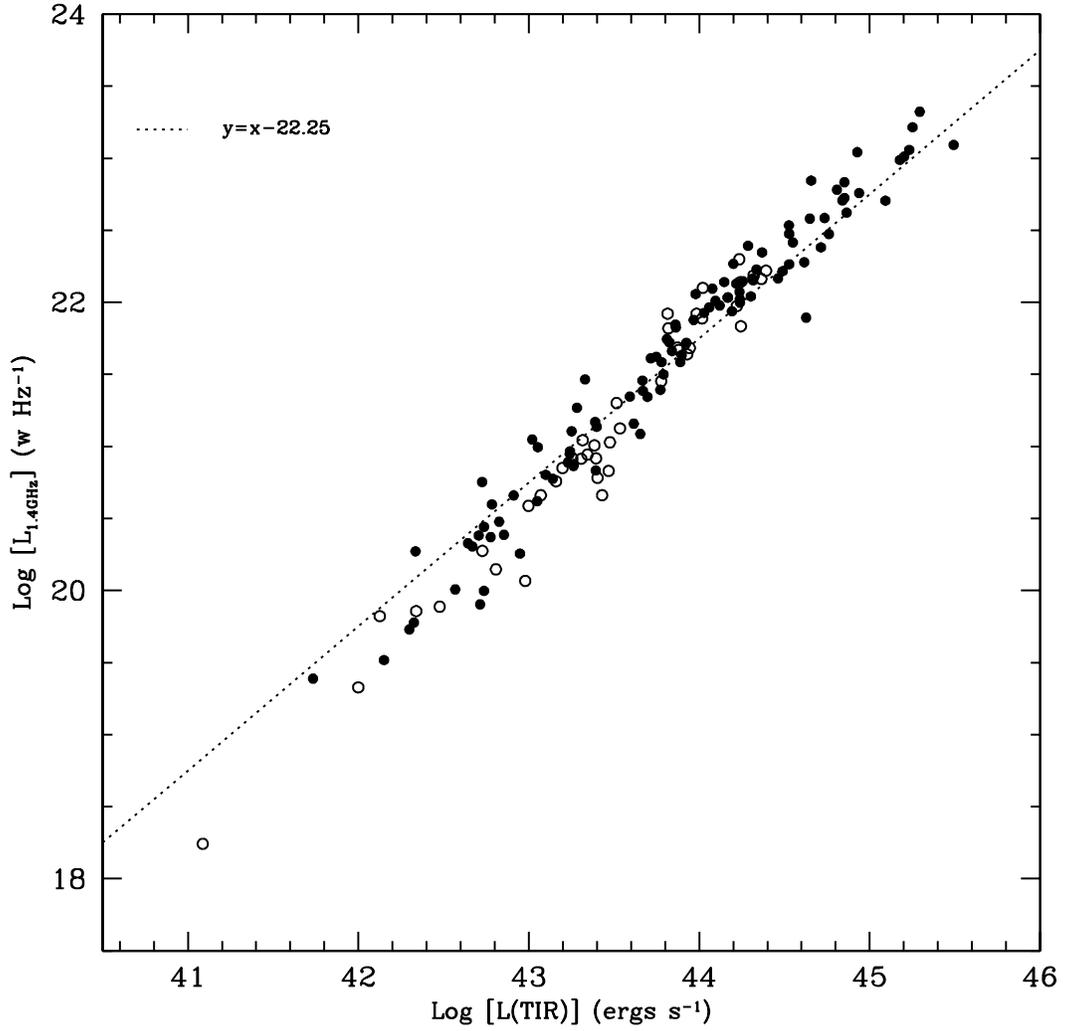}
\caption{Relationship between 1.4\,GHz radio continuum luminosity and 
total infrared (TIR) luminosity for SINGS galaxies (open circles) and
MK06 galaxies (solid circles).  The dotted line shows a linear correlation for
reference.}
\end{figure}

\clearpage

\begin{figure}
\plottwo{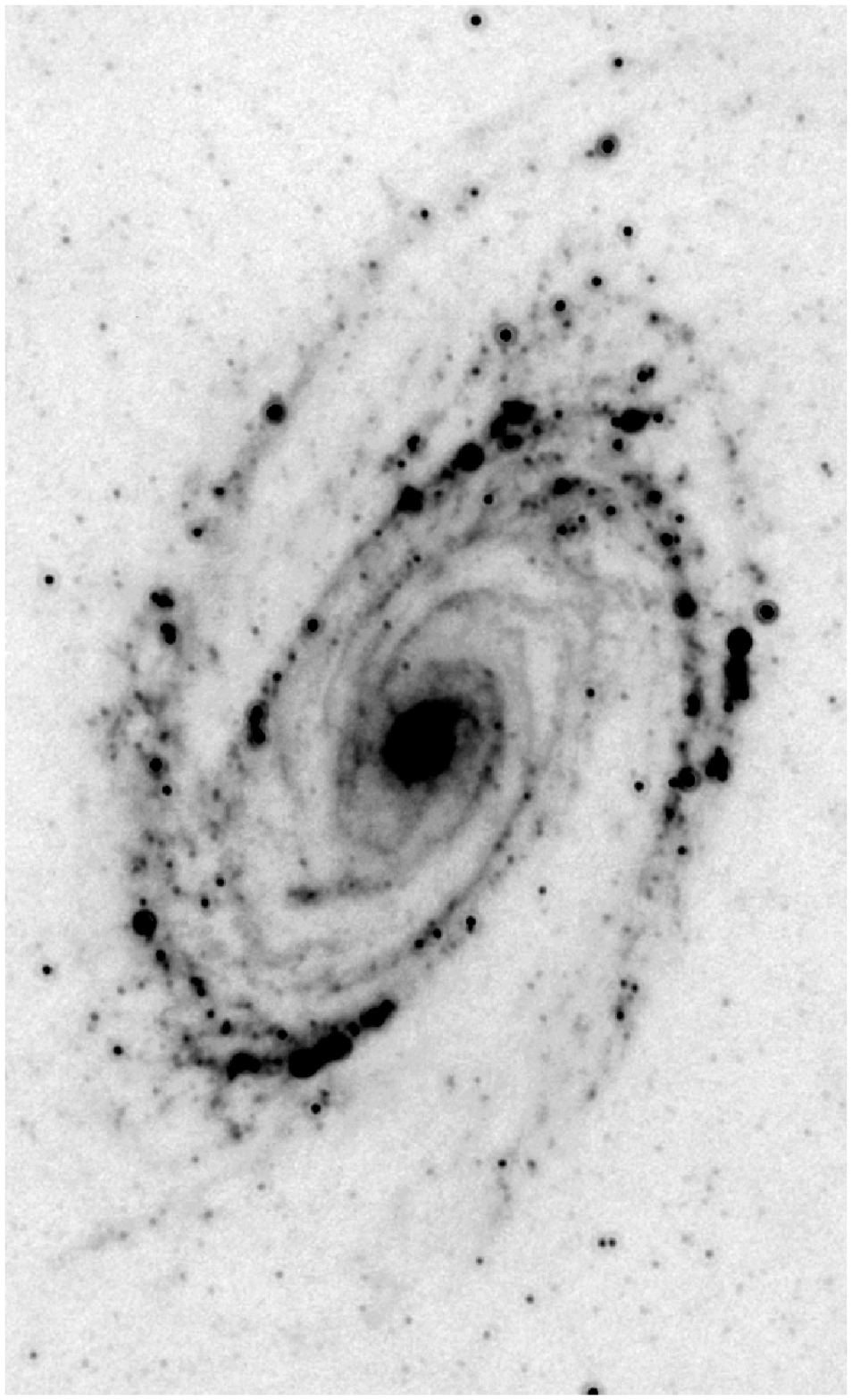}{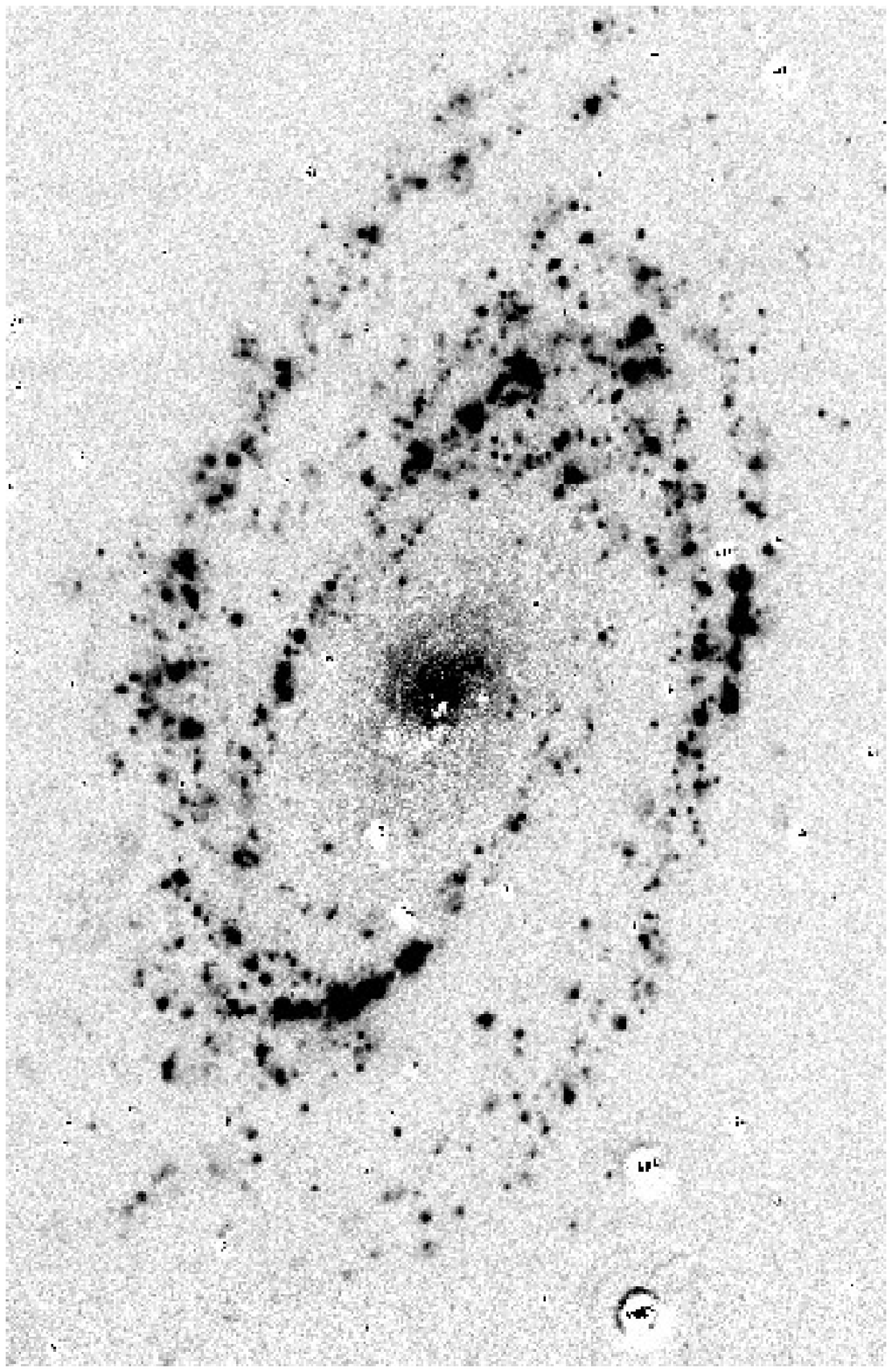}
\caption{SINGS galaxy M81 observed in the mid-infrared at 24\,\um\ 
{\it (left)}, and in \halpha\ {\it (right)}.  Note the strong spatial
correlation of bright infrared sources with optical HII regions, and 
the extensive diffuse 24\,\um\ emission in regions devoid of \halpha\
emission.}
\end{figure}

\begin{figure}
\epsscale{0.9}
\plotone{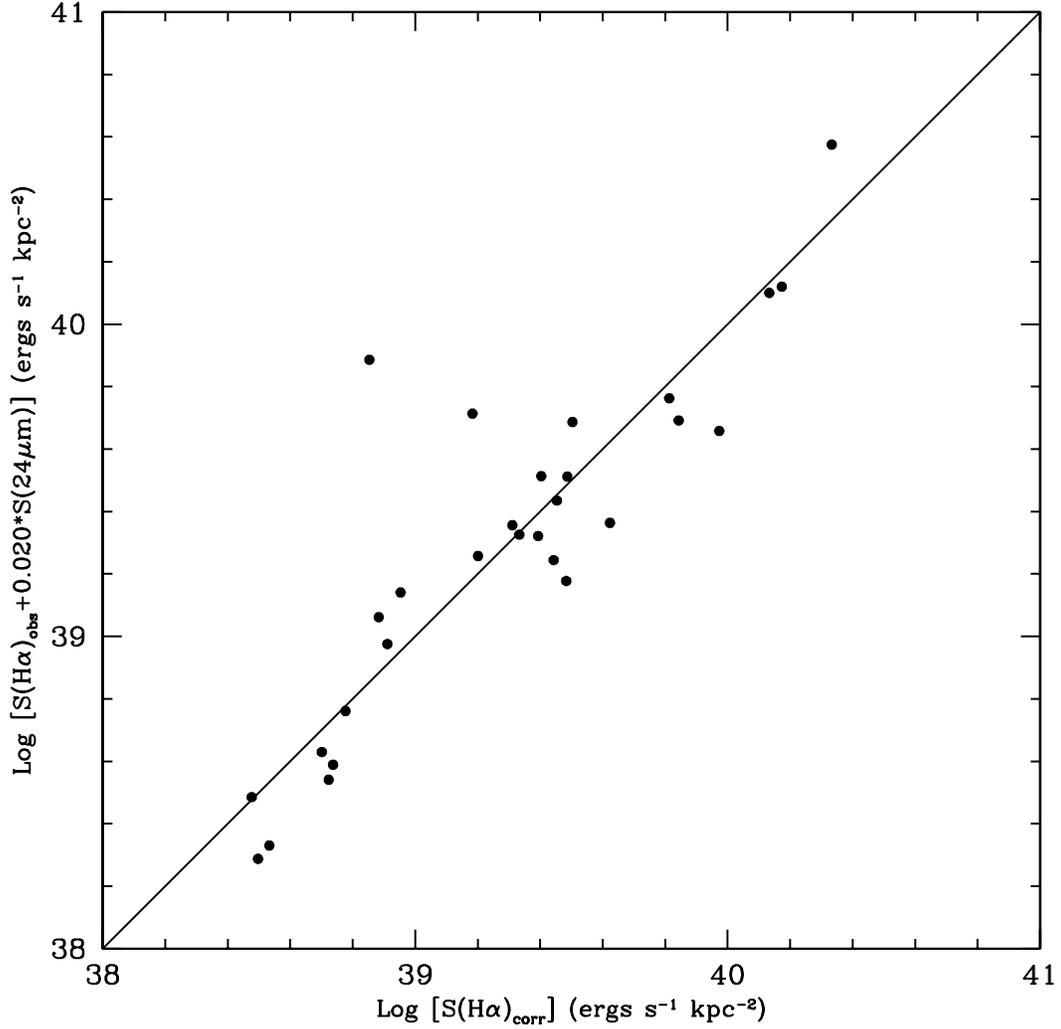}
\caption{Extinction-corrected \halpha\ surface brightnesses for the central
50\arcsec\ $\times$ 50\arcsec\ regions of 29 SINGS galaxies ($\sim$ 144\arcsec\ $\times$ 144\arcsec\ for NGC\,5194),
as derived from the observed \halpha\ and 24\,\um\ fluxes, plotted as a function of the attenuation-corrected
\halpha\ surface brightnesses as derived from the Pa\,$\alpha$/\halpha\ ratio.}
\end{figure}

\begin{figure}
\epsscale{0.9}
\plotone{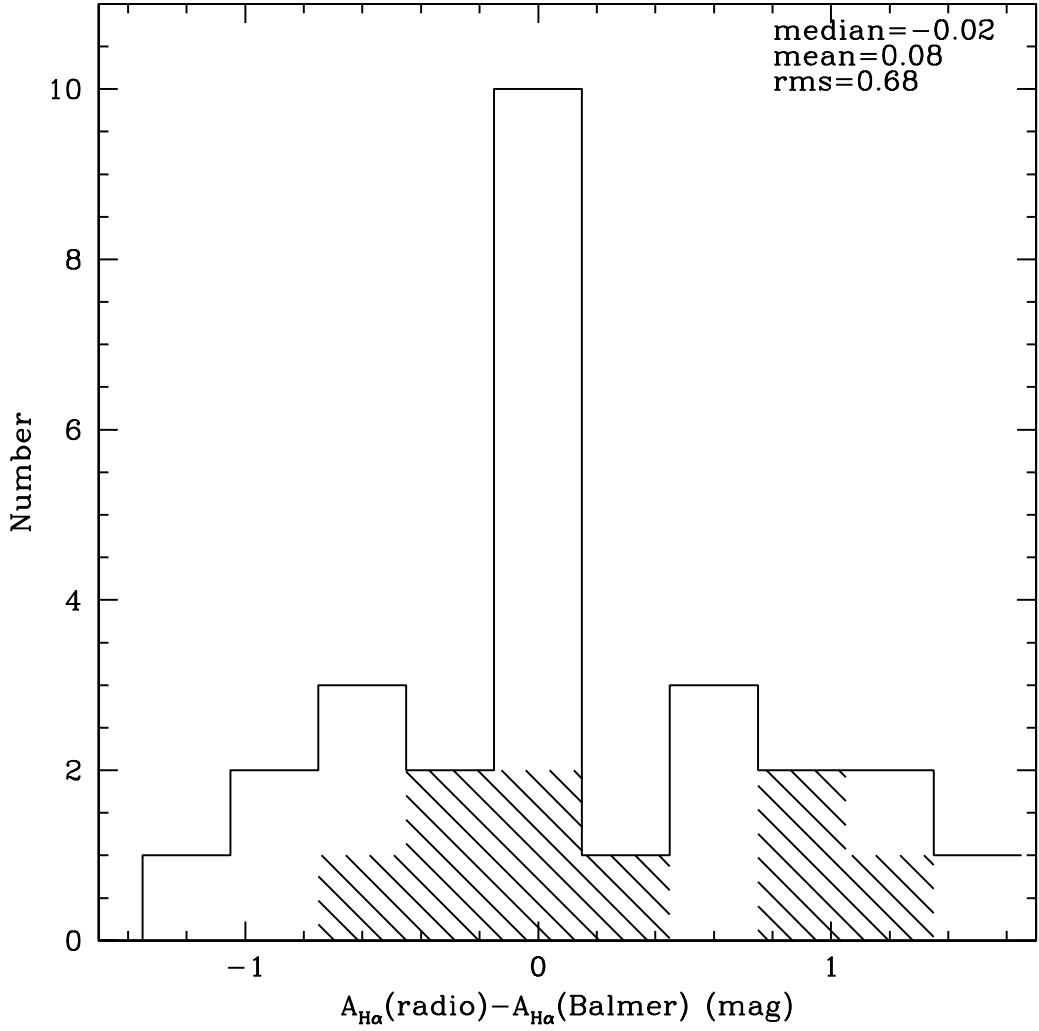}
\caption{Histogram of the differences between the H$\alpha$ attenuations estimated 
from the thermal radio to H$\alpha$ ratios and those from the H$\alpha$/H$\beta$ ratios.
The hatched histogram denotes the distribution of the nine objects whose thermal fractions are
upper limits.}
\end{figure}

\begin{figure}
\epsscale{0.9}
\plotone{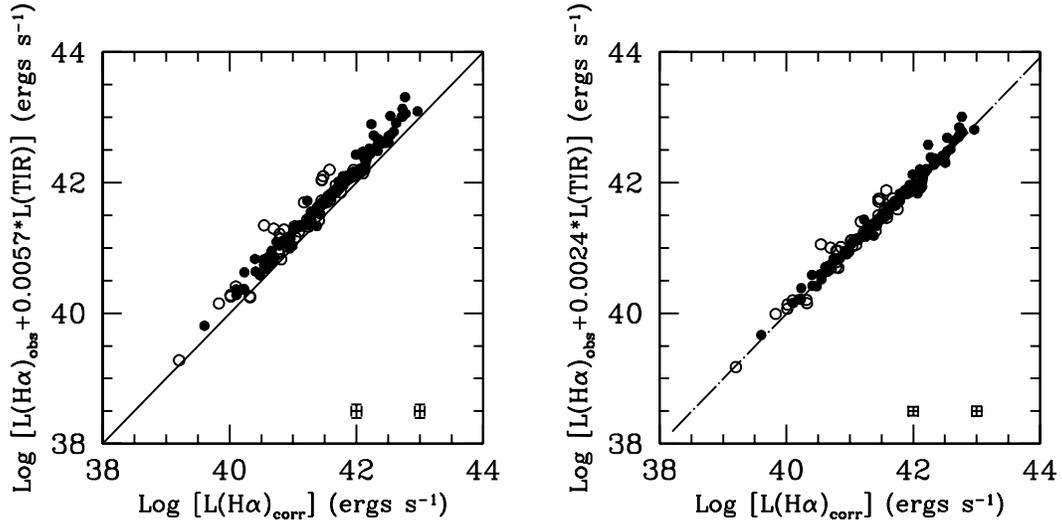}
\caption{{\it Left:}\ Comparison of attenuation-corrected \halpha\ 
luminosities from a combination of \halpha\ and TIR luminosities
using coefficients from Kennicutt (1998a) with Balmer-corrected luminosities.
The line superimposed shows the expected relation if two sets of
luminosities were equal.  {\it Right:}\  Same comparison but using
the empirically measured value for the coefficient $a$ in eq. (1).
In this case the line is from a set of dust attenuation and IR
emission models from Calzetti et al. (2007), with dust heated
by a galaxy with a constant SFR over the past 10 Gyr.  In both
panels open circles represent SINGS galaxies and solid circles denote
MK06 galaxies.}
\end{figure}

\clearpage

\begin{deluxetable}{lcccccccccc}
\tabletypesize{\scriptsize}
\rotate
\tablecolumns{8}
\tablewidth{0pt}
\tablecaption{Integrated Measurements of SINGS Galaxies}
\tablehead{
\colhead{Name} & \colhead{Distance}   & \colhead{log(H$\alpha$+[NII])}    
& \colhead{H$\alpha$/H$\beta$} & \colhead{[NII]/H$\alpha$} &
\colhead{S$_\nu$(25~$\mu{\rm m}$)}    & \colhead{S$_\nu$(60~$\mu{\rm m}$)}   &  \colhead{S$_\nu$(100~$\mu{\rm m}$)} &
\colhead{references} & \colhead{references} & \colhead{references} \\
\colhead{} & \colhead{(Mpc)} & \colhead{(erg~s$^{-1}$~cm$^{-2}$)} & \colhead{} & \colhead{} & \colhead{(Jy)} &
\colhead{(Jy)} & \colhead{(Jy)} & \colhead{(H$\alpha$/H$\beta$)} & \colhead{([NII]/H$\alpha$)} & \colhead{(IRAS)} \\
\colhead{(1)} & \colhead{(2)}   & \colhead{(3)}    & \colhead{(4)} & \colhead{(5)} &
\colhead{(6)}    & \colhead{(7)}   &  \colhead{(8)} &  \colhead{(9)} &  \colhead{(10)} &  \colhead{(11)} }
\startdata
NGC0024 & 7.3 & -11.87$\pm$0.04\phn & 2.902$\pm$0.219\phn & 0.373$\pm$0.034\phn & 0.16$\pm$0.046\phn & 1.26$\pm$0.139\phn & 3.59$\pm$0.395\phn & 3 & 3 & 103 \\
NGC0337 & 22.4 & -11.43$\pm$0.03\phn & 3.465$\pm$0.042\phn & 0.230$\pm$0.004\phn & 0.76$\pm$0.050\phn & 9.07$\pm$0.043\phn & 20.11$\pm$0.387\phn & 1 & 1 & 101 \\
NGC0628 & 7.3 & -10.84$\pm$0.04\phn & 3.215$\pm$0.225\phn & 0.345$\pm$0.046\phn & 2.87$\pm$0.060\phn & 21.54$\pm$0.045\phn & 54.45$\pm$0.229\phn & 3;5;7;8 & 5;6;7;8 & 101 \\
NGC0855 & 9.73 & -12.23$\pm$0.04\phn & 3.292$\pm$0.121\phn & 0.185$\pm$0.011\phn & \nodata & \nodata & \nodata & 3 & 3 & \nodata \\
NGC0925 & 9.12 & -11.10$\pm$0.70\phn & 3.323$\pm$0.269\phn & 0.201$\pm$0.066\phn & 0.83$\pm$0.034\phn & 7.82$\pm$0.041\phn & 21.08$\pm$0.087\phn & 7;9 & 7;9 & 101 \\
NGC1097 & 17.1 & -10.95\tablenotemark{a}  & 5.134$\pm$0.160\phn & 0.69$\pm$0.41\phn & 7.30$\pm$0.041\phn & 53.35$\pm$0.038\phn & 104.79$\pm$0.122\phn & 3 & 4 & 101 \\
NGC1482 & 23.2 & -11.80$\pm$0.05\phn & 5.682$\pm$0.202\phn & 0.692$\pm$0.010\phn & 4.68$\pm$0.044\phn & 33.36$\pm$0.060\phn & 46.73$\pm$0.068\phn & 3 & 3 & 101 \\
NGC1512 & 11.8 & -11.66$\pm$0.10\phn & 4.082$\pm$0.329\phn & 0.34$\pm$0.20\phn & 0.24$\pm$0.024\phn & 3.14$\pm$0.157\phn & 11.00$\pm$0.550\phn & 3 & 4 & 103 \\
NGC1566 & 20.4 & -10.88$\pm$0.08\phn & 4.744$\pm$1.121\phn & 0.623$\pm$0.057\phn & 3.02$\pm$0.020\phn & 22.53$\pm$0.026\phn & 58.05$\pm$0.120\phn & 3;10 & 10 & 101 \\
NGC1705 & 5.1 & -11.50$\pm$0.02\phn & 2.86$\pm$0.034\phn & 0.088$\pm$0.004\phn & \nodata & \nodata & \nodata & 3 & 3 & \nodata \\
NGC2403 & 3.13 & -10.25$\pm$0.04\phn & 3.418$\pm$0.271\phn & 0.217$\pm$0.045\phn & 6.29$\pm$0.944\phn & 51.55$\pm$7.733\phn & 148.49$\pm$22.274\phn & 5;7;24;25 & 5;6;7;24;25 & 102 \\
HoII & 3.39 & -11.27$\pm$0.04\phn & 2.86$\pm$0.202\tablenotemark{c}\phn & 0.12$\pm$0.07\phn & \nodata & \nodata & \nodata & assume & 4 & \nodata \\
DDO053 & 3.56 & -12.43$\pm$0.12\phn & 2.995$\pm$0.064\phn & 0.023$\pm$0.007\phn & \nodata & \nodata & \nodata & 1 & 1 & \nodata \\
NGC2798 & 26.2 & -11.84$\pm$0.06\phn & 4.764$\pm$0.132\phn & 0.36$\pm$0.08\phn & 3.21$\pm$0.031\phn & 20.60$\pm$0.048\phn & 29.69$\pm$0.156\phn & 3 & 2 & 101 \\
NGC2841 & 14.1 & -11.64$\pm$0.22\phn & 2.86$\pm$0.212\phn & 0.612$\pm$0.083\phn & 0.83$\pm$0.125\phn & 4.41$\pm$0.662\phn & 24.21$\pm$3.632\phn & 3;12 & 6 & 102 \\
NGC2915 & 3.78 & -11.95$\pm$0.05\phn & 2.86$\pm$0.086\phn & 0.145$\pm$0.009\phn & \nodata & \nodata & \nodata & 3 & 3 & \nodata \\
HoI & 3.84 & -12.44$\pm$0.05\phn & 2.86$\pm$0.645\phn & 0.075$\pm$0.077\phn & \nodata & \nodata & \nodata & 15 & 15 & \nodata \\
NGC2976 & 3.56 & -11.19$\pm$0.06\phn & 3.409$\pm$0.086\phn & 0.357$\pm$0.008\phn & 1.71$\pm$0.020\phn & 13.09$\pm$0.029\phn & 33.43$\pm$0.344\phn & 3 & 3 & 101 \\
NGC3049 & 23.9 & -11.93$\pm$0.07\phn & 3.914$\pm$0.189\phn & 0.404$\pm$0.014\phn & 0.43$\pm$0.055\phn & 2.82$\pm$0.169\phn & 4.24$\pm$0.297\phn & 1 & 1 & 103 \\
NGC3031 & 3.55 & -10.32$\pm$0.05\phn & 3.033$\pm$0.368\phn & 0.545$\pm$0.084\phn & 5.42$\pm$0.813\phn & 44.73$\pm$6.710\phn & 174.02$\pm$26.103\phn & 11 & 6;11;19 & 101;102 \\
HoIX & 3.7 & -13.07$\pm$0.14\phn & 2.86$\pm$0.202\tablenotemark{c}\phn & 0.04$\pm$0.02\phn & \nodata & \nodata & \nodata & assume & 4 & \nodata \\
M81DwB & 5.3 & -12.86$\pm$0.05\phn & 2.916$\pm$0.174\phn & 0.05$\pm$0.03\phn & \nodata & \nodata & \nodata & 3 & 4 & \nodata \\
NGC3190 & 26.38 & -12.68$\pm$0.05\phn & 2.86$\pm$0.392\phn & 1.532$\pm$0.197\phn & 0.35$\pm$0.084\phn & 3.19$\pm$0.351\phn & 10.11$\pm$0.506\phn & 3 & 3 & 103 \\
NGC3184 & 11.1 & -11.12$\pm$0.05\phn & 3.664$\pm$0.247\phn & 0.523$\pm$0.052\phn & 1.32$\pm$0.025\phn & 8.72$\pm$0.029\phn & 28.58$\pm$0.118\phn & 3;5;7 & 5;7 & 101 \\
NGC3198 & 13.68 & -11.40$\pm$0.04\phn & 3.447$\pm$0.160\phn & 0.304$\pm$0.018\phn & 1.08$\pm$0.029\phn & 7.15$\pm$0.041\phn & 18.44$\pm$0.103\phn & 1 & 1 & 101 \\
IC2574 & 4.02 & -11.23$\pm$0.07\phn & 3.722$\pm$0.141\phn & 0.046$\pm$0.017\phn & 0.08$\pm$0.012\phn & 2.41$\pm$0.362\phn & 10.62$\pm$1.593\phn & 15 & 15 & 102 \\
NGC3265 & 23.2 & -12.28$\pm$0.07\phn & 4.466$\pm$0.133\phn & 0.532$\pm$0.010\phn & 0.36$\pm$0.055\phn & 2.18$\pm$0.174\phn & 3.39$\pm$0.204\phn & 1 & 1 & 103 \\
Mrk33 & 22.9 & -11.70$\pm$0.01\phn & 3.596$\pm$0.020\phn & 0.270$\pm$0.002\phn & 0.95$\pm$0.057\phn & 4.68$\pm$0.281\phn & 5.32$\pm$0.319\phn & 1;3 & 1;3 & 103 \\
NGC3351 & 9.33 & -11.24$\pm$0.08\phn & 3.802$\pm$0.251\phn & 0.655$\pm$0.027\phn & 2.79$\pm$0.053\phn & 19.66$\pm$0.062\phn & 41.10$\pm$0.102\phn & 3;5;12 & 1 & 101 \\
NGC3521 & 10.1 & -10.85$\pm$0.04\phn & 4.769$\pm$0.109\phn & 0.558$\pm$0.008\phn & 5.46$\pm$0.082\phn & 49.19$\pm$0.100\phn & 121.76$\pm$0.405\phn & 1 & 1 & 101 \\
NGC3621 & 6.55 & -10.55$\pm$0.04\phn & 3.970$\pm$0.460\phn & 0.402$\pm$0.071\phn & 4.44$\pm$0.048\phn & 29.32$\pm$0.044\phn & 77.34$\pm$0.144\phn & 3;9 & 9 & 101 \\
NGC3627 & 9.38 & -10.74$\pm$0.05\phn & 4.611$\pm$0.350\phn & 0.55$\pm$0.05\phn & 8.55$\pm$0.071\phn & 66.31$\pm$0.059\phn & 136.56$\pm$0.118\phn & 3;22 & 4;22 & 101 \\
NGC3773 & 11.9 & -11.99$\pm$0.07\phn & 3.273$\pm$0.041\phn & 0.233$\pm$0.004\phn & \nodata & \nodata & \nodata & 1 & 1 & \nodata \\
NGC3938 & 13.4 & -11.25$\pm$0.00\phn & 3.978$\pm$0.876\phn & 0.42$\pm$0.25\phn & 1.23$\pm$0.020\phn & 9.18$\pm$0.029\phn & 27.50$\pm$0.077\phn & 3 & 4 & 101 \\
NGC4254 & 16.5 & -10.89$\pm$0.04\phn & 4.526$\pm$0.058\phn & 0.449$\pm$0.004\phn & 4.38$\pm$0.045\phn & 37.46$\pm$0.080\phn & 91.86$\pm$0.149\phn & 1 & 1 & 101 \\
NGC4321 & 14.32 & -11.06$\pm$0.07\phn & 5.105$\pm$1.265\phn & 0.430$\pm$0.015\phn & 3.10$\pm$0.053\phn & 26.00$\pm$0.050\phn & 68.37$\pm$0.101\phn & 1;3 & 1 & 101 \\
NGC4450 & 16.5 & -12.21$\pm$0.04\phn & 3.757$\pm$0.502\phn & 0.51$\pm$0.31\phn & \nodata & \nodata & \nodata & 3 & 4 & \nodata \\
NGC4536 & 14.45 & -11.38$\pm$0.03\phn & 3.927$\pm$0.190\phn & 0.454$\pm$0.092\phn & 4.04$\pm$0.060\phn & 30.26$\pm$0.042\phn & 44.51$\pm$0.141\phn & 22;23 & 22 & 101 \\
NGC4559 & 10.3 & -10.97$\pm$0.05\phn & 3.528$\pm$0.232\phn & 0.281$\pm$0.148\phn & 1.03$\pm$0.030\phn & 10.23$\pm$0.044\phn & 25.41$\pm$0.074\phn & 3;9 & 9 & 101 \\
NGC4569 & 16.5 & -11.53$\pm$0.22\phn & 5.042$\pm$0.498\phn & 0.992$\pm$0.056\phn & 2.06$\pm$0.070\phn & 9.80$\pm$0.065\phn & 26.56$\pm$0.173\phn & 3;23 & 1 & 101 \\
NGC4579 & 16.5 & -11.48$\pm$0.01\phn & 3.211$\pm$0.217\phn & 0.62$\pm$0.37\phn & 0.78$\pm$0.047\phn & 5.93$\pm$0.054\phn & 21.39$\pm$0.243\phn & 3 & 4 & 101 \\
NGC4625 & 9.2 & -12.03$\pm$0.06\phn & 3.537$\pm$0.116\phn & 0.525$\pm$0.014\phn & 0.19$\pm$0.040\phn & 1.20$\pm$0.132\phn & 3.58$\pm$0.250\phn & 1 & 1 & 103 \\
NGC4631 & 7.62 & -10.55$\pm$0.06\phn & 3.711$\pm$0.063\phn & 0.28$\pm$0.06\phn & 8.97$\pm$0.046\phn & 85.40$\pm$0.062\phn & 160.08$\pm$0.260\phn & 3 & 2 & 101 \\
NGC4736 & 5.20 & -10.72$\pm$0.06\phn & 3.576$\pm$0.046\phn & 0.711$\pm$0.006\phn & 6.11$\pm$0.040\phn & 71.54$\pm$0.085\phn & 120.69$\pm$0.199\phn & 1 & 1 & 101 \\
DDO154 & 4.3 & -12.76$\pm$0.05\phn & 2.86$\pm$0.217\phn & 0.05$\pm$0.03\phn & \nodata & \nodata & \nodata & 13;14 & 4 & \nodata \\
NGC4826 & 7.48 & -11.15$\pm$0.11\phn & 3.578$\pm$0.166\phn & 0.72$\pm$0.02\phn & 2.86$\pm$0.059\phn & 36.70$\pm$0.077\phn & 81.65$\pm$0.099\phn & 3 & 3 & 101 \\
DDO165 & 4.57 & -12.91$\pm$0.10\phn & 2.86$\pm$0.202\tablenotemark{c}\phn & 0.08$\pm$0.05\phn & \nodata & \nodata & \nodata & assume & 4 & \nodata \\
NGC5033 & 14.8 & -11.23$\pm$0.09\phn & 5.095$\pm$0.218\phn & 0.48$\pm$0.29\phn & 2.14$\pm$0.033\phn & 16.20$\pm$0.073\phn & 50.23$\pm$0.092\phn & 3 & 4 & 101 \\
NGC5055 & 7.8 & -10.80$\pm$0.07\phn & 4.849$\pm$1.132\phn & 0.486$\pm$0.019\phn & 6.36$\pm$0.050\phn & 40.00$\pm$0.049\phn & 139.82$\pm$0.356\phn & 5 & 5 & 101 \\
NGC5194 & 8.40 & -10.45$\pm$0.04\phn & 4.310$\pm$0.064\phn & 0.590$\pm$0.006\phn & 9.56$\pm$0.077\phn & 97.42$\pm$0.193\phn & 221.21$\pm$0.329\phn & 1 & 1 & 101 \\
Tol89 & 16.7 & -11.79\tablenotemark{b} & 3.065$\pm$0.106\phn & 0.26$\pm$0.16\phn & 0.27$\pm$0.035\phn & 1.56$\pm$0.078\phn & 2.70$\pm$0.270\phn & 3;21 & 4 & 103 \\
NGC5408 & 4.81 & -11.33$\pm$0.02\phn & 3.381$\pm$0.591\phn & 0.056$\pm$0.030\phn & 0.44$\pm$0.011\phn & 2.83$\pm$0.141\phn & 2.96$\pm$0.325\phn & 16;21 & 16;21 & 103 \\
NGC5474 & 6.8 & -11.65$\pm$0.03\phn & 2.86$\pm$0.095\phn & 0.22$\pm$0.13\phn & 0.08$\pm$0.017\phn & 1.33$\pm$0.067\phn & 4.80$\pm$0.240\phn & 3 & 4 & 103 \\
NGC5713 & 29.4 & -11.63$\pm$0.00\phn & 4.085$\pm$0.054\phn & 0.550$\pm$0.005\phn & 2.84$\pm$0.038\phn & 22.10$\pm$0.065\phn & 37.28$\pm$0.088\phn & 3 & 3 & 101 \\
NGC6822 & 0.460 & -10.54$\pm$0.04\phn & 3.252$\pm$0.228\phn & 0.048$\pm$0.032\phn & 2.46$\pm$0.369\phn & 47.63$\pm$7.140\phn & 95.42$\pm$14.300\phn & 17;18 & 17;18 & 102 \\
NGC6946 & 6.8 & -10.42$\pm$0.06\phn & 3.415$\pm$0.407\phn & 0.448$\pm$0.087\phn & 20.70$\pm$0.029\phn & 129.78$\pm$0.071\phn & 290.69$\pm$0.458\phn & 5;8 & 5;8 & 101 \\
NGC7331 & 14.52 & -11.07$\pm$0.03\phn & 3.671$\pm$1.064\phn & 0.610$\pm$0.039\phn & 5.92$\pm$0.036\phn & 45.00$\pm$0.091\phn & 110.16$\pm$0.468\phn & 12 & 6 & 101 \\
NGC7793 & 3.91 & -10.60$\pm$0.08\phn & 3.723$\pm$0.233\phn & 0.310$\pm$0.072\phn & 1.67$\pm$0.048\phn & 18.14$\pm$0.048\phn & 54.07$\pm$0.089\phn & 3;5;20 & 5;20 & 101 \\
\enddata
\tablenotetext{a}{No error was given in the original paper.}
\tablenotetext{b}{This value is from Kennicutt et al. 2009 (in preparation). No error was given.}
\tablenotetext{c}{They are low metallicity galaxies. The H$\alpha$/H$\beta$ ratio is assumed to be
2.86 so that the estimated extinction is zero under the Case B assumption made in this paper. 
The error is the median error of the remaining galaxies.}
\tablerefs{
(1)MK06; (2)Kennicutt 1992; (3)Moustakas et al. 2008; (4)[NII]/H$\alpha$--M$_B$ relation from Kennicutt et al. (2008)
(5) McCall et al. 1985; (6) Bresolin et al. 1999; (7) van Zee et al. 1998; (8) Ferguson et al. 1998;
(9) Zaritsky et al. 1994; (10) Hawley \& Phillips 1980; 
(11) Garnett \& Shields 1987; (12) Oey \& Kennicutt 1993; (13) van Zee et al. 1997;
(14) Kennicutt \& Skillman 2001;  (15) Miller \& Hodge 1996;
(16) Stasinska et al. 1986;  (17) Peimbert et al. 2005;
(18) Lee et al. 2006; (19) Stauffer \& Bothun 1984; (20) Webster \& Smith 1983;
(21) Terlevich et al. 1991; (22) Kennicutt (unpublished); (23) this work; 
(24) Garnett et al. 1997; (25) Garnett et al. 1999;
(101)Sanders et al. 2003; (102) Rice et al. 1989; (103) Moshir et al. 1990 }
\end{deluxetable}
\begin{deluxetable}{lcccccccc}
\tabletypesize{\scriptsize}
\rotate
\tablecolumns{9}
\tablewidth{0pt}
\tablecaption{Integrated Measurements of MK06 Sample}
\tablehead{
\colhead{Name} & \colhead{Distance}   & \colhead{H$\alpha$}
& \colhead{H$\beta$} & \colhead{S$_\nu$(25~$\mu{\rm m}$)}    & \colhead{S$_\nu$(60~$\mu{\rm m}$)}   
& \colhead{S$_\nu$(100~$\mu{\rm m}$)} & \colhead{S$_{\rm 1.4GHz}$} & \colhead{references} \\
\colhead{} & \colhead{(Mpc)} & \colhead{($10^{-15}$~erg~s$^{-1}$~cm$^{-2}$)} & 
\colhead{($10^{-15}$~erg~s$^{-1}$~cm$^{-2}$)} & 
\colhead{(Jy)} & \colhead{(Jy)} & \colhead{(Jy)} & \colhead{(mJy)} & \colhead{(IRAS)} \\
\colhead{(1)} & \colhead{(2)} & \colhead{(3)} & \colhead{(4)} & \colhead{(5)} &
\colhead{(6)} & \colhead{(7)} & \colhead{(8)} & \colhead{(9)} }
\startdata
ARP256  &  113.9 &  1099.0$\pm$5.4\phn & 254.48$\pm$3.21\phn & 1.20$\pm$0.055\phn & 7.48$\pm$0.048\phn & 9.66$\pm$0.138\phn &  \nodata & 101 \\
NGC0095  &  74.9 &  884.1$\pm$7.3\phn & 214.79$\pm$4.73\phn & 0.19$\pm$0.053\phn & 2.20$\pm$0.154\phn & 5.28$\pm$0.476\phn & 36.7$\pm$2.0\phn & 103 \\
NGC0157  &  24.0 & 4311.4$\pm$24.4\phn & 1014.76$\pm$15.10\phn & 2.17$\pm$0.042\phn & 17.93$\pm$0.048\phn & 42.43$\pm$0.103\phn & 171.0\tablenotemark{a} & 101 \\
NGC0245  &  56.5 & 1262.9$\pm$9.0\phn & 296.42$\pm$5.59\phn & 0.56$\pm$0.090\phn & 4.22$\pm$0.295\phn & 8.68$\pm$0.521\phn & 38.2$\pm$1.9\phn & 103 \\
IC0051   &  24.8 & 1470.1$\pm$8.5\phn & 358.08$\pm$5.52\phn & 0.20$\pm$0.046\phn & 2.21$\pm$0.133\phn & 4.69$\pm$0.328\phn & 25.3$\pm$1.5\phn & 103 \\
NGC0278  &  12.1 & 6362.0$\pm$17.1\phn & 1579.58$\pm$10.59\phn & 2.65$\pm$0.021\phn &  25.03$\pm$0.040\phn &  44.46$\pm$0.418\phn & 138.0\tablenotemark{a} & 101 \\
NGC0337  &  23.5 & 3800.6$\pm$17.3\phn & 1096.79$\pm$12.22\phn & 0.76$\pm$0.050\phn & 9.07$\pm$0.043\phn & 20.11$\pm$0.387\phn & 106.5$\pm$3.9\phn & 101 \\
IC1623   &  84.1 & 2764.6$\pm$5.8\phn & 666.57$\pm$2.79\phn & 3.65$\pm$0.050\phn & 22.93$\pm$0.062\phn & 31.55$\pm$0.113\phn & 248.5$\pm$9.8\phn & 101 \\
MCG-03-04-014 & 139.9 & 490.0$\pm$3.4\phn & 86.62$\pm$1.48\phn & 0.90$\pm$0.036\phn & 7.25$\pm$0.060\phn & 10.33$\pm$0.136\phn & 43.8$\pm$1.4\phn & 101 \\
NGC0695   &  135.4 & 1002.5$\pm$4.1\phn & 183.09$\pm$2.01\phn & 0.83$\pm$0.041\phn & 7.59$\pm$0.031\phn & 13.56$\pm$0.167\phn & 74.8$\pm$3.1\phn & 101 \\
\enddata
\tablecomments{Table 2 is published in its entirety in the electronic edition of the
Astrophysical Journal.}
\tablenotetext{a}{Radio fluxes are obtained from Condon (1987). No uncertainties of the fluxes was given in that paper.}
\tablerefs{
(101)Sanders et al. 2003; (102) Soifer et al. 1989; (103) Moshir et al. 1990}
\end{deluxetable}
\begin{deluxetable}{lcccccccc}
\tabletypesize{\scriptsize}
\rotate
\tablecolumns{9}
\tablewidth{0pt}
\tablecaption{Measurements of Centers of SINGS Galaxies}
\tablehead{
\colhead{Name} &  \colhead{H$\alpha$\tablenotemark{a}}
& \colhead{H$\beta$\tablenotemark{a}} & \colhead{[OII]\tablenotemark{a}} & \colhead{S$_\nu$(3.6~$\mu{\rm m}$)} &
\colhead{S$_\nu$(8~$\mu{\rm m}$)}
& \colhead{S$_\nu$(24~$\mu{\rm m}$)} \\
\colhead{} & \colhead{} & \colhead{} & \colhead{} &
\colhead{(Jy)} & \colhead{(Jy)} & \colhead{(Jy)} \\
\colhead{(1)} & \colhead{(2)} & \colhead{(3)} & \colhead{(4)} & \colhead{(5)} &
\colhead{(6)} & \colhead{(7)} }
\startdata
NGC0024 & 91.5$\pm$2.7\phn & 31.15$\pm$2.15\phn & 74.1$\pm$6.7\phn & 0.0127$\pm$0.0013\phn & 0.0194$\pm$0.0019\phn & 0.0167$\pm$0.0007\phn \\
NGC0337 & 552.6$\pm$4.3\phn & 153.60$\pm$2.94\phn & 318.5$\pm$8.2\phn & 0.0198$\pm$0.0020\phn & 0.0858$\pm$0.0086\phn & 0.1567$\pm$0.0063\phn \\
NGC0628 & 69.0$\pm$2.8\phn & 18.37$\pm$2.03\phn & 10.1$\pm$4.2\phn & 0.0382$\pm$0.0038\phn & 0.0461$\pm$0.0046\phn & 0.0433$\pm$0.0017\phn \\
NGC0855 & 379.3$\pm$3.3\phn & 114.83$\pm$2.32\phn & 292.5$\pm$6.7\phn & 0.0179$\pm$0.0018\phn & 0.0325$\pm$0.0033\phn & 0.0618$\pm$0.0025\phn \\
NGC0925 & 281.8$\pm$3.1\phn & 84.67$\pm$2.19\phn & 184.6$\pm$6.1\phn & 0.0122$\pm$0.0012\phn & 0.0365$\pm$0.0037\phn & 0.0355$\pm$0.0014\phn \\
NGC1097 & 2626.5$\pm$17.8\phn & 514.28$\pm$10.15\phn & 357.2$\pm$18.0\phn & 0.2725$\pm$0.0272\phn & 1.1612$\pm$0.1161\phn & 3.2867$\pm$0.1315\phn \\
NGC1482 & 726.8$\pm$4.6\phn & 104.88$\pm$2.30\phn & 107.1$\pm$5.9\phn & 0.1559$\pm$0.0156\phn & 1.3957$\pm$0.1396\phn & 3.3599$\pm$0.1344\phn \\
NGC1512 & 423.1$\pm$5.4\phn & 108.28$\pm$4.16\phn & 98.4$\pm$9.4\phn & 0.0703$\pm$0.0070\phn & 0.1163$\pm$0.0116\phn & 0.1496$\pm$0.0060\phn \\
NGC1705 & 816.7$\pm$4.7\phn & 282.53$\pm$3.88\phn & 538.7$\pm$11.0\phn & 0.0121$\pm$0.0012\phn & 0.0084$\pm$0.0009\phn & 0.0219$\pm$0.0009\phn \\
NGC2403 & 92.5$\pm$2.9\phn & 34.05$\pm$2.50\phn & 52.8$\pm$5.5\phn & 0.0267$\pm$0.0027\phn & 0.0415$\pm$0.0041\phn & 0.0403$\pm$0.0016\phn \\
DDO053 & 110.6$\pm$1.7\phn & 39.30$\pm$1.25\phn & 88.5$\pm$4.5\phn & 0.0006$\pm$0.0001\phn & 0.0017$\pm$0.0002\phn & 0.0177$\pm$0.0007\phn \\
NGC2798 & 1031.4$\pm$5.5\phn & 191.72$\pm$2.87\phn & 205.6$\pm$6.1\phn & 0.0826$\pm$0.0083\phn & 0.5908$\pm$0.0591\phn & 2.4276$\pm$0.0971\phn \\
NGC2915 & 743.7$\pm$6.5\phn & 282.68$\pm$7.10\phn & 736.1$\pm$27.1\phn & 0.0128$\pm$0.0013\phn & 0.0092$\pm$0.0009\phn & 0.0285$\pm$0.0011\phn \\
NGC2976 & 235.6$\pm$2.8\phn & 76.16$\pm$2.08\phn & 147.0$\pm$5.8\phn & 0.0175$\pm$0.0018\phn & 0.0431$\pm$0.0043\phn & 0.0886$\pm$0.0036\phn \\
NGC3049 & 648.0$\pm$4.0\phn & 171.30$\pm$2.44\phn & 220.5$\pm$6.6\phn & 0.0129$\pm$0.0013\phn & 0.0817$\pm$0.0082\phn & 0.3646$\pm$0.0146\phn \\
M81DwB & 74.4$\pm$2.1\phn & 24.73$\pm$1.48\phn & 57.7$\pm$5.5\phn & 0.0011$\pm$0.0001\phn & 0.0007$\pm$0.0001\phn & 0.0026$\pm$0.0001\phn \\
NGC3184 & 159.1$\pm$2.5\phn & 37.38$\pm$1.78\phn & 31.0$\pm$4.7\phn & 0.0215$\pm$0.0022\phn & 0.0514$\pm$0.0051\phn & 0.1181$\pm$0.0047\phn \\
IC2574 & 82.4$\pm$2.3\phn & 26.61$\pm$1.78\phn & 69.4$\pm$6.1\phn & 0.0011$\pm$0.0001\phn & 0.0012$\pm$0.0002\phn & 0.0022$\pm$0.0001\phn \\
NGC3265 & 341.1$\pm$3.1\phn & 79.58$\pm$1.98\phn & 114.1$\pm$5.0\phn & 0.0208$\pm$0.0021\phn & 0.0935$\pm$0.0093\phn & 0.2726$\pm$0.0109\phn \\
Mrk33 & 1545.6$\pm$6.3\phn & 445.26$\pm$3.49\phn & 989.0$\pm$9.8\phn & 0.0197$\pm$0.0020\phn & 0.1190$\pm$0.0119\phn & 0.8359$\pm$0.0334\phn \\
NGC3351 & 1810.3$\pm$7.7\phn & 441.89$\pm$5.39\phn & 219.5$\pm$12.4\phn & 0.1507$\pm$0.0151\phn & 0.4430$\pm$0.0443\phn & 1.5081$\pm$0.0603\phn \\
NGC3773 & 782.7$\pm$3.9\phn & 250.70$\pm$2.44\phn & 560.3$\pm$7.1\phn & 0.0110$\pm$0.0011\phn & 0.0393$\pm$0.0039\phn & 0.1271$\pm$0.0051\phn \\
NGC3938 & 104.0$\pm$3.0\phn & 27.86$\pm$2.34\phn & 10.6$\pm$5.7\phn & 0.0308$\pm$0.0031\phn & 0.0471$\pm$0.0047\phn & 0.0495$\pm$0.0020\phn \\
NGC4254 & 427.7$\pm$4.6\phn & 87.60$\pm$3.22\phn & 45.8$\pm$6.0\phn & 0.0786$\pm$0.0079\phn & 0.2545$\pm$0.0255\phn & 0.3373$\pm$0.0135\phn \\
NGC4321 & 996.3$\pm$7.9\phn & 213.21$\pm$5.30\phn & 126.2$\pm$11.8\phn & 0.1015$\pm$0.0102\phn & 0.4269$\pm$0.0427\phn & 0.7201$\pm$0.0288\phn \\
NGC4536 & 1004.8$\pm$5.7\phn & 147.05$\pm$2.87\phn & 176.6$\pm$6.4\phn & 0.1399$\pm$0.0140\phn & 0.9692$\pm$0.0969\phn & 2.6835$\pm$0.1073\phn \\
NGC4559 & 180.9$\pm$3.0\phn & 48.86$\pm$2.16\phn & 83.1$\pm$5.8\phn & 0.0217$\pm$0.0022\phn & 0.0582$\pm$0.0058\phn & 0.0627$\pm$0.0025\phn \\
NGC4625 & 155.1$\pm$2.3\phn & 43.47$\pm$1.66\phn & 58.6$\pm$4.7\phn & 0.0127$\pm$0.0013\phn & 0.0399$\pm$0.0040\phn & 0.0393$\pm$0.0016\phn \\
NGC4631 & 236.4$\pm$3.2\phn & 54.89$\pm$1.99\phn & 157.2$\pm$5.5\phn & 0.1039$\pm$0.0104\phn & 0.5595$\pm$0.0560\phn & 1.0009$\pm$0.0400\phn \\
Tol89 & 1013.0$\pm$5.2\phn & 318.30$\pm$3.45\phn & 657.3$\pm$13.2\phn & 0.0051$\pm$0.0005\phn & 0.0044$\pm$0.0004\phn & 0.0055$\pm$0.0002\phn \\
NGC5713 & 629.2$\pm$3.4\phn & 138.70$\pm$2.18\phn & 159.7$\pm$5.6\phn & 0.0596$\pm$0.0060\phn & 0.4412$\pm$0.0441\phn & 1.2356$\pm$0.0494\phn \\
NGC6946 & 887.4$\pm$8.7\phn & 109.12$\pm$5.73\phn & 29.8$\pm$15.5\phn & 0.1779$\pm$0.0178\phn & 1.5100$\pm$0.1510\phn & 4.8580$\pm$0.1943\phn \\
NGC7793 & 343.7$\pm$4.6\phn & 98.16$\pm$4.26\phn & 140.8$\pm$10.6\phn & 0.0219$\pm$0.0022\phn & 0.0614$\pm$0.0061\phn & 0.0715$\pm$0.0029\phn \\
\enddata
\tablenotetext{a}{The fluxes of H$\alpha$, H$\beta$ and [OII] are in units of $10^{-15}$~ergs~s$^{-1}$~cm$^{-2}$.}
\end{deluxetable}
\begin{deluxetable}{lccccccc}
\tabletypesize{\scriptsize}
\tablecolumns{8}
\tablewidth{0pc}
\tablecaption{Summary of Coefficients}
\tablehead{
\colhead{} & \multicolumn{3}{c}{adopted} &  \colhead{}  & \multicolumn{3}{c}{dispersion} \\
\cline{2-4} \cline{6-8} \\
\colhead{Relation} & \colhead{SINGS} & \colhead{MK06} &
\colhead{SINGS+MK06} &  \colhead{}  & \colhead{SINGS} & \colhead{MK06} &
\colhead{SINGS+MK06}}
\startdata
L(H$\alpha)_{\rm obs}$ + a*L(24~$\mu$m) & 0.015$\pm$0.004\phn & 0.021$\pm$0.005\phn & 0.020$\pm$0.005\phn & & 0.140 & 0.108 & 0.119 \\
L(H$\alpha)_{\rm obs}$ + a*L(TIR) & 0.0020$\pm$0.0005\phn & 0.0025$\pm$0.0006\phn & 0.0024$\pm$0.0006\phn &  & 0.131 & 0.067 & 0.089 \\
L(H$\alpha)_{\rm obs}$ + a*L(8~$\mu$m) & 0.010$\pm$0.003\phn & \nodata & 0.011\tablenotemark{b}$\pm$0.003\phn & & 0.112 & \nodata & 0.109 \\
L(H$\alpha)_{\rm obs}$ + a*L$_{\rm 1.4GHz}$ & 0.41$\pm$0.13\phn & 0.39$\pm$0.10\phn & 0.39$\pm$0.10\phn & & 0.146 & 0.087 & 0.099 \\
L([OII])$_{\rm obs}$ + a*L(24~$\mu$m) & \nodata & 0.029$\pm$0.005\phn & \nodata & & \nodata & 0.136 &  \nodata \\
L([OII])$_{\rm obs}$ + a*L(TIR) & \nodata & 0.0036$\pm$0.0006\phn & \nodata & & \nodata & 0.089 & \nodata \\
L([OII])$_{\rm obs}$ + a*L(8~$\mu$m) & 0.016\tablenotemark{c} & \nodata & \nodata &  & 0.159 & \nodata & \nodata \\
L([OII])$_{\rm obs}$ + a*L$_{\rm 1.4GHz}$ & \nodata & 0.54$\pm$0.10\phn & \nodata & & \nodata & 0.122 & \nodata \\
\enddata
\tablenotetext{b}{This coefficient is derived based on the 20''X20'' centers and the entire galaxies of SINGS sample.}
\tablenotetext{c}{Because the number of data points with both [OII] and 8$\mu$m flux measurements is small, this coefficient
was not derived by fitting to the data. Instead it was obtained from the combinations of H$\alpha$ and 24~$\mu$m, H$\alpha$
and 8~$\mu$m and [OII] and 24~$\mu$m.}
\end{deluxetable}

\vfill
\end{document}